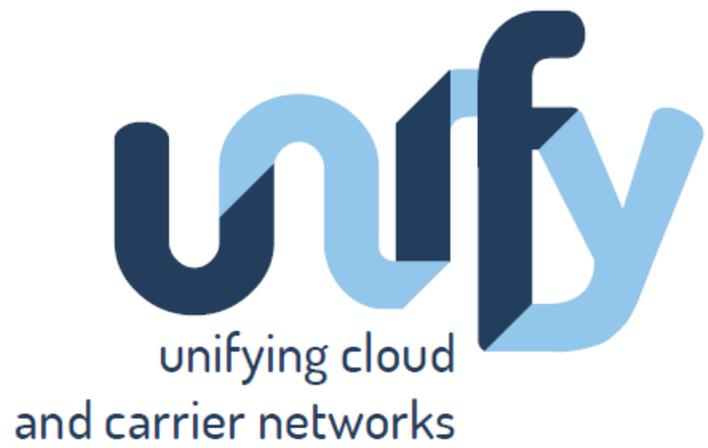

unifying cloud
and carrier networks

# D4.3

Updated concept and evaluation results for SP-DevOps

| | |
|---|---|
| Dissemination level | PU |
| Version | 1.2 |
| Due date | 30.04.2016 |
| Version date | 14.10.2016 |

This project is co-funded
by the European Union

# Document information

## Editors and Authors:

Editors: Guido Marchetto (Polito), Riccardo Sisto (Polito), Wolfgang John (EAB)

Contributing Partners and Authors:

| | |
|---|---|
| ACREO | Pontus Sköldström, Bertrand Pechenot |
| BME | Felicián Németh, István Pelle |
| DTAG | Juhoon Kim |
| EAB | Xuejun Cai, Chunyan Fu, Wolfgang John, Catalin Meirosu |
| EICT | Kostas Pentikousis |
| iMinds | Sachin Sharma |
| OTE | Ioanna Papafili |
| POLITO | Guido Marchetto, Riccardo Sisto, Serena Spinoso, Matteo Virgilio |
| SICS | Rebecca Steinert, Per Kreuger, Shaoteng Liu, Jan Ekman |
| TI | Antonio Manzalini |
| TUB | Apoorv Shukla |

## Project Coordinator

Dr. András Császár

Ericsson Magyarország Kommunikációs Rendszerek Kft. (ETH) AB

KONYVES KALMAN KORUT 11 B EP

1097 BUDAPEST, HUNGARY

Fax: +36 (1) 437-7467

Email: andras.csaszar@ericsson.com

## Project funding

7th Framework Programme

FP7-ICT-2013-11

Collaborative project

Grant Agreement No. 619609

# Table of contents

## List of figures

## List of tables

## Abstract

This document reports on the results of the SP-DevOps related activities in the UNIFY project. In particular, it has a first focus on the final definition and assessment of the SP-DevOps concept. The SP-DevOps concept is realized by a combination of various functional components facilitating integrated service verification, efficient and programmable observability, and automated troubleshooting. Our assessment shows that SP-DevOps can help providers to reach a medium level of DevOps maturity (level 3 on a scale from 1-5) and a reduction in OPEX with respect to time investments by 32%-77% by applying our technical SP-DevOps solutions. Secondly, this document focuses on the evaluation of the proposed SP-DevOps tools. The set of tools proposed supports operators and developers across all stages, with a focus on the deployment, operation and debugging/troubleshooting phases. Another focus of this document is on automation of operation tasks, showing how the individual SP-DevOps tools and functions can activate automated processes and workflows for verification, observability and troubleshooting in software defined environments like NFV. Finally, concrete use cases are presented in order to showcase selected processes, which lead to prototyped demonstrators allowing a final evaluation of SP-DevOps in close-to-real scenarios.

## Reading guide for this report:

After the introduction and motivation of the SP-DevOps work in Section 1, readers which would like to learn more about the scope and previous results of UNIFY as a whole should continue with Section 2. However, readers with a main interest in the general ideas behind SP-DevOps are referred directly to Section 3, which gives a high-level overview of how and where in an NFV architecture SP-DevOps might solve specific problems and bring benefits in terms of OPEX savings. In particular, Section 3.2 acts as an index to Section 4 and points to the detailed descriptions and evaluation results for individual SP-DevOps tools in case they are of specific interest for the reader. In other words, Section 4 can be omitted by readers with only a general interest in the SP-DevOps concept as a whole. Section 5 puts the individual tools into a more concrete context by defining management processes serving different purposes, scopes and roles, while Section 6 describes the subset of SP-DevOps processes that have actually been implemented together with UNIFY use-cases. Theses sections are of relevance for readers who want to get a more concrete understanding of the SP-DevOps processes after the summarized concept in Section 3. Finally, the concluding section provides a summary together with the main highlights and lessons learned. To sum up, the core of this document consists of Section 1, 3 and 7. The large Section 4 acts as a catalog of the developed SP-DevOps tools, as indexed by Section 3.2. Section 5 and 6 are relevant for practitioners who would like to learn the details about the definitions and implementation of SP-DevOps processes SP-DevOps workflows and processes

# 1 Introduction

The current growing interest for SDN and NFV paradigms (actually re-proposing well-known principles as decoupling software from hardware and virtualization) is likely motivated by the novelty of the overall techno-economic context, which is making these paradigms really exploitable and sustainable with the required level of performance. In fact, in the last two decades we have witnessed a number of drivers, maturing and converging: for example, the impressive diffusion of fixed and mobile ultra-broadband, the increasing performance of chipsets and IT/hardware architectures, the ever-growing availability of open source software, the availability of more and more powerful terminals and the overall costs reductions (determined also by a shift in how IT services are provided).

It is argued that thanks to these drivers and due to the emerging techno-economic context, SDN and NFV will soon impact not only current telecommunications networks (in all the segments, core, distribution, access) but also service and application platforms [1] [2] [3] [4]. In fact, SDN and NFV, together with Cloud, Edge and Fog Computing, can be seen as facets of a broad innovation wave (called Softwarization) accelerating the on-going migration of "intelligence" towards the Users Equipment (UE).

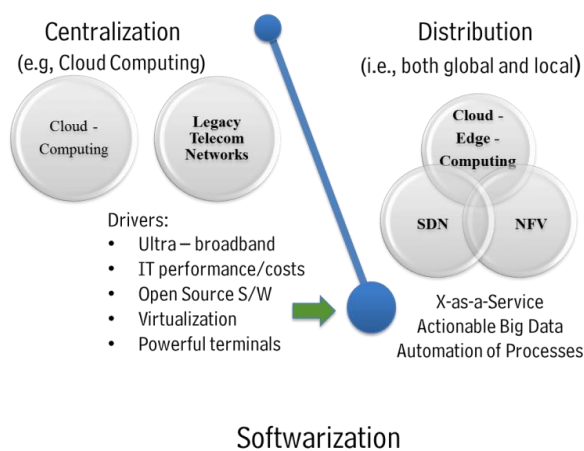

*Figure 1 Softwarization as a systemic trend in Telecommunications and ICT.*

Softwarization will be a radical change of paradigm [5]. Current telecommunications infrastructures have been exploited with purpose-built equipment designed for specific functions. In the future, network functions and services will be virtualized software processes executed on distributed horizontal platforms mainly made of standard hardware resources.

Softwarization of Telecoms will accelerate the substantial convergence process towards infrastructures supporting a wide variety of virtualized network and service functions. This is likely to bring to a split between Operators owning and operating the converged infrastructure and the Service Providers. Retailing of traditional telecoms services is going disappear, becoming said services packaged with other ones (e.g., voice with Internet access, premium TV). Telecoms retailing is likely to join with OTT, while there will be the merging between the wholesale Telecoms Providers and Data Centre hosting.

Control, Orchestration, Management and Policy (COMP) will require solutions based on agile approaches such as DevOps allowing to deal with a sheer number of software processes, rather than closed pieces of equipment. In this sense it is strategically important to extend the COMP capabilities across domains (e.g., access and distribution networks), both fixed and mobile. In view of this transformation, industries will have to reshape themselves and turn into ecosystems that rely on large horizontal platforms. A growing number of these industries will as a consequence seek to adopt "agile" operation approaches inspired by DevOps principles, capable of hosting a wide variety of modes of production by involving start-ups, social enterprises and communities in creating user-generated content.

In UNIFY, we took a first step towards agile operations of network services and infrastructure with the development of the SP-DevOps concept. The SP-DevOps concept collectively refers to a set of technical components and automated workflow definitions in support of integrated and automated operations tasks. The final results documented in this report are a natural continuation of the guidelines described in our intermediate deliverable D4.2 [6]. In this final report, we have given particular focus to the enrichment of specific monitoring, troubleshooting, and verification processes, with emphasis on automation. Automation is in fact a must for a proper realization of DevOps, as functions and tools should be able to operate without or limited human intervention, clearly for the sake of feasibility and scalability. These processes are strictly related to the programmability opportunities and the orchestration procedures described in D3.3 [7].

The main purpose of this document is to report on the final SP-DevOps concept and the resulting SP-DevOps tools. We put specific focus on automation, showing how tools and functions can activate automated processes for monitoring, verification, and troubleshooting in software defined infrastructures, exposed to both Developer and Operator roles. In order to do that, we define activity workflows showing the specific interactions between the available functions and/or the proper modules in charge of handling the provided data. SP-DevOps processes cover verification of service graph functionality integrated with the orchestration, software-defined monitoring approaches providing key quality indicators with minimal overhead, and developer-friendly troubleshooting through an automated debugging processes. We also highlight the benefits of the SP-DevOps concept in terms of operational time and cost savings, and sketch possible actions for a gradual exploitation plan.

Another main target of the document is to report on the final experimental activities. In particular, we cover recent developments, the integration work, and final evaluation results of the SP-DevOps concept and its implementing tools, also considering the final realization of the public SP-DevOps toolkit. The document also introduces some use cases to show real SP-DevOps operation. These are used as final demonstrations of the main SP-DevOps processes to further prove the validity and the effectiveness of both the concept and its realization in UNIFY.

The document is organized as follows. Section 2 describes the relevant developments in other UNIFY workpackages. Section 3 presents the SP-DevOps concept in its final form, considering its functional components and highlighting its benefits and some possible actions for a gradual exploitation plan. Section 4 contains details about each individual SP-DevOps tool and function. This includes a summary of the research problems and resulting concepts behind the tools, as well as the final evaluation results. Section 5 covers the final definition and description of SP-DevOps processes, highlighting how the individual SP-DevOps tools can be combined to create automated processes with a wide range of scopes and purposes. Section 6 reports on the use-case scenarios and prototype implementations used to demonstrate one selected process for each category of verification, observability, and troubleshooting. Finally, Section 7 concludes the document, also highlighting some possible future work.

# 2 Overview of UNIFY and relevant developments in other WPs

As discussed above, the overarching goal of UNIFY is to break up traditional, rigid network control approaches, which limit the flexibility and manageability of speedy service creation [8]. For instance, it is not unusual that an average service creation time cycle exceeds 90 hours, whereas given recent advances in virtualization and Softwarization (including technology advances such as SDN and NFV), one would be interested in service creation times in the order of minutes [9] if not seconds. The UNIFY project pursues full network and service virtualization to enable rich and flexible services and operational efficiency, thereby overcoming today's limits in flexibility of service creation. In four technical work packages, we research, develop and evaluate means to orchestrate, verify, observe and deliver end-to-end services from home and enterprise networks through aggregation and core networks to data centres. While focusing on enablers of this unified production environment, we

- develop an automated, dynamic service creation platform, leveraging software defined networking technologies;

- create a service abstraction model and a service creation language, enabling dynamic and automatic placement of networking, computing and storage components across the unified infrastructure;

- develop an orchestrator with novel optimization algorithms to ensure optimal placement of elementary service components across the unified infrastructure;

- research new management technologies (Service Provider DevOps (SP-DevOps)) and operation schemes to address the dynamicity and agility of new services;

- and evaluate the applicability of a universal network node based on commodity hardware to support both network functions and traditional data center workloads.

Before this deliverable will document the main results of SP-DevOps, we use the following subsections to briefly summarize the results of the other parallel activities performed in UNIFY.

## 2.1 Summary of Use Cases and Architecture

As part of these project wide activities, we first collected and categorized requirements for modern service creation from the perspective of service providers and consumers. Second, we designed the architecture of the UNIFY framework based on the discovered requirements. Third, we defined various use cases that sketch examples of creating network services or handling potential issues (e.g., scalability). Fourth, we integrated prototypes implemented by other partners to a large integrated demonstrator (IntPro), including the outcomes of the SP-DevOps prototyping activities (DevOpsPro). Finally, we performed a techno-economic analysis of the UNIFY framework. This section summarizes developments made within these activities in the order of the resulting deliverable.

D2.1 [10] defines the important design principles of the UNIFY framework and provides the initial proposal of the UNIFY architecture based on the review of the state-of-the-art with over 30 aspects that support the development of UNIFY, identification of 46 requirements collected from the technical work packages, and 3 wider use-case groups (including 11 detailed use-cases). Furthermore, D2.1 identifies 6 fundamental processes that describe the behavior of the UNIFY architecture, i.e., boot–strapping, programmability, verification, observability, troubleshooting, and VNF development. As the result of putting these identifications and definitions together, the initial proposal of the UNIFY architecture was described in two steps in D2.1, i.e., overarching architecture and a functional architecture.

D2.2 [11] mainly addresses the detailed interface definitions of the architecture. The deliverable describes use cases and presents the reference architecture by illustrating the main components, abstract interfaces and primitives, which reflect the priorities of design, functional and business requirements. In order to create a resilient, trustworthy and energy–efficient architecture supporting the complete range of services and applications, the UNIFY architecture supports the full flexibility of NFV, isolation among service requests (Section 3.2.7 in D2.2), policy enforcement per service consumer (Section 3.2.8 in D2.2), embedded monitoring services (Section 3.2.9 in D2.2) to provide enablers for Quality of Service (QoS) support, and service resilience at any level of the virtualization hierarchy.

For the validation of UNIFY's techno–economic efficiency, a simulation–based cost analysis in different business environments is performed. More precisely, the cost of creating a IP VPN service by network function chaining using the state-of-the-art solutions is assessed and to be documented in a forthcoming deliverable [12].

The project has developed and demonstrated a prototype integration platform that is built on a set of existing operating system and development tools such as Mininet, Click, and NETCONF and POX components, namely, the Extensible Service ChAin Prototyping Environment [13]. Currently, we are working towards implementing the integrated prototype (IntPro) and validating the cost efficiency of the UNIFY system, which will be documented in another forthcoming deliverable [14].

## 2.2 Summary of results on Service Abstraction, Programmability and Orchestration

One UNIFY workpackage focused on describing a programmability framework and defining the necessary functionalities at this framework. A subset of this framework and functionality is selected and implemented as the Service Programming, Orchestration and Optimization Prototype.

The work has started by analyzing the existing resource description languages, open source APIs, protocols such as OpenFlow, ForCES, NETCONF and languages such as Frenetic/NetCore, HILTI and Nettle to map their functionality to the initial framework in order to identify the missing parts. Based on this gap analysis, a service programmability framework with relevant process flows, interfaces, information models and orchestration functionality for the UNIFY architecture was proposed [15]. UNIFY programmability is focused on the interface between different layers in the architecture, i.e., the Service Layer and the Orchestration Layer. The heart of this programmability model is the Network Function–Forwarding Graph (NF–FG) which is used for i) exposing (virtualized) network and cloud resources and ii) mapping a service request to the exposed resources.

As a next step, the functional description and algorithms which were required for the UNIFY orchestration framework were identified [16]. These identified challenges were in different areas: 1) programmability interfaces and data models, 2) service decomposition, 3) orchestration algorithms, 4) VNF scalability and resiliency, and 5) a scalable orchestration framework.

The initial NF-FG model was updated based on the feedback from prototyping and the ongoing work of other projects. The service decomposition is the process of breaking down a high-level service into more elementary (atomic) blocks to better utilize the resources and simplify the construction of new services. A detailed analysis of pros and cons and possible trade-offs of service decomposition was performed in which the atomic blocks could be as: i) hardware, ii) executable-based software, or iii) Click modular router components. Initially a service request is represented as a service graph which in transformed to an NF-FG which needs to be mapped to the infrastructure resources. However, embedding an NF-FG on the available resources is not a trivial problem. Different approaches including optimal solutions (ILP-based algorithms in particular) and heuristic approaches with/without combining the embedding with the decomposition were proposed. Oriented towards elastic, scalable and resilient network services, a distributed state transfer mechanism was proposed which resulted in substantial gains in migration time and required bandwidth [16].

The programmability framework for the UNIFY architecture was updated to be in line with the other work packages, i.e. SP-DevOps and the Universal Node [7]. All relevant components of the service programming and orchestration framework have been implemented in ESCAPE, the main proof of concept prototype for orchestration [17]. ESCAPE relies on Mininet, Click, NETCONF and POX which implements all 3 layers of UNIFY architecture (i.e., Service Layer, Orchestration Layer and Infrastructure Layer). ESCAPE was designed for both intra and inter-workpackage integration which supports efficient integration of different modules and the easy re-use and integration of available tools. To have the integrated prototype, it supports the orchestration on top of Universal Node and the integration of SP-DevOps verification, monitoring and troubleshooting tools. A specific use case, Elastic Router, was selected to be demonstrated in ESCAPE while it is orchestrated in the Universal Node and different tools of verification, monitoring and troubleshooting are used [18].

## 2.3 Summary of Universal Node development and benchmarking

A final workpackage was focusing on the development of the Universal Node (UN) to enable the design of a flexible and easily programmable network. The UN aims to provide this by designing a software architecture optimized for executing dataplane-oriented workloads on standard high-volume servers, such as Intel-based blades. Depending on the specific use case, applications can either be executed on bare hardware or through lightweight virtualization primitives such as Linux Containers or Docker, as well as in a fully virtualized environment. This technical approach opens up wider possibilities not only for reducing OPEX and CAPEX but also for building up new services or reengineering existing ones with a far more flexible approach.

The work was started by assessing the requirements for the UN and the impact on the key aspects of the data plane with respect to network virtualization and resource sharing, as well as switching and traffic steering [19]. In parallel, the first functional specification of the UN data plane involving three modules for management, control and execution was introduced. A more detailed and mature UN architecture was presented early in the project in terms of more concrete definitions of architectural blocks and their interactions [20]. The key element of the UN architecture is the Universal Resource Manager (URM), where compute and networking resources are commonly managed. Discussions between the SP-DevOps and UN teams have included issues about the architectural support for locally deployed Monitoring functions (MFs) and access to various data sources (e.g. counters) on the UN. This has led to an extension of the UN architecture in terms of a Monitoring Management Plug-in (MMP). The introduction of the MMP essentially enables the orchestrator to interact with the control plane of a particular MF and observability point (OP) through a uniform interface, which is crucial for achieving increased observability in a scalable manner. Section 4.2.3.1 of this deliverable includes a discussion of the architectural aspects of the MMP, while Section 6.2 provides an overview of the MMP implementation for the Elastic Router demo and describes how it is used for the observability process of the elastic router realized on a UN.

As a next step, UN prototyping implementation details and phases activities have been described [21]. In this context, the implementation details of the MMP have been discussed between SP-DevOps and UN teams as part of the work towards a project-wide integrated prototype with the aim of showcasing aspects of the SP-DevOps concept (i.e. Observability and Troubleshooting processes, see Section 5.3 and 5.4). The actual UNIFY Universal Node prototype, which finally implements the architecture defined in D5.2, was documented in D5.5 [22]. The prototype includes  as main components the local orchestrator with different computing and network plugins, the softswitch, and the northbound interface implementing the NF-FG as defined by the service abstraction work. The UN architecture is modular and supports different execution environments and different softswitches. This enables to exploit different technologies and to adapt to the necessity of different use cases. In the demo scenarios for SP-DevOps observability and troubleshooting processes (section 6.2), a UN version with Docker as VNF Execution environment and OVS for the virtual switching environment has been selected.

In terms of UN Benchmarking activities, there have been two major results during the final year of the project. First, an initial benchmarking documentation [23] presented the results of the UN performance characterization and development activities. Specifically, the initial results characterize the performance of the major individual building blocks of the UN, i.e. virtual switching engines, hypervisors, as well as first complete test setups. Second, a final deliverable was presenting UN performance characterization on the actual UN prototype [24].The results include benchmarks on VNFs realized by different virtualization methods (VMs, Containers, DPDK processes, native functions), different softswitch versions (OVS, ERFS; xDPd), different host/VNF interfaces (KNI, IVSHMEM, virtio, vhost-user, SR-IOV, PCI-PT), as well as VM-to-VM communication.

# 3 Final SP-DevOps concept

In this section we present the final SP-DevOps concept, evolved from the previous iterations documented in D4.1 [25], M4.1 [26] and D4.2 [6]. Recall that SP-DevOps derives from DevOps, a concept born by the fast software development practices of major software companies such as IBM and HP.

Through a number of tools and support functions, generally referred to as the SP-DevOps tools, the UNIFY SP-DevOps framework maintains the deployment and re-deployment cycle of production through four major processes, namely *VNF Developer support, Verification, Observability, and Troubleshooting.* These four processes are mapped on the general service lifecycle of UNIFY as shown in Figure 2. They are effectively dealing with the four crucial principles of the general DevOps concept (i.e., *monitoring and validation of the operational quality, development and testing against production-like systems, reliable and repeatable deployment,* and *amplifying feedback loops*) [4]. These four SPDevOps processes have been described in D4.1. Since then, however, less emphasis has been put on the *VNF developer support*, since we concluded that this process is to a large degree supported by a combination of the remaining three processes. We thus provide more detailed descriptions only for concrete Verification, Observability, and Troubleshooting processes supporting different purposes in Section 5, while actual implementations of these three process categories for selected use-cases are documented in Section 6.

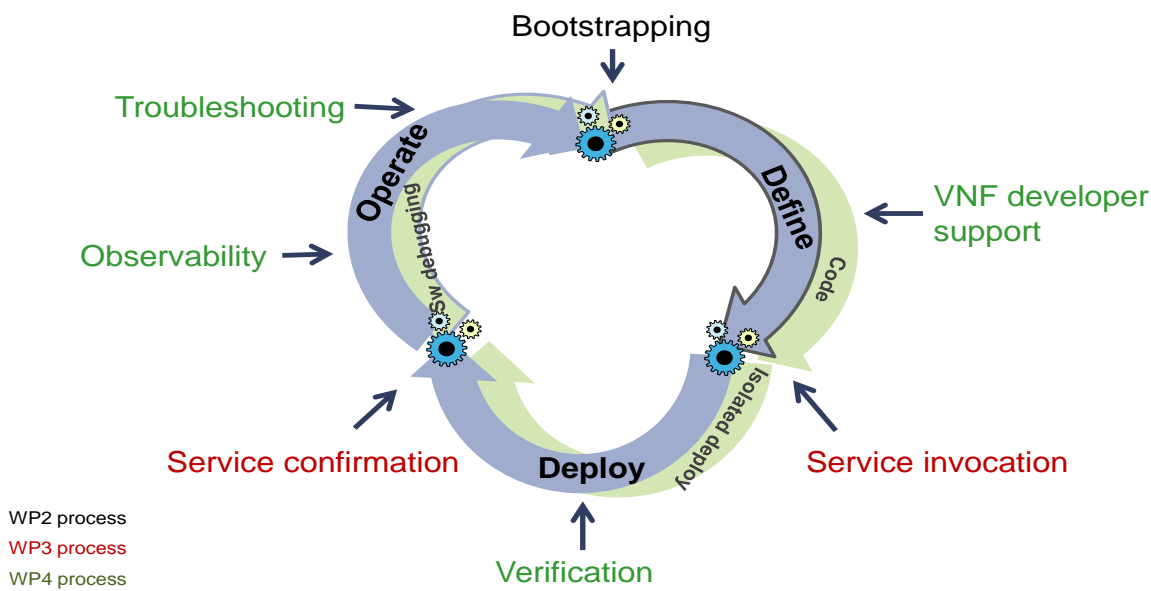

*Figure 2 SP-DevOps processes mapped on the UNIFY service lifecycle.*

In the following section, we will first map individual SP-DevOps tools onto the UNIFY architecture to indicate how integrated workflows through UNIFY deployment and orchestration functions and the SP-DevOps tools will naturally emerge. After a brief introduction of the five main categories of SP-DevOps functional components, we will discuss whether the final SP-DevOps concept reaches the desired DevOps maturity model level specified in [25]. Finally, the actual business benefits of applying SP-DevOps in a UNIFY scenario are assessed with the help of several recent surveys regarding operator business and operations practices.

## 3.1 UNIFY SP–DevOps concept

In this section, we will give an overview of how the SP–DevOps framework operates in the functional architecture presented in D2.2 [11] and describe the key components forming SP–DevOps processes. The UNIFY functional architecture is depicted in Figure 3, including service layer, (recursions of) orchestration layers, and infrastructure layer, which consists of domain specific controllers and the actual infrastructure resources (e.g. an SDN controller with OpenFlow switches; a Universal node with Local Orchestrator; or an OpenStack controlled datacentre).

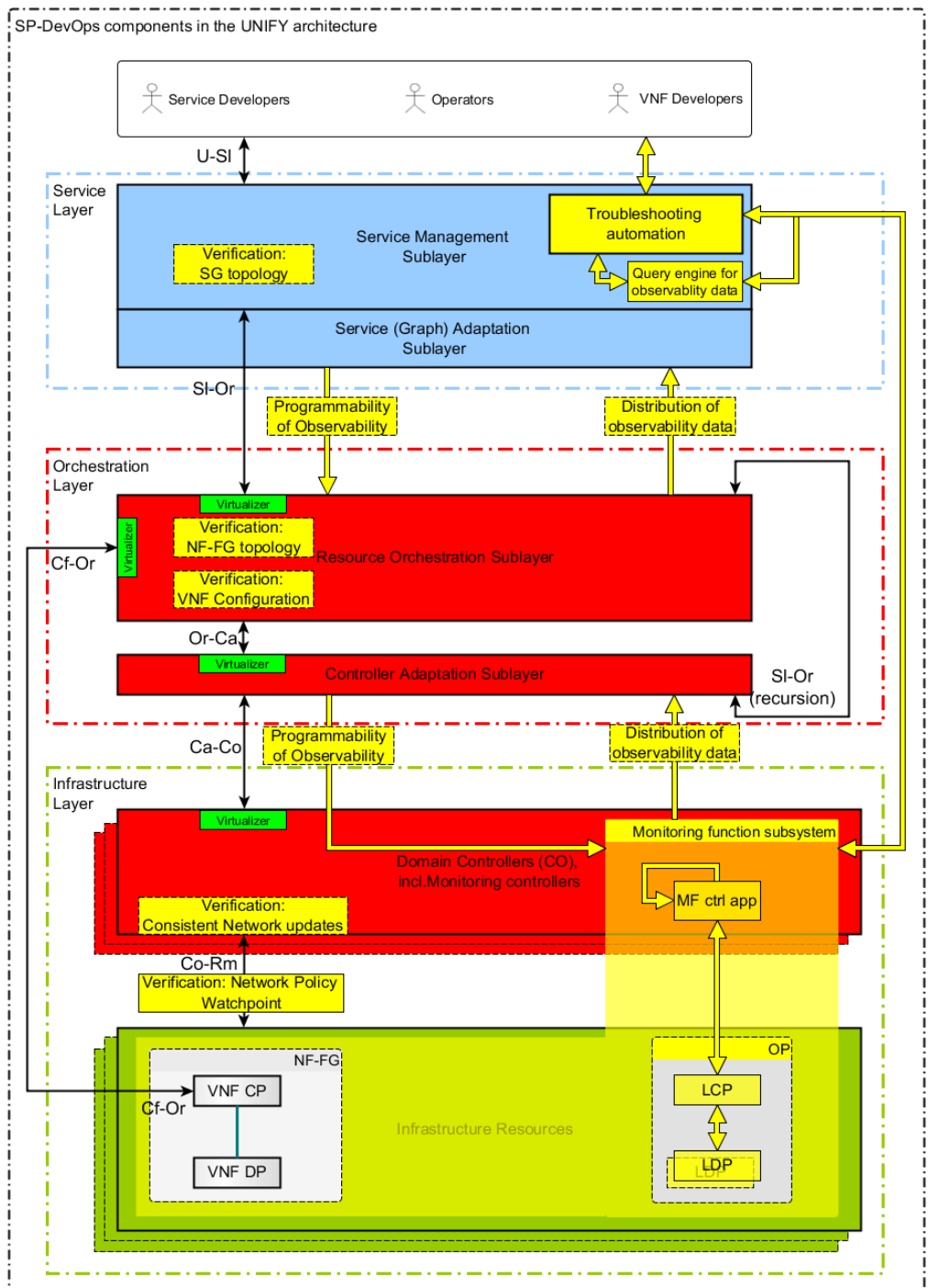

*Figure 3 Architectural mapping of functional components of SP–DevOps (yellow) on the UNIFY architecture.*

In this figure, we superposed the architecture with SP-DevOps components in yellow colour, to give a global overview of their architectural mapping and their interaction within a service lifecycle. Verification functions for service programming are integrated in all layers above the infrastructure resources, depicted by the yellow boxes on the left hand side of the figure. Cross layer programmability features for observability are realized as annotations to the NF-FGs definition crossing the interfaces Sl-Or and Cf-Or, as well as Ca-Co. Additionally, Service-, Orchestration- and Controller sublayers require integration of modules to handle these annotations alongside any manipulations done to the NF-FG, such as decomposition or scoping of VNFs (not shown explicitly in this figure). The Monitoring Function (MF) subsystem in the infrastructure layer as depicted on the lower right-hand side in the figure is realizing monitoring and verification functions for the actual infrastructure and VNF resources. Efficient distribution of observability data is supported by messaging and querying systems both within MFs, but also cross-layer, allowing dissemination of aggregated monitoring results all the way up to the service layer (depicted by yellow components on the right part of the figure). Finally, automated troubleshooting is supported by a modular framework which can instrument both UNIFY and 3rd party monitoring, verification, or debugging tools. In the figure, these supporting functions are depicted as service layer components, available for both developer and operator roles.

## 3.2 Functional components of UNIFY SP-DevOps

In the following, we will give a short overview of the main functional components of UNIFY SP-DevOps, grouped by 5 functional categories:

### 3.2.1 Integrated verification of service programming

Many important properties of services expressed as graphs can be verified before actual deployment, which is an essential task in environments where reconfiguration of services is expected to be triggered frequently. As in any development process, identification of problems early in the service lifecycle can significantly reduce times and costs spent on complicated debugging and troubleshooting processes. Misconfiguration of (V)NFs, violation of specified network policies, or artificial insertion of malicious network functions are just examples of issues that an automated services deployment solution should handle in order to ensure service uptime and preserve network integrity and reliability. In UNIFY SP-DevOps, functions supporting continuous verification of service programming instructions and configurations are integrated into all layers above the actual infrastructure resources.

*SG and NF-FG pre-deployment verification*: In the Service Layer (SL), a module verifies the Service Graph (SG) topology against basic reachability properties. In the Orchestration Layer (OL), first an NF-FG topology verification step takes place in order to guarantee that required topology properties are still satisfied after the translation of the SG into the richer NF-FG representation. This basic verification step if followed by a novel verification of the actual VNF configuration, which considers reachability properties in conjunction with all the constraints imposed by the complete model of the network, including each VNF forwarding behavior and its configuration. More details about the verification modules integrated into SL and OL for pre-deployment verification can be found in Section 4.1.

*Consistent network updates*: For the networking part, the verification toolset includes also controller-level verification functionality ensuring consistent network updates. Specifically, policy updates from Controllers to OpenFlow switches are guaranteed to experience synchronicity and consistency, thereby avoiding out of order updates and ensuring waypoint enforcement and weak loop freedom. More details about our consistent network update algorithms can be found in the section about verification of path functionality (Section 4.2)

*Network policy watchpoint:* UNIFY SP-DevOps proposes a watchpoint function supporting network policy enforcement by intercepting control traffic events on a Co-Rm interface realized by OpenFlow. The Watchpointer tool is able to catch policy violations (e.g. in the form of conflicting OpenFlow rules) and trigger various actions, such as notifications, packet drops, etc. Note that the watchpoint tool can also be used as a debugging and troubleshooting tool for network configuration. The Watchpointer tool has been demonstrated in the first-year UNIFY review as well as at IM'15 [27]. It is documented in its final form in D4.2 [6].

### 3.2.2 Programmability of observability

A crucial part of SP-DevOps is to support programmability of monitoring and verification functions across the UNIFY architecture. Specifically, UNIFY makes two contributions in this respect: a generic, high-level programming interface of programming observability features for software defined infrastructures; and a standardized data-model for common interfacing of legacy OAM tools within the infrastructure resources.

*Cross-layer programmability of observability.* These functions and their interface definitions enable automated deployment and configuration of MFs by translating high-level KPI and requirement definitions into concrete MF instances including placement, configuration and data aggregation. Specifically, we propose MEASURE for expressing measurement intents, states and reactions as annotation to NF-FG definitions on Sl-Or, Ca-Co and Cf-Or interfaces. The MEASURE language and the respective handling functions in the various architecture layers are described in Section 4.3.

*Programmability of legacy OAM tools.* As the adoption of programmable nodes will increase over time, the transition from current infrastructures to software-defined infrastructures and networking requires interfaces capable of bridging the gap between legacy equipment and programmable network entities. In this respect, most legacy devices (router, switches, etc.) have some OAM capabilities like BFD, LSP Ping, TWAMP, etc., that are usually configured rather statically via proprietary interfaces, which limits flexible and dynamic instantiation across diverse infrastructures. As part of UNIFY, we therefore have been involved in a standardization action in order to define a common TWAMP data model as an instance of standardized support for increased programmability of legacy OAM tools. More details on the data model and the mapping of TWAMP into the UNIFY monitoring function concept can be found in Section 4.4.

### 3.2.3 Infrastructure and VNF Monitoring and Verification

In UNIFY, we refer to monitoring and verification capabilities for infrastructure resources and VNFs collectively as Monitoring Functions (MFs). In the spirit of DevOps, these functions can be instrumented by both operator and developer roles not only for continuous observability, but also for troubleshooting purposes.

MFs typically implement functionality for collecting IT resource (e.g., CPU, memory, and storage) and network performance metrics (e.g., bandwidth, delay, jitter, and packet loss), but can also implement management tasks beyond pure monitoring, such as verification of configurations, etc. MFs may not only collect data, but also pre-process monitoring information (e.g., aggregate, filter, etc.) from other MFs across one or several infrastructure domains (e.g. UNs, SDN, or OpenStack domains). An MF subsystem is implemented as one or several Observability points (OP) and an MF control app. The following components form one MF (see Figure 3, Monitoring functions subsystem box):

- An MF control app is the software defined monitoring equivalent to SDN control apps, i.e., a logically centralized control application taking care of the configuration of one or multiple OPs and parts of the MF-related processing operations.

- An OP is a MF component that runs locally on an infrastructure node. In general, the implementation of OP capabilities encompasses measurement or verification mechanisms, node-local aggregation and analytics, depending on the type of the MF (for instance, node-local operations are not necessarily performed by all types of MFs). An OP operates in terms of the local control plane (LCP) and local data plane (LDP), and is managed by an MF control app.

  o An LCP is splitting certain control functionality from the MF control app for scalability and resource consumption purposes. It reflects essentially a local monitoring controller which provides functions for retrieving data from the LDP; processes obtained data; and controls the monitoring behavior (e.g., measurement intensity).

  o An LDP is basically any kind of data source in the UN (e.g., statistics from logical switches, resource metrics from the VNF executing environment, hardware counters, meters, injected packets, log data, etc.), retrievable from the virtualized environments (VNF execution environment (EE) or virtual switches environment (VSE)), the UN operating system, or specific monitoring VNFs.

UNIFY SP-DevOps provides a set of example implementations of MFs, including both observability as well as verification purposes. The tools and their purposes are briefly summarized below:

*Scalable and efficient network observability*: Introducing a split control between MF control app and LCP offers the operator a trade-off between monitoring fidelity on one hand and monitoring overhead on the other hand. We provide concrete examples of MFs utilizing LCPs in Sections 4.5 to 4.7 in the form of scalable network performance measurement tools, i.e. novel methods to estimate network rate, delay and loss metrics.

*Service specific QoS monitoring*: As an example of a service specific monitoring function, UNIFY SP-DevOps includes an IPTV service quality monitor tool (Section 4.8). The IPTV Quality Monitor extracts relevant parameters (e.g., video/audio codecs, bitrates, timestamps, etc.) from IPTV traffic in order to indicate the quality of the service in terms of packet losses, startup delay and estimated audio/video quality experienced by the user.

*Run-time network verification*: Since verification actions after the deployment are typically performed in the infrastructure layer, we realize run-time verification tools inline with the MF concept. One example of a MF with applicability in verification processes is AutoTPG (see Section 4.9). AutoTPG checks the correctness of flow tables by actively probing the FlowMatch part of OpenFlow Rules. Another example of a run-time verification tool is the path verification tool (Section 4.2), which applies efficient tagging and sampling mechanisms for verification of configuration and policies in the data plane during service operations time.

### 3.2.4 Cross-layer observability data distribution

In this subsection, we introduce support functions that allow monitoring data to be disseminated within the infrastructure domain as well as towards receivers in higher layers (service and orchestration layers) in an efficient manner, once observability and verification results are created.

*Fast and transport agnostic monitoring API*: UNIFY SP-DevOps proposed DoubleDecker (DD) as a messaging system to enable information exchange between LCPs within one domain and between LCPs and their MF control apps. In DD, clients only need to have a minimal knowledge of the actual architecture, but need to follow a standardized protocol to communicate with the messaging system and have to define their own sub-protocol to communicate with other clients. While it is based on the popular existing ZeroMQ, it provides several required carrier-grade features. More details about the technology choices behind the DoubleDecker messaging system developed in UNIFY can be found in Section 4.10.

*Recursive querying of monitoring data*: UNIFY SP-DevOps includes a query engine that interprets a specifically defined query language to efficiently collect and aggregate monitoring information from distributed software-defined infrastructure. This allows both developer and operator roles to query monitoring metrics on high level of abstraction, corresponding to NF-FG and SG abstractions in higher layers of the UNIFY architecture. The queries are resolved by the query engine in a recursive way (following NF-FG definition throughout the hierarchical UNIFY architecture) until primitive infrastructure metrics are available in actual monitoring databases residing in infrastructure domains. More details on the recursive query language can be found in Section 4.11.

### 3.2.5 Troubleshooting automation

This category includes a function that supports automated troubleshooting through the definition of troubleshooting workflows, thereby instrumenting virtually any type of UNIFY or 3[rd] party monitoring, verification, or debugging tools (including most tools and functions from the four categories above).

*Troubleshooting framework*: UNIFY's EPOXIDE is a troubleshooting framework that enables integrating and automating troubleshooting tasks by troubleshooting graphs. It enables the developer to define sequences of tools to be used for debugging particular problems with the network or simple virtual network functions, while having the results presented in the same terminal window. Details about EPOXIDE can be found in Section 4.12.

## 3.3 Assessment of the UNIFY SP-DevOps concept

With the UNIFY SP-DevOps concept, we set out to define a first set of tools to support specifically service providers in their adoption of DevOps principles. In the following subsections, we will first assess how the results of the SP-DevOps work at the end of the UNIFY project can be positioned on DevOps maturity models introduced in D4.1 [25]. We will then provide a model based on publicly available reports and numbers, allowing us to assess the potential savings to be gained by SP-DevOps assuming technologies as outlined in our research results would be hardened and implemented on a wide scale on software-defined infrastructure. Finally, we reason about required actions to realize gradual exploitation of SP-DevOps at Operators.

### 3.3.1 SP-DevOps maturity level

In D4.1, we outlined the ambition of the UNIFY SP-DevOps work to achieve an overall "Defined" level of the HP DevOps maturity model [28]. *Defined* is the 3rd level of this model, which describes five different levels of DevOps maturity. For each level, three dimensions are examined, i.e. Process maturity; Automation maturity; and Collaboration maturity. In summary, we argue that we have successfully shown the way for a Service Provider to achieve a *Defined* maturity level in terms of DevOps. Below, we provide our qualitative assessment by discussing what UNIFY SP-DevOps provides in terms of functional components and process definitions with respect to each of the three dimensions of the HP DevOps maturity model.

**Process maturity**:  In a *Defined* level, "Processes are well characterized and standardized across projects. These standard processes are used to establish consistency throughout the organization" and "the organization needs to be able to establish end-to-end release processes eliminating the silos, and avoiding the duplication of functions" [28]. In D4.1, we interpreted release management as ability that "enables successive improvements of service graphs in a UNIFY production environment. Such improvements result by defining new categories or classes of services that include certain constraints or policies by default, or changing the composition of the service graph to include newly-developed virtual functions".

The final UNIFY SP-DevOps concept supports this aim by:

- Components allowing integrated verification of service programming. Specifically, SP-DevOps provides pre-deployment verification of service chains integrated with orchestration, verification algorithms ensuring consistent network update, and network policy watchpoints (Section 3.2.1).
- Components allowing programmable observability, specifically automated deployment and configuration of monitoring and verification functions (Section 3.2.2).
- Advanced components for infrastructure and VNF Verification in form of novel run-time network verification tools (Section 3.2.3).

**Automation maturity:** In a *Defined* level, "there is central automated infrastructure that supports overarching enterprise processes built for the overall organization, instead of silo automations tailored for specific application or services, environments, or even tasks" [28]. In D4.1, we envision identical self-service one-click automated build, orchestration and deployment processes in all environments (development, test, production).

Eventually, the final UNIFY SP-DevOps concept supports this aim by:

- Components allowing programmable observability, specifically automated deployment and configuration of monitoring and verification functions (Section 3.2.2).
- Supporting functions for cross-layer and cross-environment data distribution of monitoring and verification results in an efficient manner (Section 3.2.4).
- A troubleshooting framework allowing both Dev and Ops personnel to create automated workflows (Section 3.2.5).

**Collaboration maturity**:  In a *Defined* level, "Collaboration is established between the teams. It becomes an essential part of the established processes and tool chains enabling idea sharing, better visibility and faster feedback.  All team members belong to a single system with shared accountability, frequent communication and mutual trust" [28]. In D4.1, we pointed out that in the scope of UNIFY we will address this area only through technical means, i.e. through a common tool chain by providing tools that could be used by cross-functional team members addressing both development and operations scope.

Finally, the UNIFY SP-DevOps concept supports this aim by:

- Components allowing programmable observability, specifically for automated observability workflows that can be defined for production environments (primarily Ops), but also test/debugging slices (primarily Devs) (Section 3.2.2).
- A troubleshooting framework allowing both Dev and Ops personnel to create automated workflows (Section 3.2.5).
- Set of tools with interfaces friendly for both Dev (mainly REST APIs) and Ops (mainly CLI and/or GUI), including integrated verification tools as well as infrastructure and VNF monitoring functions (Sections 3.2.1 and 3.2.3).
- Monitoring, troubleshooting and verification processes for various purposes, scopes and receivers (Sec. 5).

### 3.3.2 OPEX saving potential of SP-DevOps

A 2015 survey from PuppetLabs (arguably one of the most popular survey in the DevOps community with almost 5000 respondents worldwide) indicated that enterprises that are high performers in implementing DevOps would typically have the ability to recover from about 52% of the incidents in less than one day, but only about 15% of the incidents are recovered in less than 15 minutes [29]. This is in contrast with the low performing enterprises that are able to recover from only about 5% of incidents in less than 15 minutes and about 22% of incidents in less than a day. This is consistent with DZone's 2015 Continuous Delivery Survey [30] that reports mean time to recovery between 0-2 hours for 38% of the respondents, while the mean time between incidents was reported to be less than 24 hours for 25% of the 900 respondents. Given that we do not have inside knowledge on the PuppetLabs survey methodology, we observe that the characteristics indicated for high performance enterprises seem to be equivalent to the "Measured" maturity level from the HP model.

Today, very few telecom operators implement DevOps current infrastructures. However, due to the stringent focus on high-availability, incidents appear to be handled faster using current technology and operational practices. A HeavyReading survey [31] performed in 2015 on more than 70 operators showed that 54% of the respondents encounter "often" incidents that require less than 15 minutes for solving, while incidents with duration between 4 and 24 hours are "rarely" seen by 59% of the respondents.

Could the ongoing transformation, virtualizing many of the network functions, automating a significant portion of the operations and introducing some of the DevOps techniques in telecom networks improve the situation such that operators may answer "rarely" to 15-minute incidents and "never" to 4-24 hours-lasting incidents to such survey questions in the near future? On the other hand, a reduction of the recovering time of incidents is just one of the potential capabilities of DevOps for reducing outages.

The HeavyReading report [31] identifies network congestion as the second most common cause of outages and the most common cause of service degradations. Configuration issues are also occupying a forth position as most common cause of outages and is classified third when it comes to service degradations. 24% of the respondents indicate that significant costs are associated to dealing with congestion in their networks. Operational expenses to fix the problem is only the second most important component of the cost, while an increased rate of subscriber churn comes on the first position. This indicates that it is important to first prevent congestion and when it occurs that the situation is identified and fixed as soon as possible in order to minimize the number of affected subscribers.

The RateMon congestion predictor (Section 4.5) would address the congestion problem in a unified cloud and transport network infrastructure. When coupled with additional aggregation features of the UNIFY monitoring system as shown in our paper [32], the method makes it possible to predict (with a network overhead of only a few MBits/s) even short transient congestions that are difficult to detect otherwise on infrastructures that scale to hundreds of thousands of ports. Assuming that the unified infrastructure automation may be able to react to such congestion indication (for example, Ericsson mobile core network functions handle signalling storms to keep congestion under control [33], automated scaling as shown in the UNIFY Elastic Router (Section 6.2), etc.), significant operational costs and lost revenue could be saved.

An IT administrator of virtualized workloads in a data center environment is able to handle 3 – 10x [34] [35] more workloads or servers than in the case of an IT administrator working with physical servers. This is partly due to automation via DevOps tools such as Chef and Puppet and partly due to the inherent automation features that come with orchestration software such as OpenStack (for VMs) or Kubernetes (for containers). The combined WP3 and WP4 work in UNIFY shows the way operators may use orchestration and DevOps-inspired technology in their networks. There is no indication that the gains might be less, although due to the spatial distribution of the nodes it is likely that the increase in the number of nodes handles by an operations person might be at the middle range of the spectrum for the most remote nodes (mainly in the RAN) while typical datacenter rates are likely to apply for the mobile core network nodes as well as nodes that support Cloud RAN functions.

Alcatel-Lucent argued in [36] that the use of the Cloudband platform in conjunction with a virtualized network function would reduce the healing process by 88% on average for application and operating system failures compared to a traditional management process that involves many tools and several people to identify and solve the problem. When translating the scenario in a UNIFY environment, we observe that arguably similar scaling capabilities together with integration of SP-DevOps components for verification, programmable observability, and automated troubleshooting with orchestration are available. Furthermore, if we assume that a business interface between the Operator and Developer of the network function exists that allows our troubleshooting framework (Section 4.12) to be deployed and troubleshooting workflows executed, the solving of both application and OS failures could be performed significantly faster, probably in a matter of minutes instead of the hour-level intervals of Cloudband.

In the annual ENISA survey of significant telecom network incidents [37], 44% of the incidents with the Mobile Internet service were caused by software bugs, 19% by network overload, and 10% respectively by each of faulty software changes or updates and faulty policies or procedures. We use these percentages as well as the average duration of the incidents reported in the study for a modelling exercise that tries to determine the impact of SP-DevOps assuming technologies as outlined in our research results would be hardened and implemented on a wide scale on software-defined infrastructure. The data from the study is incomplete with respect to the precise root causes. Therefore, we assume in the following that the average duration of the incidents could be shortened between 10% and 80%. In D2.3, we include a discussion of the cost savings as part of the techno-economic analysis report. In this deliverable, we keep the discussion related to technical parameters. Figure 4 a) shows the baseline incident durations from the ENISA report and Figure 4 b) shows the results of our modelling of potential gains in terms of incident lifetime.

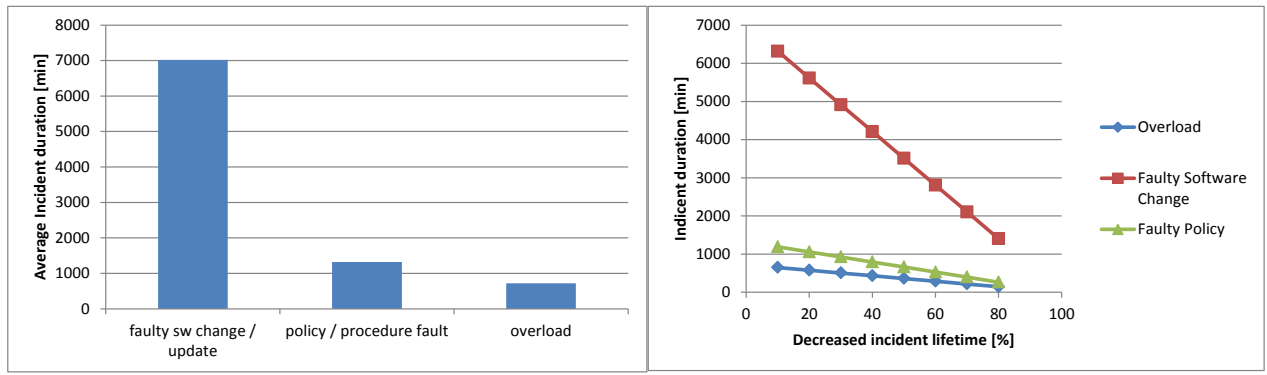

*Figure 4 (a – left): Average incident duration per category; (b-right): Potential gains per category.*

To estimate the potential for reducing the total amount of incidents, we define two mixes of savings per-type of incident. This corresponds to staged deployments (where gains would be partial and concentrated on the newer technology). It also allows us to discuss different assumptions with respect to how many of the incidents could actually be avoided by using the SP-DevOps technology. Not knowing for sure the exact cause of the incidents and what other characteristics (such as automatic handling of overload) might be implemented in the various network functions deployed it is very difficult to make definitive statements in this respect. However, the two scenarios we consider allow us to determine certain limits for the potential savings. The two mixes of savings are defined as follows:

*Conservative mix*: SP-DevOps could allow the operator to avoid 10% of the software bugs, 50% of the overload, 30% of the faulty software changes and 30% of the faulty policies

*Optimistic mix*: SP-DevOps could allow the operator to avoid 20% of the software bugs and 80% of each of the overload, faulty software changes and faulty policy incidents

Figure 5 shows a comparison of a normalized number of incidents in the original ENISA study [37] as well as our *Conservative* and *Optimistic* mixes.

We observe that the *Optimistic mix* might permit the operator to avoid about 50% of the incidents, while the Conservative mix might allow to avoid about 25% of the incidents. The remaining incidents in these three categories would also have their lifecycle shortened in line with the discussions from Figure 4.

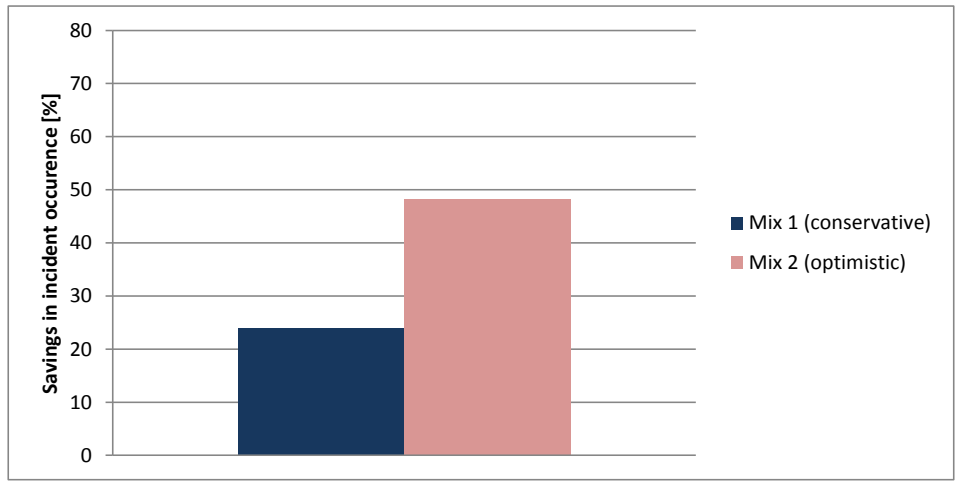

*Figure 5 Modelled savings in incident occurrence.*

We further define two models for mixing the amount of reduction in incident times in line with Figure 4:

- *Conservative mix*: 20% reduction on average incident duration for all overload, faulty software change and faulty policy and no time reduction for software bugs

- *Optimistic mix*: 60% reduction on average incident duration for faulty software changes, 80% reduction for overload and faulty policy incidents and 10% reduction for software bugs

Figure 6 shows results for the combined models for incident reductions both in terms of occurrence rate and duration. The conservative mix would provide about 32% savings, while the optimistic mix would translate onto about 77% of savings in terms of time spent handling incidents.

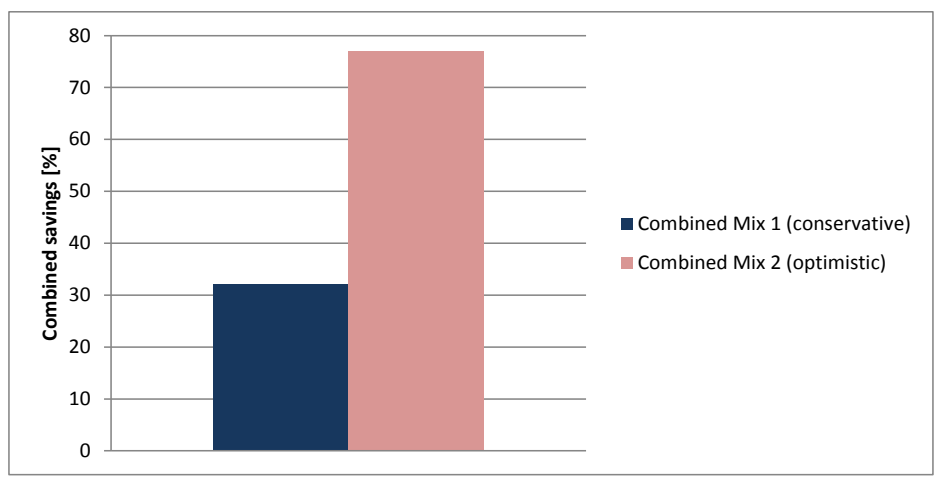

*Figure 6 Combined modelled savings in incident occurrence and duration.*

It should also be mentioned that. another area where SP-DevOps may attract growing attention is the increase of flexibility in service developments and provisioning and the consequent reduction of the Time To Market (TTM).

### 3.3.3 Gradual exploitation scenarios

Softwarization will be a radical Digital Business Transformation for Service Providers. It is likely that Softwarization will be exploited by Service Providers by adopting two alternative strategies: inertial and a bi-modal strategy, through two innovation cycles: The former is one relatively smooth, looking at a seamless introduction of some SDN-NFV features into current network infrastructures; the latter is much faster and disruptive, where Softwarization will pave the way to the deployment of a new parallel integrated Network and Service software virtualized platforms.(e.g., AT&T project CORD). In this last case, the innovation produces a radical transformation in the Operations, by exploiting approached a la DevOps.

These two cycles strategies will probably even co-exist for some time bringing to the deployment of "sandboxes", operated with IT-style processes and capable of providing specific end-to-end services. These "sandboxes" will be interconnected also through legacy interconnection links. In this sense, a key point will be interoperability, which should be pursued since the beginning, by design. This is very challenging but fundamental.

It is argued that "sandboxes" should embed interoperability features and capabilities. Orchestration, management and control should be "external" but recursively embedded into the "sandboxes". An interesting first approach in this direction is being pursued, for example, by the EC funded project FELIX [38].

Flexibility and automation of the Operations processes in these "sandboxes" will dramatically reduce the Time To Market (TTM) compared to the legacy infrastructures. In these domains it will be possible, for certain classes of services, to embrace full adoption of the API-economy paradigms (e.g., pursuing automatic provisioning and services transactions). When standardization efforts will not be fast enough (e.g., due to lock-in and disagreements) to support this fast exploitation, then "standards-de-facto" will emerge and be adopted, upon markets requests and dynamics.

The ways SP-DevOps could be exploited will follow the two main strategies being adopted by Service Providers for their Digital Business Transformation. SPs adopting the bi-modal strategy will deploy DevOps in a parallel to virtualised Network and Service platform. This will produce a radical transformation in the Operations (and the culture of the Company), whilst keeping a certain level of interoperability with the legacy infrastructure to provide e-to-e services (i.e., across sandboxes). On the other hand, SPs adopting the inertial strategy will gradually deploy DevOps in the legacy infrastructure being virtualised. This process will be slow as it will be delayed by a deeper integration of DevOps processes with legacy OSS/BSS ones.

# 4 SP–DevOps tools and functions

In the following subsections, we provide the final development and integration status of the SP–DevOps tools. How these individual components relate to the SP–DevOps concept has been outlined above (Section 3.2), and will be discussed in more details in Section 5 with respect to processes defined in support of integrated verification, programmable observability and automated troubleshooting. The relation of the tools with respect to the research challenges defined in D4.1 [25] is summarized in Annex 1. Note that the Watchpointer tool is considered to be finalized and has not been further developed since its documentation in D4.2 [6], thus it will not be described in a separate subsection. This section closes with a description of the open–sourced SP–DevOps toolkit based on a subset of all our tools. Table 1 summarizes the tools and serves as index into the individual subsections.

*Table 1: Summary of SP-DevOps tools. OpenSource tools are marked as being part of the SP-DevOps toolkit, whereas "public" indicates code available in a non-finalized state.*

| SP–DevOps Tool or Function | Problem solved | Section | Availability |
|---|---|---|---|
| VeriGraph | Formal verification tool for complex networks containing active NFs. Able to automatically verify NF–FGs and their configuration within a matter of few seconds (1.2 sec in our evaluation scenario). | 4.1 | Part of toolkit |
| Path Verification and Consistent Network Updates | Verification of the operational dataplane paths by a tagging approach, popularly known as traceback approach. An additional tool is able to guarantee either waypoint enforcement or loop freedom with limited update latency overhead. | 4.2 | |
| MEASURE | Automatic deployment and configuration of monitoring functions as well as local aggregation points for increased scalability of the observability system. | 4.3 | (Public) |
| TWAMP Data Model | Data model to support cross–vendor and cross–technology programmability of legacy OAM tools. | 4.4 | Part of toolkit |
| Ramon | Provides precise statistical estimates of link utilization in a distributed manner to offer congestion indications. | 4.5 | Part of toolkit |
| Probabilistic delay monitoring | Capable of self–adaptively setting the appropriate number of samples for detecting changes. Can capture changes with an expected probability while keeping the false alarm rate as low as required. | 4.6 | |
| EPLE | An efficient packet loss estimator for SDN based on OpenFlow. Induces zero dataplane overhead, and reduces even control and management plane overhead significantly compared to active measurement tools. | 4.7 | |
| IPTV Quality Monitor | A standardized IPTV quality monitoring mechanism (ITU P.1201.2). We show that such service specific functionality can be realized as a virtualized function (i.e. VNF) deployed as part of a NF–FG. | 4.8 | |
| AutoTPG | A tool that actively probes the FlowMatch part of OpenFlow rules to identify bugs in the forwarding path when aggregate flow descriptors are used. For networks where bandwidth availability is high, verification can be achieved in time intervals of a few seconds | 4.9 | Part of toolkit |
| DoubleDecker | A flexible, carrier–grade communication services between virtual monitoring functions and controllers, orchestrators, and network management systems. | 4.10 | Part of toolkit |
| Recursive Query Language | An engine (RQE) that interprets the recursive query language (RQL) to efficiently collect and aggregate monitoring information from distributed software–defined infrastructures. | 4.11 | |
| EPOXIDE | Enables integrating and automating troubleshooting tasks by troubleshooting graphs. Allows developers to define sequences of tools to be used for debugging particular problems with the network or VNFs. | 4.12 | Part of toolkit |
| Network Watchpoints | Intercepting control traffic events on an OpenFlow interface. Able to catch policy violations and trigger actions. Can also be used as a debugging and troubleshooting tool for network configuration. | D4.2 [6] | |

## 4.1 VeriGraph – Pre-deployment Verification

### 4.1.1 Purpose and problem statement

VeriGraph aims to check certain properties related to the UNIFY service models (e.g., network node reachability, node isolation and node traversal) before the actual service deployment, so as to reduce the risk of introducing critical and erroneous network configurations. In order to achieve this goal, the verification process is based on formal methods, i.e. mathematically founded methods that can be used to prove that the involved models (SG, NF-FG, middleboxes configurations) fulfil certain properties. Such formal techniques should allow the verification process to be completed in a reasonable amount of time and with fair processing resources (e.g. CPU, memory and so on). Given these requirements, we decided to explore Satisfiability Modulo Theories (SMT) based techniques since they seem to be more promising than traditional model checking methods [39] [40], which usually suffer memory scalability problems. We considered in particular networks that contain active network functions implemented as VNFs, i.e., functions that dynamically change the traffic forwarding path according to local algorithms and an internal state based on traffic history. Examples of those active functions (or middleboxes) are NATs, load balancers, packet marking modules, intrusion detection systems and more. Those are not well supported by the state-of-the-art verification tools (see [25] and [6] for more details) and hence require proper attention in the research work in this area.

### 4.1.2 Brief technical description

The verification process exploits an approach similar to the one recently proposed at UC Berkeley [41]. In order to achieve high performance, the verification engine exploits Z3 [42], a state of the art SMT solver, and also addresses network scenarios with multiple active network functions connected together to form a complex network graph. Z3 is instructed by VeriGraph to solve reachability problems thanks to the translation of these problems into SAT problems. The model of all the involved middleboxes and the overall network behavior are represented as a set of first order logic formulas and completed with other formulas that express the properties to be verified, for example reachability properties between two nodes in the network, in such a way that the satisfiability of the formulas implies the truth of the specified properties. See [6] for more details on the VeriGraph internals and for some network modeling and verification examples.

### 4.1.3 Recent improvements concerning the tool implementation

As already mentioned in the previous Deliverables, various verification processes take place in the Unify architecture before deployment:

- Service Layer (SL) verification on the Service Graph (SG) topology: basic reachability properties[1] are verified by only looking at the given topology;

- Orchestration Layer (OL) verification on the NF-FG topology: when the SG is translated into a richer representation, another topology verification process takes place in order to guarantee that certain required topology properties are still satisfied. This step is needed since topology level errors could be detected early in this stage without the burden of a complete verification with Z3;

---

[1] In this section we use the terms property and verification policy interchangeably.

- OL verification on the NF-FG at flow-level: when the NF-FG is ready to be deployed with all the VNFs attached to it, the corresponding configurations must be properly defined for each VNF. For example: an ACL for a firewall or the list of blocked domains for a DNS filter and so on. When all these configurations are available, we run another complex verification task, which considers reachability properties in conjunction with all the constraints imposed by the complete model of the network, including each VNF forwarding behavior and its configuration.

The whole pre-deployment verification process is performed by the synergistic collaboration of different modules that interact with one another by means of well-defined interfaces, according to the verification workflow (see section 5.2). Since the whole verification process is split into multiple sub processes, we designed a modular architecture to dispatch and execute the work accordingly. In order to modularize the design of the overall tool and ease the integration with the reference prototype implementation of the Unify architecture (Escape v2 in [7]), we designed the components shown in the Verification Module in Figure 7 to perform the above-mentioned tasks.

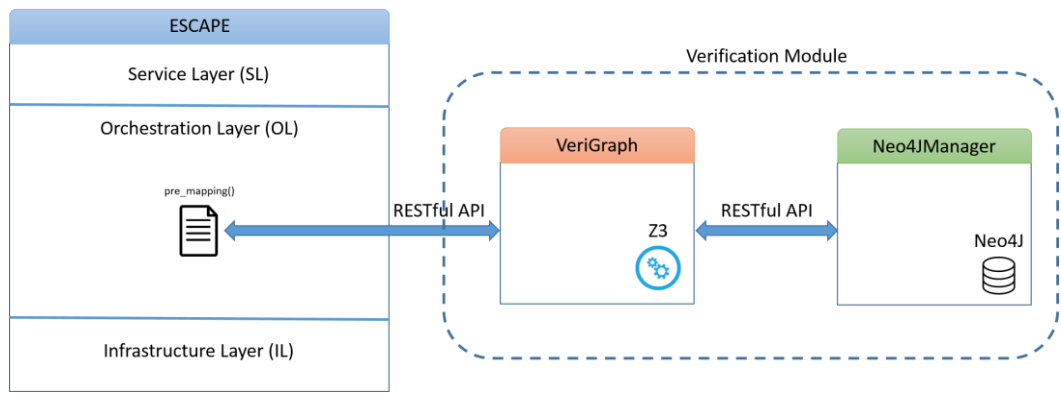

*Figure 7 Verification Tool Architecture.*

As it can be seen, an externally exposed RESTful interface offers access to a set of services able to perform different types of verification. The two components named Neo4JManager and VeriGraph, are responsible for the following activities: (i) Neo4JManager provides an API to easily store, manage and delete network graphs into the Neo4J database. In this context, a graph is a set of network nodes connected by directed arcs. Included in its API, the component exports methods to test, for example, topological reachability from one point to another in the graph and also to compute all the paths from a source to a given destination node. This operation is particularly useful for other steps of the verification process, as we will describe in the following; (ii) VeriGraph, which is the front-end of the verification tool based on a SMT solver. It provides a service to perform the NF-FG verification, including all the involved VNF configurations. This task is run by considering not only the topology but also a model for the forwarding behavior of each VNF. This model is expressed by means of first-order logic formulas that represent both the network (including the models of the VNFs) and also the reachability properties to be checked, as already reported in D4.2 [6]. In order to exhaustively verify all the possible service chains available in a given NF-FG, VeriGraph exploits Neo4JManager to store a graph on the Neo4J database and to extract all the paths from the source node to the destination node. After that, VeriGraph generates a model for each extracted chain, including the configurations, and performs a complete verification step.

To be more specific, we document the most important REST methods exposed by Neo4JManager for managing SGs and NF-FGs and for performing path searches in the graphs:

1. METHOD:        POST
   URL:           ${BASE_URL}/graphs
   DESCRIPTION:   This method creates a new resource representing the provided graph
   REQ. BODY:     Representation of the graph
   RESP. BODY:    In case of success, the service replies with an OK message and the ID of the newly created graph

2. METHOD:        GET
   URL:           ${BASE_URL}/graphs/{graphId}/topologyProperty
   DESCRIPTION:   This method checks whether the topological property specified by means of query parameters is satisfied or not
   PARAMETERS:    type -> a string specifying the type of properties. It can be: REACHABILITY, ISOLATION or NODE_TRAVERSAL
   sourceNode -> the source node,
   destNode -> the destination node,
   middleBox -> the node that must be traversed (used if type is NODE_TRAVERSAL or ISOLATION)
   RESP. BODY:    The verification results

3. METHOD:        GET
   URL:           ${BASE_URL}/graphs/{graphId}/paths
   DESCRIPTION:   This method retrieves all the possible paths from one node to another.
   PARAMETERS:    sourceNode -> the paths root,
   destNode -> the destination node
   RESP. BODY:    An array containing all the paths between the two nodes

For what concerns VeriGraph, the following API has been defined to expose the verification service:

1. METHOD: POST
   URL: ${BASE_URL}/graphs
   DESCRIPTION: This method creates a new graph
   REQ. BODY:  Description of the graph
   RESP. BODY: In case of success, the service replies with an OK message and the ID of the newly created graph

2. METHOD: GET
   URL: ${BASE_URL}/graphs/{graphId}/nodes
   DESCRIPTION: This method returns all the nodes of a single graph
   RESP. BODY: List of available graphs

3. METHOD: GET
   URL: ${BASE_URL}/graphs/{graphId}/nodes/{nodeId}/neighbours
   DESCRIPTION: This method returns the neighbours of a given node
   RESP. BODY: List of neighbours

4. METHOD: GET
   URL: ${BASE_URL}/graphs/{graphId}/property
   DESCRIPTION:    This method checks whether the property specified by means of query parameters is satisfied or not
   PARAMETERS:    type –> a string specifying the type of properties. It can be: REACHABILITY or ISOLATION
   sourceNode –> the source node,
   destNode –> the destination node,
   middleBox –> the node that must be traversed (used if type is ISOLATION)
   RESP. BODY:    The verification results

Concerning operation 2 of Neo4JManager and operation 4 of VeriGraph, only the reachability policy has been implemented. On the other hand, instead, the node traversal policy can be obtained by slightly extending the existing tool, while the isolation policy can be trivially obtained as the negation of the node traversal property.

An important aspect that has been improved w.r.t. previous deliverables, is that the internal modules of VeriGraph performing the translation of a given network verification problem into equivalent FOL formulas, have been completely improved and rewritten from scratch in the Java programming language, thanks to the Z3 Java API bindings. In this way, the previous core of the tool, based on the Python library developed at Berkeley, has been completely eliminated from our tool.

It is worth noting that VeriGraph can perform different kinds of checks. In particular, we distinguish two classes of properties, namely Topology-level and Flow-level properties. These two classes are verified through the aforementioned tools (i.e., respectively, Neo4JManager and VeriGraph). For any property that has to be checked, the verification tool must know two nodes in the graph, namely source and destination nodes, plus a middlebox in case of isolation or node traversal properties.

Topology-level properties are verified by looking at topology aspects only, i.e. by exploiting graph theory. Neo4JManager implements these verifications and offers, through its REST API, the possibility to check three kinds of properties in both the SG and the NF-FG:

- Reachability: the tool checks that at least one path exists between the source and destination nodes. However, the satisfiability of reachability properties at topology-level does not imply that all packets from the source are correctly forwarded in the path to the destination. Hence further verification processes are needed in order to be sure about graphs correctness;

- Isolation: the tool checks that no packets between source and destination nodes can pass through a certain middlebox in the graph (e.g., packet from a web client to a web server must not traverse an anti-spam function). Neo4JManager implements the isolation property verification by checking that all the paths between the source and destination nodes do not contain the undesired middlebox;

- Node-Traversal: this property is the opposite of the isolation property, because it is satisfied when all packets between source and destination nodes traverse a specific middlebox (e.g., all packets from client A to server B must traverse a firewall). Here Neo4JManager can verify that all paths that connect the source and destination nodes include that middlebox.

On the other hand, Flow-level properties consider the way each middlebox in the graph processes and forwards traffic. VeriGraph implements these verifications and offers, through its REST API, the possibility to check almost the same properties available at the Topology-level, but of course, they have a different meaning:

- Reachability: the tool builds a formal model of each middlebox involved in the path between source and destination and checks if at least one packet from the source node can arrive at the destination. In this case we are sure that if the reachability property is not satisfied, there is no connection between the two nodes in the graph;

- Isolation: the tool checks that no packet that goes from source to destination passes through a certain middlebox. In practice, this is equivalent to verifying that the source cannot reach this middlebox in all the paths toward the destination node.

It is worth noting that Flow-level properties cannot be verified on the Service Graph, due to its nature of high-level graph, where tenants do not have the vision on which middleboxes compose the graph. Hence VeriGraph must verify such properties on a Network Function – Forwarding Graph, thanks to its peculiarity of being a middle-level graph in the UNIFY architecture.

Further investigations have been done to improve the potential of a pre-deployment verification. In particular, from additional analysis of the current literature, other formalisms have emerged for the definition of the network services offered to the final users (i.e., SGs and NF-FGs). For example, forwarding policies are an alternative to describe how traffic flows (identified by means of a set of network fields) should be forwarded into the network, by traversing different Service Function Chains. In this context, it is possible to verify further types of properties, in addition to the afore-mentioned ones.

In particular, we can distinguish several classes of properties to be checked before deploying a graph in order to verify the correctness of the required service when forwarding policies are used instead of SGs or NF-FGs. For each of these classes, it is possible to define a set of properties that will be checked against a set of forwarding policies. In particular, these classes are:

- Single-Field policies check the presence of errors that involve one network field in a single forwarding rule (e.g., VLAN identifier is greater than its maximum values);

- Pair-Field class involves two network fields of the same forwarding rule (e.g., source and destination IP addresses are equal).

- The Node Ordering policies aim at verifying an ordering constraint on all the service function chains that a flow could potentially traverse in the graph. In particular, such ordering constraints are related to the position of the network functions within a SFC in a forwarding rule (e.g., a user would verify that his web traffic always traverses a firewall as last node of the chain).

- Chain Constraint properties are related to the case that a traffic flow is allowed to traverse more than one path in the graph. Let us consider an NF-FG containing a load-balancer to distribute mail traffic on multiple anti-spamming instances: in this case, at run-time, each packet will traverse only one of the outgoing paths after the load-balancer, based on the forwarding choice taken by the function. In this context, it could be useful to verify whether each possible network path of the mail traffic does not violate a specified condition (e.g., each path must contain a monitoring function).

- Sub-Optimization policy class regards those forwarding policies that refer to different sets of packets, which are forwarded to the same SFCs (e.g., in case of forwarding rule duplication). In this case, we will not have traffic forwarding errors at run-time, but an under-optimized exploitation of the network resources (i.e., memory usage in network nodes).

- Conflicting policies are two forwarding rules that manage the same traffic flow but they refer to different set of chains, generating forwarding errors at run-time.

More information about this approach can be found in [43]. The verification of these classes of properties is not integrated into the pre-deployment verification framework developed within UNIFY, since the formalism of forwarding policies is not integrated into the UNIFY architecture, because final-users are allowed to define their network graph by exploiting the well-known formats of SG and NF-FG rather than forwarding policies. Future work beyond UNIFY may implement this alternative approach.

### 4.1.4 Recent improvements concerning the tool integration

We integrated the already described components and their logic within the ESCAPEv2 framework. In order to accomplish this integration, we exploited the extensibility features of ESCAPE. Specifically, we implemented our custom `AbstractMappingDataProcessor`[2] where the under deployment graph undergoes the verification stage in the overriding of the *pre_mapping_exec()* method. The verification related operations of the pre-mapping phase can be summarized in the following way:

1. receive a NF-FG as input from ESCAPE;

2. receive some policies to be checked from upper layers;

3. store the graph and all of its nodes into VeriGraph by means of the REST API (the *NF-FG database creation sub-process* in Figure 9 and Figure 10);

4. extract all the possible chains to/from the nodes involved in the provided policies (*NF chain extraction*);

5. generate a verification scenario for each chain extracted, including all the necessary FOL formulas to model the encountered middleboxes (*VNF chain formula creation*);

6. run all the verification tasks generated in the previous step (*Z3 running*);

7. raise an exception if any error occurred during the previous step.

---

[2] https://sb.tmit.bme.hu/escape/util/mapping.html#escape.util.mapping.AbstractMappingDataProcessor

In order to test the complete verification workflow and its integration within ESCAPE we setup a comprehensive demo to show the described modules in action. Further details can be found in Section 6.1.

### 4.1.5 Evaluation results and conclusions

In Section 5.10.4 of D4.2 [6] we defined a possible scenario to validate our verification approach. The presented results were mainly obtained by feeding the Z3 SMT solver with the proper formulas to model the considered graph and the corresponding policies. In this section, we further refine our validation by considering the newly implemented components, i.e. VeriGraph with Neo4JManager and the complete toolchain including the automatic chains extraction and verification scenarios generation. The aim is to understand the impact of each phase of the verification on the overall verification. To do this, in Figure 8 we briefly report the scenario already presented in [6].

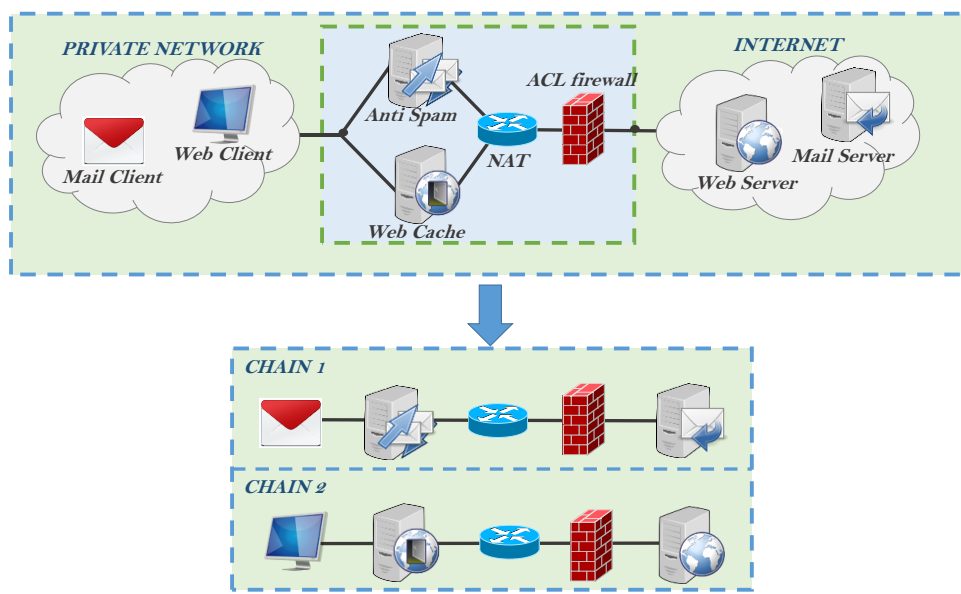

*Figure 8 VeriGraph verification scenario.*

Basically, we have a graph where two types of traffic are handled and correctly forwarded over two different sequences of middleboxes:

- **Chain 1** is designed to handle the mail traffic and contains an antispam function;

- **Chain 2** is designed to handle the web traffic and contains an HTTP cache.

Notice that the NAT and Firewall functions are included in both chains so that they can be shared. In order to characterize the runtime behaviour of our tool, we perform a verification task by configuring the functions in such a way that a reachability property from the client to the server is successfully checked. This can be easily obtained by disabling filters on the Antispam and on the Firewall.

In the case of **Chain 1**, the obtained results show that the verification modules take an overall time of 3.824 seconds (evaluated on 5 test runs). This total execution time has been split also into the partial times required by the different subtasks performed by VeriGraph, as depicted in Figure 9. The reported results are expressed as a percentage over the total verification time.

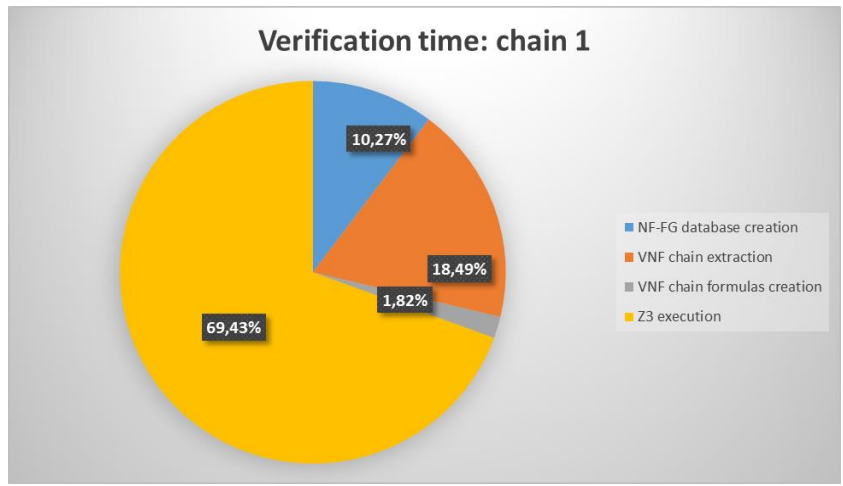

*Figure 9 VeriGraph phases (Chain1).*

As it can be seen, a large part of the computation is done by Z3 to actually solve the FOL problem (about 70%) and by Neo4JManager (about 20%) which has the important task of analysing the provided graph to extract the paths needed to check the required policies (the *VNF chain extraction* sub-process).

Similar results are achieved in the verification of the Chain 2, which are shown in Figure 10. The total execution time elapsed for verifying the correctness of the second chain is 1.576 seconds, evaluated over 5 test runs. Also in this verification scenario the retrieval of the second chain (i.e., implemented by the Neo4JManager sub-process) and the execution of the SMT solver are the most expensive sub-tasks of the verification module.

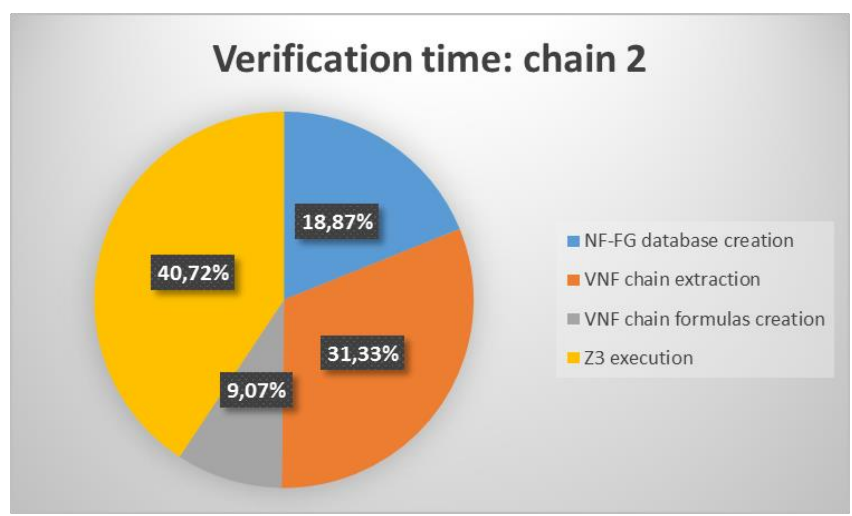

*Figure 10 VeriGraph phases (Chain2).*

## 4.2 Verification of Path Functionality

### 4.2.1 Purpose and problem statement

Over the last years, much research went into the development of formal methods to ensure a correct and "safe" operation of virtualized and programmable distributed systems, like targeted in UNIFY. In particular, the ability to formally specify, verify and compose policies is arguably one of the key benefits of Software Defined Networking (SDN): unlike failed "innovative" networking paradigms like Active Networking, where packets carry arbitrary code and are hence hard to verify, SDN's match-action paradigm gives us a principled approach which naturally allows for a formal reasoning.

However, today, most work focusses on offline resp. deployment time verification in the control plane (or middleware), e.g. through programming languages like Frenetic and Pyretic [44]), and verification schemes like VeriFlow or the UNIFY pre-deployment verification tool VeriGraph, as described in Section 4.1. Much less is known about how to perform verification of network policies or service chains at runtime. To fill this gap, we are exploring mechanisms that allow us to verify functionality in the data plane during run-time. In particular, we aim to verify network flows and paths.

There are many reasons why verification in the control plane and at deployment time need to be complemented by run-time verification methods:

- A programmable network is essentially an asynchronous distributed system: the installation of new policies as well as the update of existing policies need to be communicated from the control plane to the dataplane elements (switches, universal nodes, appliances, etc.). This transmission is performed over an asynchronous and unreliable network, and hence, updates may actually arrive out of order.
- Errors: There are many reasons why an update or flowmod may not have the effects it was expected to have. For example, different switches from different vendors may react differently to certain events, or may not support all the expected features, or not in the expected form.
- Misbehavior: over the last years, hackers have repeatedly demonstrated to be able to hack into routers. Without verification, a misbehaving router or network function may have negative consequences.
- Long runtime: often, after deployment, a network runs for many years, and runtime verification is the only possible solution.

We have hence also put efforts in developing a run-time path verification framework. This includes a) mechanisms to verify consistent network updates, i.e. whether the intended network changes actually took effect (e.g., using barrier requests) and network operation and updates are consistent, despite the fact that commands from the controller are transmitted over an asynchronous network; and b) an implementation of a tagging and efficient sampling mechanism for run-time verification of datapath forwarding behavior.

The novelty of our approach not only lies on the runtime verification and dataplane focus, but also in that it takes into account the specific needs introduced by network function virtualization: while classic verification techniques like trajectory sampling work fine in network, in the presence of middleboxes and NFVs, as well as service chains, these classic solutions cannot be applied: as middleboxes often change the header fields.

This is a challenge indeed: implementation of the configuration and policies need to be verified on the data plane level as well. With the advent of NFV, there has been the nice possibility to dynamically scale out the network functionality like Router, Firewall, Load Balancer, IDS etc and at the same time with a lot of placement flexibilities.

### 4.2.2 Brief technical description

Our run-time path verification framework consists of two complementing approaches, i.e. ensuring consistent deployment of the paths configured for individual service chains during network updates; and verification of the resulting, operational dataplane paths by a tagging approach, popularly known as traceback approach.

**Consistent Network Updates**: Based on the update scheduling algorithms namely Wayup and Peacock presented in [45] and respectively, we divide the policy updates from controller to OpenFlow switches in the form of rounds. Each round culminates in the controller sending the Barrier requests to each OpenFlow switch receiving updates in that round and waits for acknowledgement in the form of Barrier reply. Once done with receiving all Barrier replies, the controller initiates the next round of policy update. Therefore, we ensure there is synchronicity and consistency avoiding out of order updates and ensuring waypoint enforcement [45] and weak loop freedom [46]. More work on multiple policies can be found on [47] and [48]. This tool will be demonstrated in [49] and [50].

**Path verification**: The basic idea for our path verification tool is to tag the packets to save path information on the packets. This information is carried in the form of unique per link tag stored in the available Layer 2 (MPLS or VLAN ID field) or Layer 3 (DSCP bits) space in packet in such a way that it does not alter the data plane behavior of the packet. There are two possible approaches for path verification: either use the controller application or it can be done directly in the data plane by the OpenFlow switches.

a. **Controller application approach**: The traffic hits the first switch and is mirrored and sampled i.e. 1 out of n packets (using sflow or Netflow) and then sent to controller application for inserting unique tag in into MPLS, VLAN or DSCP bit (whichever does not cause change in behaviour of traffic) based on inport, outport and switch number (uniquely identifies a link). Then, the tagged packet is re-injected to the datapath and follows the same procedure until it encounters a link to a middlebox where a copy of the packet is sent to the data plane collector which parses the packet and assigns them to a queue for analysis. This way, we prevent the mangling of inserted tags or header because of post-traversal flow re-association problem of middleboxes. At the collector level, analysis can be done on the tags to determine the trajectory of the packet of a particular flow. As depicted in Figure 11, the controller application residing the SDN controller tags the packets.

b. **Data plane approach**: The traffic hits the first switch and there is rule with a soft configurable timeout that has action push unique tag into MPLS, VLAN or DSCP bit (whichever does not cause change in behavior of traffic). The unique tag is based on inport, outport and switch number so it uniquely identifies a link. As soon as the tagged packet is about to go to the middlebox, it is duplicated with the help of group table and sent to the collector for similar treatment like mentioned previously in controller application approach. As shown in Figure 11, the switches S1, S2 and S3 tag packets.

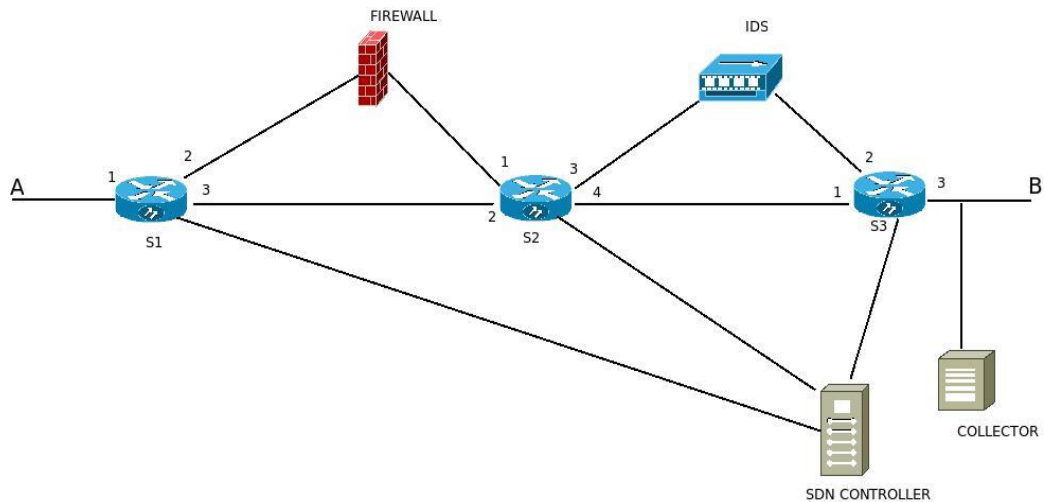

*Figure 11 Depiction of controller application and data plane approach for tagging. A Controller application on top of the SDN controller tags packet in the former whereas the switches (S1–3) tag packets in the latter approach.*

### 4.2.3 Evaluation results and conclusions

### 4.2.3.1 Consistent network Updates Tool

In terms of consistent network updates, we have implemented a prototype in mininet based on a RYU controller. We have implemented wayup [45] and peacock [46] algorithms successfully and have been running our evaluations with respect to the update time of flow tables in OpenFlow switches. As shown in Figure 12, the test setup for consistent network updates tool consists of reference single network update round for both Wayup and Peacock algorithms. The topology consists of 12 nodes or OpenVswitches in mininet. From Figure 12 to Figure 17, the edges having a solid line, build the old route through the network. The edges having a dashed line, build the new route through the network.

For Wayup, the node filled with black colour i.e. S6 is the waypoint.

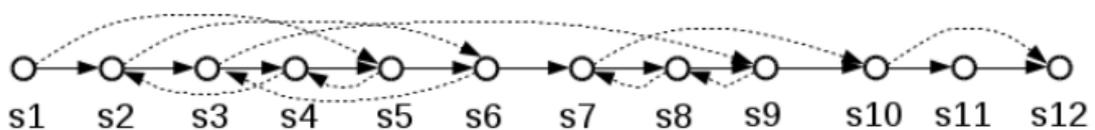

*Figure 12 Wayup and Peacock reference setup with single network update round.*

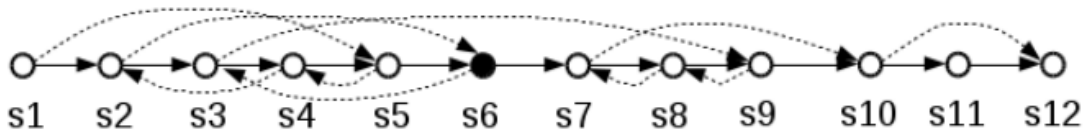

*Figure 13 Wayup2 with two rounds of network update.*

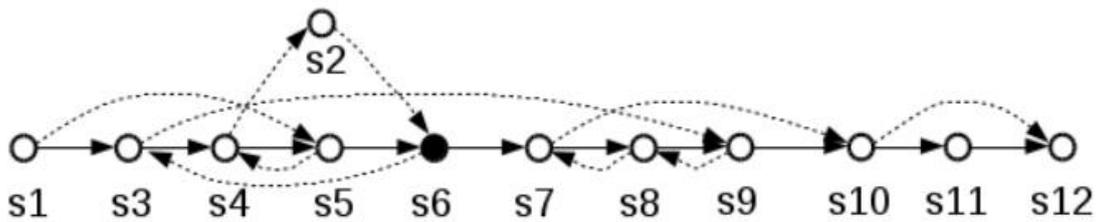

*Figure 14 Wayup3 with three rounds of network update.*

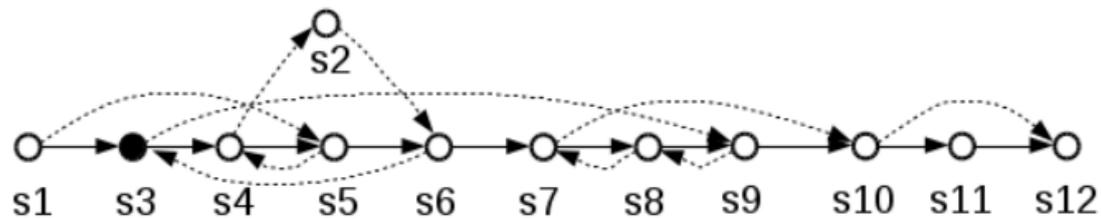

*Figure 15 Wayup4 with four rounds of network update.*

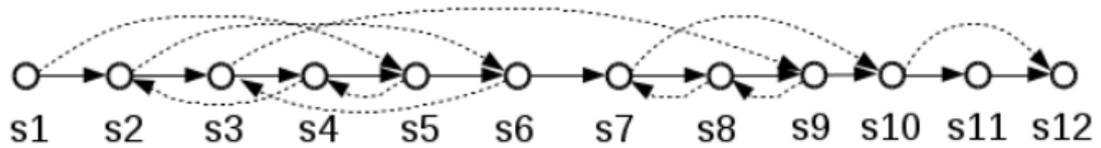

*Figure 16 Peacock3 with three rounds of network update.*

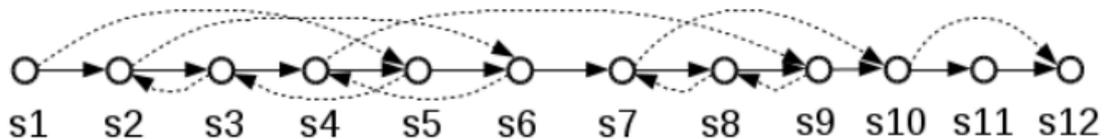

*Figure 17 Peacock5 with five rounds of network update.*

**Results Wayup:** As shown in Figure 12, the Wayup Algorithm in the consistent network update tool was conducted with original one round update named as reference, Figure 13 shows 2 round update (Wayup2), Figure 14 shows 3 round update (Wayup3) and Figure 15 depicts four round update (Wayup4). The corresponding update duration in seconds is depicted in the Figure 18 shows that time taken for network update in more than one round is more but allows waypoint traversal. The duration of the update performed during this measurement consists of time taken by the controller to parse the REST request, calculate an update schedule, send the OpenFlow messages to the switches, the OpenFlow messages need to travel to the switches, the switches need to apply the new forwarding rules and the Barrier Reply messages need to travel back to the controller. The results show the tradeoff between update duration and consistency property of network update which is waypoint traversal. The update duration is a bit more than standard one round network update which is expected but ensures consistency in network update.

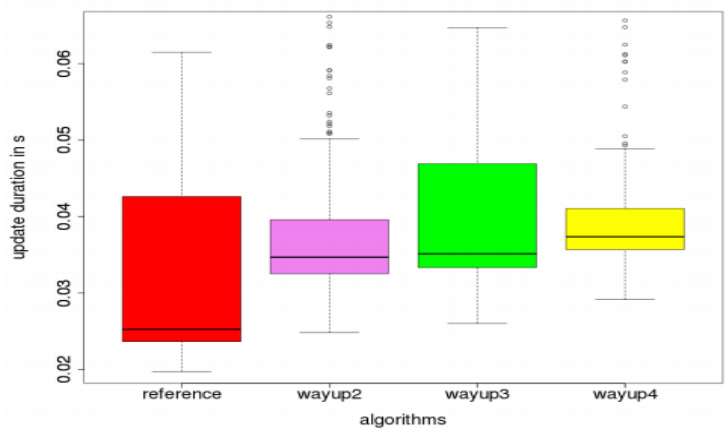

*Figure 18 Update Duration of Wayup Algorithms.*

**Results Peacock:** As shown in Figure 19, the Peacock Algorithm in our consistent network update tool was conducted on same topology with original one round update named as reference and then with 3 round update (Peacock 3) and five round update (Peacock 5). The corresponding update duration in seconds is depicted in the Figure 12 where Wayup algorithm with four rounds of updates is mentioned as Wayup4. The results show the tradeoff between update duration and consistency property of network update which is loop freedom. The update duration is a bit more than standard one round network update which is expected but ensures consistency in network update.

The demo video of consistent network updates can be found at http://tinyurl.com/zf4v7qo.

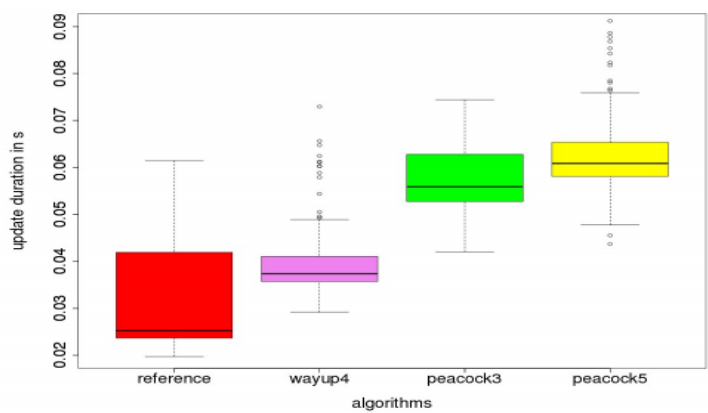

*Figure 19 Update Duration of Peacock Algorithms.*

#### 4.2.3.2 Path Verification tool

In terms of path verification, we implemented the above mechanism in the RYU controller, and a first test was successful as tags were detected by a wireshark packet sniffer. The future work involves mirroring of the traffic and sending the mirrored traffic to controller application to avoid disruption of regular production traffic. There is also a decision being taken on the sampling mechanism e.g., sending one in every one thousand packet to controller application for tagging (probabilistic tagging). We are also looking for ways to limit the number of mirrored packets being sent to the controller.

As shown in Figure 20, our test setup consisted of a traffic generator h1 that generates traffic. The traffic with destination port 23 is supposed to be forwarded to the port 3 of the switch that goes into middlebox h3, h1 is the generic host and h4 is the collector of tagged packets on the data plane. The tag is inserted in the DSCP bit of IP header and tag inserted has value zero and is a unique tag for the link between switch s1 and middlebox h3 and is based on inport: port1, outport: port 3 and switch number: s1.

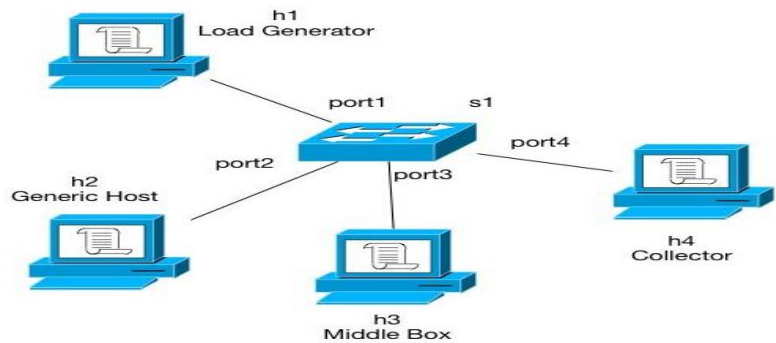

*Figure 20 Testbed topology.*

**Results:** As shown in Figure 21, the untagged packet is received at the middlebox h4. The group tables have actions to tag the packet with tag 0 and send it to the collector. As shown in Figure 22, the DSCP tags are collected at the data plane collector with the tag 0 in place. Moreover, the tags did not cause any alteration in the data plane behavior of the packet.

These results were according to the expectations and show that the tagged packets with all path information are collected at the collector just before entering the middlebox as a middlebox normally is a layer 4 to layer 7 device and it modifies the layer 2 or layer 3 header. The packets are duplicated and duplicated packets are tagged by group table rule so that there is no altering or latency introduced in the production traffic. The reason for using the group table is to avoid duplicate actions and reducing rule complexity, we need to keep track of fewer rules. The group table and flow table entry of switch s1 is depicted in Figure 23.

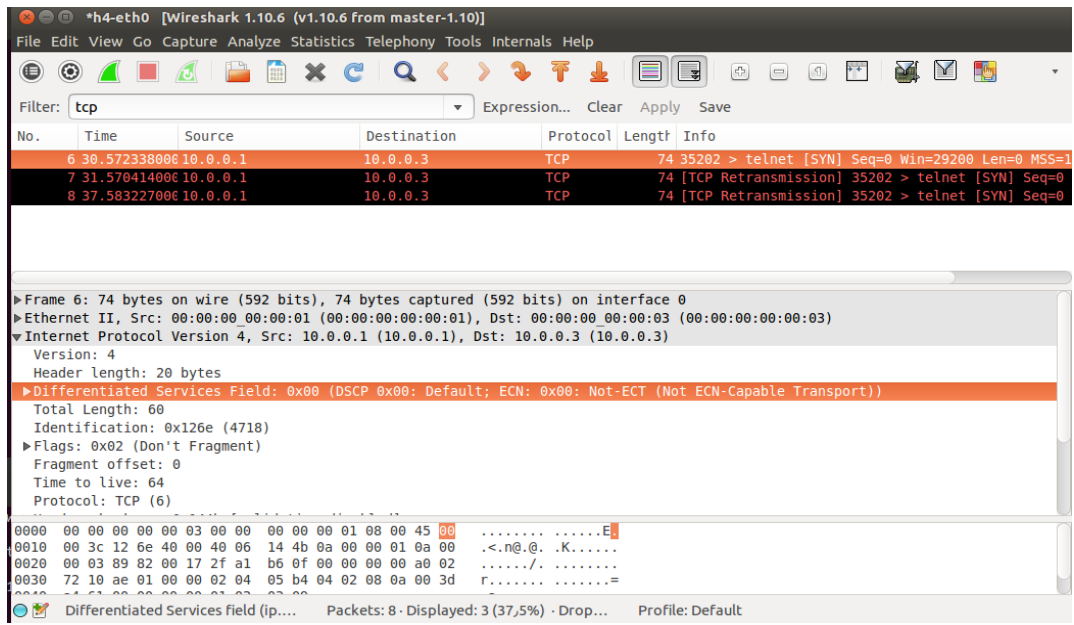

Figure 21 Untagged packets received at Middlebox.

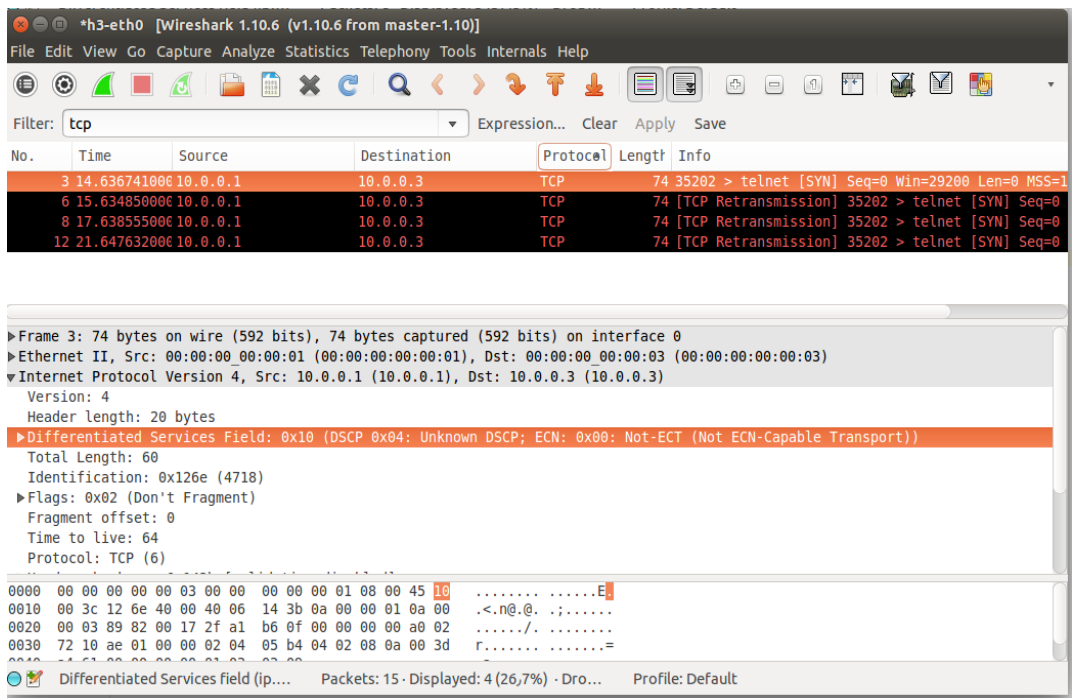

Figure 22 Tagged packets received at Collector.

*Figure 23 Flow table and Group table entries in the switch s1.*

## 4.3 MEASURE – Expressing Measurement Intents, States and Reactions

### 4.3.1 Purpose and problem statement

The MEASURE language was introduced in M4.1 [26] as a way of providing:

> "Machine-readable descriptions of the capabilities and their configuration parameters need to be developed. They specify which measurement functions should be activated, what and where they should measure, how they should be configured, how the measurement results should be aggregated, and what the reactions to the measurements should be."

The MEASURE concept is at the core of programmable observability work envisioned by SP-DevOps. It supports SP-DevOps observability to be more reactive, more automated and easier to maintain. One of the main overall goals of UNIFY SP-DevOps is observability of heterogeneous and dynamic environments. Several monitoring or verification functions are available but they need to be instantiated, configured and the results exploited in terms of appropriate (re)actions. To do so in a generic manner, we introduced MEASURE for Measurements, States and Reactions. At the core lies the MEASURE annotation, defined by a simple grammar to define the metrics to be monitored, information about the expected values and what to do it those values reach pre-defined thresholds (see Figure 24 for an example). To be more specific, the Measurements part defines the metrics and a minimum set of parameters to configure them. States are defined as zones - as long as the value of one metric or a logical combination of several metrics stays in the interval, the zone remains the same. A state-machine like mechanism reacts to the migration form one zone to another and contains information on the reaction to adopt, e.g. notify an orchestrator or control entity that resources are running short.

The MEASURE language contains three main sections:

1. **Measurement definitions** describe *which* measurement function to activate, and *where* the particular measurement should be taken, and *how* the measurement should be configured. These are defined like functions in a programming language, with parameters for measurement locations (e.g. a particular port or VNF), and MF specific parameters such as measurement frequency, identifiers etc.

2. **Zone definitions** that specify how to *aggregate* measurement results as well as thresholds for the aggregate values. Zones definitions result in variables that are either true of false. Zones represent an abstract state of the system according to measurement above defined.

3. **Reactions** that specify what *actions* to trigger when moving between zones, moving to and from zones, and while staying in a zone. Generally, actions consist of sending notifications, aggregated measurement results, and configuration to other components in the UNIFY architecture.

```
measurement {
  m1 = oneway_latency(SAP1, SAP2);
  m2 = cpu_load(FW1);
} zones {
  z1 = Avg(m1, '5 minutes') > 10.0;
  z2 = Avg(m1, '5 minutes') < 10.0;
  z3 = Avg(m2, '1 minute') < 90%;
  z4 = Avg(m2, '1 minute') > 90%;
} reaction {
  z3->z4: Publish(topic=alarm, msg="Warning CPU");
  z2->z1: Publish(topic=alarm,msg="Warning latency");
}
```

*Figure 24 Example MEASURE description: expected metrics are latency and CPU load of NF-FG components; the aggregation logic for the metrics are given with temporal averages; reactions to zone movements are specified.*

MEASURE typically annotates a NF-FG, specifying how monitoring for this particular service should be performed. However, MEASURE could also define infrastructure monitoring by replacing the virtual components of a NF-FG (such as virtual ports and links) with their infrastructure counterparts.

Within the MEASURE description, the components to monitor are referred to by abstract entities, such as nodes, ports, links or NF instances as specified in the corresponding NF-FG (NF-FG format see D3.3 [7]), rather than to what they are finally instantiated as in the infrastructure (e.g. a VXLAN tunnel or a Docker container). This allows the system to remain agnostic of the actual components it has to monitor until they have been instantiated. Once instantiated, the orchestrator informs the system about the mapping from abstract entities to actual implementations, which then configures monitoring functions accordingly.

The main goal of the local aggregation system is to increase scalability of network management by reducing the overhead of monitoring processes. We achieve this in two ways: i) by utilizing the hierarchical DoubleDecker messaging system (described in section 4.1, and ii) by aggregating data as locally as possible. For this reason, a local Aggregation Point receives raw monitoring results from MFs. The monitoring controller dynamically configures the aggregation logic, based on the MEASURE description (specifically the Zones and Reactions parts). Acting as a sink for monitoring results, an aggregator continuously evaluates the results and upon traversing the defined zones sends or published events to interested components, typically local network functions, orchestrations or control layer modules, or network management systems. It is then up to these entities to take appropriate response, e.g. by requesting additional resources or handling failures

### 4.3.2 Recent improvements concerning the tool implementation
The work on MEASURE is still actively ongoing and in planned to be continued beyond UNIFY in a national Swedish project involving two UNIFY partners. The work on MEASURE is relating to all technical UNIFY work packages: WP3 to understand how to integrate the monitoring framework within the NF-FG definition and how it will be decomposed through all the layers, WP4 for the integration with monitoring functions, and WP5 for a practical deployment within the Universal Node (UN) domain.

We have currently developed four components, depicted in Figure 25, as a reference implementation of the MEASURE concept in an UN. These components are:

i)      A monitoring management plug-in (MMP) acting as a monitoring controller in an UNIFY UN, performing actions according to MEASURE intents required for the Elastic Router use-case.

ii)     A parser for the MEASURE annotation (MEASUREParser), capable of producing a parse-trees as Python data structures, as XML, as JSON, and as YAML.

iii)    A Monitoring Function - Information Base (MF-IB) to catalogue MF implementations for various domains and allowing translation from metrics to associated functions providing the metrics. MFs must be registered in this database, which contains what configuration parameters and their types are expected, and what metrics the MF produces, and the data types of these metrics.

iv)    An in-stream processing Aggregator to demonstrate the efficiency of local aggregation of monitoring data.

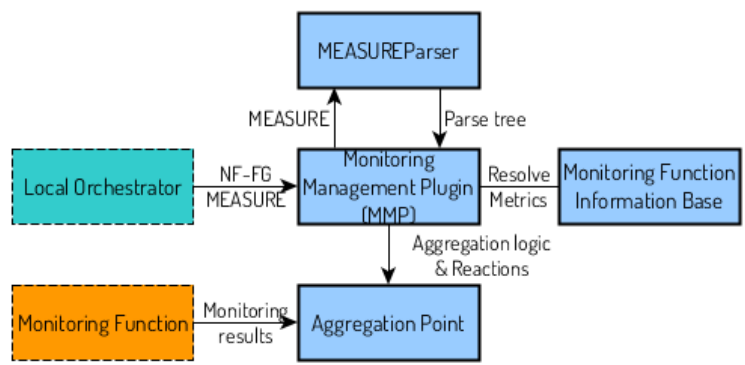

*Figure 25 Developed components in the MEASURE system.*

The **MMP** uses the DoubleDecker as its main messaging system, however, the MMP is also connected to the local deployment system, e.g. in the case of a Universal Node this component is the Docker daemon. The MMP supports JSON-RPC for receiving mapping information from the local Orchestrator, together with the MEASURE descriptions. This is called whenever an NF-FG is deployed, modified, or removed, in order to instantiate, update, or remove the associated monitoring functions.  This triggers the MMP to parse the MEASURE description, using the MEASUREParser described below. The first step is to lookup the appropriate Monitoring Function to generate the requested metric; this is done by querying the MF-IB on the current node. The MMP then generates configuration commands, which implements the Zone aggregation logic described in the MEASURE description, and sends them to the Aggregator, The next step is to start the Monitoring Functions if not already running or re-configure them otherwise. This step is very domain dependant; the process to start a virtual function being very different in all the domains considered in UNIFY. As an example, in the Universal Node this process involves the creation of new Docker containers using the Docker daemon interface. The MMP has to configure the containers at start-up for some parameters but also send runtime parameters via the DoubleDecker once the containers are running.

The **MEASUREParser** is a module implementing a parser for the MEASURE description. This parser takes the MEASURE annotation provided as meta-data along with the NF-FG and produces several outputs suited for each target. These different outputs are used to transform the MEASURE description in to a data model, or to configure the local monitoring data aggregation. This is a key component to the MEASURE prototype as it defines the syntax used for expressing monitoring intents in the service layer and gives a total abstraction of the actual deployment in the infrastructure layer.

The **Aggregator** is a generic component used at several layers to aggregate monitoring data, following the recursive architecture envisioned in UNIFY. It can be deployed locally to aggregate data as close as possible from the monitoring functions and reduce the amount of data to be transmitted to higher layers. This is even more important as the number of monitoring functions deployed for a NF-FG is increases and each MF generates multiple metrics in high frequency. A discussion of the potential overhead savings of local aggregation can be found in [32]. An aggregator furthermore has the utility to trigger automated reactions based on zones defined for compound metrics. The aggregation rules are generated by the MMP, which extracts the semantic meaning of the MEASURE annotation and then transformed into arithmetic operations that the Aggregator can execute. If several aggregators are connected through several layers the aggregations rules will be different at each layer, each layer presenting a more abstract view of the resources usage.

The **MF-IB** provides a mechanism to resolve which tool can provide a certain metric. These tools are registered in a local database and are domain specific. The MMP queries the MF-IB for a specific metric to obtain the list of available tools and their configuration parameters. It allows the decoupling of the metric from the implementation of a monitoring function, a tool can provide several metrics and one metric can be provided by several tools; the MF-IB provides the mapping. The MF-IB can be centralised and provide the same monitoring functions to several nodes independently. This is where the monitoring process becomes intent-based as the initial request doesn't have to include the exact implementation but only the desired metric.

### 4.3.3 Recent improvements concerning the tool integration

The integration of MEASURE is an ongoing process, benefiting from the Elastic Router use-case (more on this use-case in Section 6.2). We used the concrete work done with other partners on the integrated prototype to improve MEASURE and test it in a real environment. The integration work was divided in two main tasks, the integration of the MEASURE components with each other and the integration with the other prototypes and the environment itself.

As described earlier, MEASURE takes the form of four major independent components. They all have to communicate with each other so it was our first step of integration. The MMP is the central point; it receives the NF-FG with the MEASURE annotation and needs to, in order, understand it, configure the monitoring functions and finally configure the aggregation points used as destination for the monitoring data. The MEASUREParser is used directly inside the MMP and extracts the semantic meaning. Once this is done, the MMP has to effectively exploit this information. The MEASURE annotation consists of a list of metrics to be collected, intervals of values for those metrics which define states of the system and actions to be taken when the state of the system changes. The first step is to create the functions to collect those metrics. To do so the MMP queries the MF-IB and decides which function to use. We had to design a protocol to communicate between those two components and a transport mechanism. The obvious solution for the transport was to use the DoubleDecker. It allows a good abstraction of the network so the MF-IB can be located on a remote node but still accessible as easily as if it was local. For the protocol we agreed on using JSON-RPC for querying the available tools. JSON-RPC woks similarly to a REST API with commands and parameters except that no exposed IP address is needed, the commands can be sent via the DoubleDecker. One the MMP knows the available monitoring functions it can start and configure them. The aggregator is then configured to receive the monitoring data. The MMP has to make sure all the components are connected to the DoubleDecker.

The integration of MEASURE with other components mostly happened within the work on the integrated prototype and the elastic router use-case. This scenario is envisioned to take place is an Universal Node, result from WP5, which is based on a Docker containers running environment. We had to adapt the MMP to start and configure Docker containers and communicate with the Docker daemon to retrieve runtime data. In the case of a different domain, the same work would have to be done according to the execution environment. Within the UN, the MMP has to communicate with the local orchestrator to exchange information about the NF-FG and its mapping in the infrastructure. Again, the DoubleDecker was chosen to be responsible for the transport. The MF-IB was configured to expose the monitoring functions used in this scenario: cAdvisor [51] and the Rate Monitoring (section 4.5). These two monitoring functions can be configured at both start-up and run-time so we had to integrate these capabilities in the MMP. The aggregator also had to be modified in order to support the two databases used in the scenario, namely OpenTSDB and PipeDB.

### 4.3.4 Evaluation results and conclusions

Our current integration has been tested in the scope of the Elastic Router demo with the Rate Monitoring and PipelineDB (see Section 6.2). The goal was to deploy the Rate Monitoring and an aggregator, collect raw data and apply arithmetic calculations to reduce the volume of data to be stored or send to other components. The key problem is to provide a code to perform the calculation that is generic enough to handle any sort of raw data. The aggregation rules come from the MEASURE annotation, transformed by the parser, but the parser does not create directly executable code. The aggregator has to be able to create its own code and extract information from the data send by the monitoring function to perform the aggregation. We have demonstrated this possibility using PipeDB, a stream based DB. The aggregator creates insert rules as in a SQL INSERT, but this insert contains rules to aggregate the data combining arithmetic functions such as max, min or average and temporal filtering e.g. data received in the last 10 minutes. The results obtained so far with the Rate Monitoring are very promising as we can greatly reduce the amount of data after aggregation while still sampling at a high rate. Not reducing the sampling allows for instance more accuracy in case of problem; one of the reactions defined by the MEASURE annotation can be to reduce the level of aggregation in some cases.

The implementation of the MEASURE concept is still recent and needs more maturity to be evaluated in details. Besides the elastic router, we considered applying MEASURE also the UNIFY FlowNAC use case (D3.5 [18]), which revealed several shortcoming of the current MEASURE language. Specifically, the current MEASURE definition lacks the possibility to define non-observed metrics or messages as a condition, and furthermore does not allow to integrate MFs that are not under direct control of the monitoring controller (i.e. MMP), like VNF internal states published through the VNF itself.

In the FlowNAC demo, monitoring results, which represent the number of connected End Users, are not generated by an MF but by the VNF itself. Monitoring result variables in MEASURE are defined as *variable = metric(parameters)*, for example *m1 = delay(from=port1,to=port2)*. When processed this triggers the start and/or configuration of an MF able to provide the metric, for example an MF performing *ping*, this needs to be extended to support the case of VNFs generating monitoring results. One solution could be to define a special metric keyword that refers to a VNF in the NF–FG associated with the MEASURE description, together with a description of the generated monitoring result. This could for example be *m1 = fromVNF(VNF, {'users':'int'})* showing that *VNF* will generate results containing the integer metric, 'users'. This information would be enough for the MMP to resolve the name of the instanced VNF and configure the aggregation point. No other change to MEASURE would be needed to support VNFs that generate monitoring results.

The FlowNAC demo also relies on heartbeats to indicate that a VNF is alive, reactions are triggered when no heartbeats are received within a certain time. MEASURE on the other hand assumes that reactions should be triggered based on the data arriving in monitoring results. One way to remedy this would be way to express temporal aspects as part of MEASURE's zone logic which describes when a metric has moved into a separate zone, which in turn can have reactions associated with it, and logic in the Aggregation point to maintain message arrival times. The MEASURE description of this solution could be as shown in Figure 26, to the right.

| measurements { | measurements { |
|---|---|
| m1 = liveness(from=FNC_1, param = {'users':'int'}, timeout=30s); | m1 = fromVNF(FNC_1, {'users':'int'}); |
| } zones { | } zones { |
| z1 = m1 > 0; | z1 = m1.age > 30s; |
| z2 = m1 < 1; | z2 = m1.age < 30s; |
| } action { | } action { |
| z1->z2 = Publish(alarm,..); | z2 -> z1 = Publish(alarm,..); |
| } | } |

*Figure 26 Liveness solution with an external liveness MF (left), and with an internal age parameter (right).*

We plan to work on integration with other prototypes to continue exploiting the results and learning in other NFV/SDN related projects. The challenging part of MEASURE is to provide a very generic mechanism but with meaningful aggregation adapted to any metric. According to the SP-DevOps concept, service providers should be able to develop their own VNFs and MFs and simply plug them in the architecture. This can only happen if the interfaces are clearly defined but not too restrictive.

## 4.4 TWAMP – Legacy OAM protocol support

### 4.4.1 Purpose and problem statement

As explained in D4.2 [6], SP-DevOps can benefit from programmability enablers to existing protocols and implementations, such as standardized data models. A case in point introduced in [6] is the standardization of a YANG data model for the Two-way Active Measurement Protocol (TWAMP) thus complementing other SP-DevOps tools and mechanisms, such as RateMon and EPLE, with programmable active network measurements using an existing standard protocol.

### 4.4.2 Brief technical description

Sections 5.4.1 and 5.4.2 of [6] introduce the TWAMP YANG model and map it to the UNIFY functional architecture. In short, the TWAMP functional blocks Control-Client and Server are mapped to the Local Control Plane (LCP) of the TWAMP instance Observation Point (OP), while the TWAMP Session-Sender and Session-Reflector are mapped to the Local Data Plane. The standardized TWAMP data model [52] facilitates the management and administration of active network performance measurements by decoupling the protocol implementation from its vendor-specific management platform. In effect, a UNIFY SP-DevOps platform will be able to automate pertinent active measurements from deployed equipment with TWAMP support and feed the results to the distributed monitoring DB for further use e.g. by the recursive query engine (RQL).

### 4.4.3 Recent improvements concerning the tool implementation

The first version of the TWAMP data model was introduced in the IP Performance Metrics (IPPM) working group (WG) in March 2015, based on the work of an author team from UNIFY and major vendors. Since then, the document was updated and presented in IETF 93 and IETF 94 as an individual draft (draft-cmzrjp-ippm-twamp-yang). During this period, this line of work received significant commentary and support on the mailing list as well as during the IPPM meetings. In consultation with the WG Chairs and the Area Director it was deemed that the current WG charter permitted the addition of a new milestone document which could be addressed by the proposed data model. As a result, the IPPM WG issues a call for adoption which ended successfully in favour of the draft. As a result, further updates based on the call for adoption call review round were addressed and the new WG-adopted document was issued (draft-ietf-ippm-twamp-yang) and presented at IETF 95. We note that the intended status of the draft is Standards Track.

### 4.4.4 Evaluation results and conclusions

Performance measurements related to the YANG model are beyond the scope of UNIFY. Interested readers will find in the corresponding IETF datatracker page [53] the history of the TWAMP data model as well as the results of the official IETF YANG model validation results (affirmative as of March 2016).

## 4.5 RateMon – scalable congestion detector

### 4.5.1 Purpose and problem statement

The link utilization and rate monitoring MF is aimed at implementing a scalable congestion detector [54] based on the analysis of the traffic rate distribution on individual links at different time scales.

### 4.5.2 Brief technical description

As described in D4.2 [6], we employ a statistical method for node-local analytics based on the use of two byte counters for storing the first and second statistical moment (s1 = $\sum x1 / n$ , s2 = $\sum x2 / n$) observed under $n$ time intervals $\Delta t$. Assuming a log-normal distribution $f(\mu, \sigma)$ for the observed rates, the parameters $\mu$ and $\sigma$ of the distribution can be estimated from the statistical moments and used for representing the observed traffic rates in a compact form. By inspecting the percentiles of the cumulative density function $F(C; \mu, \sigma)$ corresponding to the link capacity $C$, we can assess the congestion risk, $p_r = 1 - F(C; \mu, \sigma)$. When applied at short intervals (seconds or milliseconds), we argue that this metric is significantly more informative compared to the common practice of assessing the 5-min average traffic rates, in which short congestion episodes cannot be observed [55], [56] [57]. Further, we have theoretical results indicating the benefits of node-local congestion analytics in terms of a total message reduction of a factor of at least 3000 compared to when using a similar monitoring function based on SNMP [32].

### 4.5.3 Recent improvements concerning the tool implementation

Since D4.2 [6] we have refined the congestion detector by implementing an autonomous sampling frequency scaling mechanism that samples the empirical rate distribution at high rate, only when there is an elevated risk of congestion. The underlying estimation method is the same as described above, but since the accuracy of the estimate depends on the number of samples used to produce each estimate, and we aimed for a mechanism which reports congestions based a highly accurate risk estimate but whished to reduce the sampling cost as much as possible, we decided to go for detector with variable frequency.

For any given (e.g. initial) sampling frequency, we maintain one distribution estimate, but inspect two risk levels, one to indicate the congestion itself, normally the risk of observing rates above the link capacity over any periods longer than one or a few buffer depths, and one used to scale the sampling frequency up or down depending on the risk of exceeding another threshold, e.g. 90% of the link capacity estimated at the current frequency.

We have identified a single continuous function (see Figure 27) that maps observed risks to sample intervals in a range of 10ms to 300s that works well as an autonomous scaling mechanism. The mechanism contributes to achieve high detection rates while keeping the sampling cost (i.e. the number of samples) low during a monitoring session. The function itself is modelled on the distribution function of a bounded Pareto distribution with three parameters: upper and lower bounds, and slope. . The upper bound corresponds to the longest allowable interval, the lower to the shortest and the slope parameter $\zeta$ how quickly the period decreases with elevated risk levels. In practise the lower bound has minor impact, as long as it is shorter than, and on the same order of magnitude, as the buffer depth. Exploration of the parameter space of this function, which we believe is a representative selection of real world packet trace, together with randomly generated traffic simulations has allowed us to determine a small range of parameter values that appears to be robust for almost all studied cases. The only remaining trade-off is between sampling cost at very low risks, and the ability to scale up fast enough to capture very short isolated congestion events.

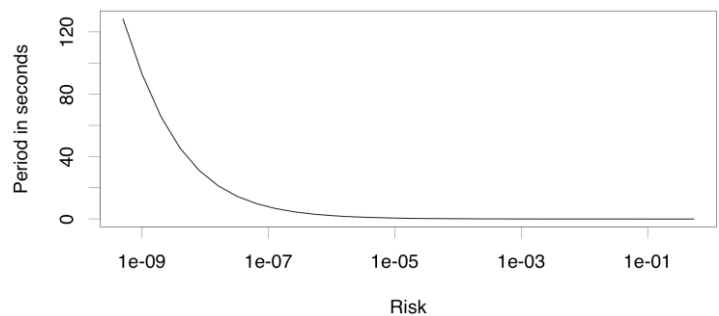

*Figure 27 Mapping risks to sampling intervals, Log scale on X-axis.*

Note that the initial approach based on using only a single counter mentioned in [58] was abandoned as it turned out to be too inaccurate for estimation purposes. . However, a single bit modelling approach has been introduced for compensating for measurement jitter caused by partially arrived packets, as the raw byte counters are normally updated at packet rate, rather than bit rate. For land lines with fixed nominal buffer read-out rates and the total flow on such links, this problem can be remedied by (conceptually) observing the momentary rates at a duration corresponding to the time taken to transfer a single bit [59].

### 4.5.4 Recent improvements concerning the tool integration

The self-scaling mechanism is implemented as an extension to the prototype described in [54] [6] [60] [27], but has not been integrated into the released version of the rate monitoring function (also known as RAMON) as part of the SP-DevOps toolkit. The latter is under integration with other SP-DevOps tools for demonstrating aspects of the outlined monitoring and troubleshooting processes [58].

### 4.5.5 Evaluation results and conclusions

The evaluation includes assessment of detection rate for different congestion buildup times $T_\rho$, and values of the slope parameter $\zeta$ (Figure 28) and sampling interval upper bounds $hb$ in milliseconds (Figure 29) and sampling cost (relative to a sample interval corresponding to a buffer depth of 30ms) and $\zeta$=1.8 (Figure 30). In general, larger slope parameter settings can improve the detection ratio but at a higher sampling cost. This is because the estimation period scales down faster with a larger $\zeta$, and thus can adapt quicker to increased risks. This is most obvious for the fast case ($T_\rho$ = 10s), where the detection ratio is enhanced by 6% with the increase of $\zeta$ from 1.2 to 3.0 (Figure 28 and Figure 30). Specifically, in Figure 28 we can see that increasing $\zeta$ above 1.8 yields only minor improvement for all but the very fast $T_\rho$. Moreover, the larger the upper bound (hb), the longer the "sleep" period of the congestion detector, and thus the higher possibility of missing short term congestion events (Figure 29). From the perspective of sampling cost, we can see in Figure 30 that increasing $\zeta$ from 1.8 to 2.4 and 3.0 significantly increases the estimation cost, especially with slow $T_\rho$.

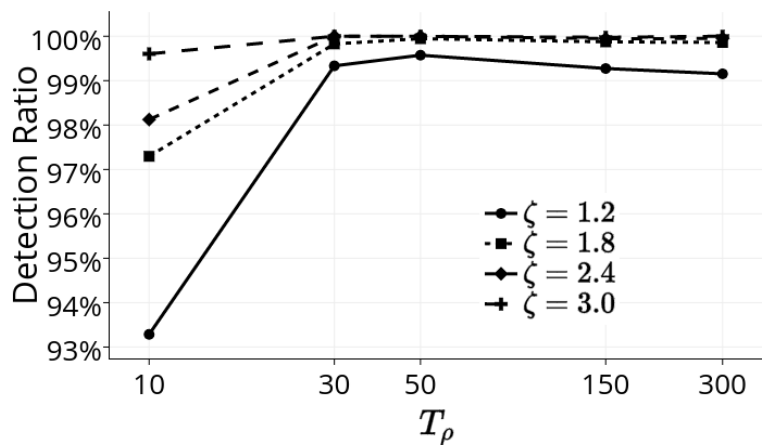

*Figure 28 Detection ratio for different slope mapping function slope parameter $\times$ and congestion buildup times $T_\rho$ in seconds. Upper bound (hb) 30 seconds.*

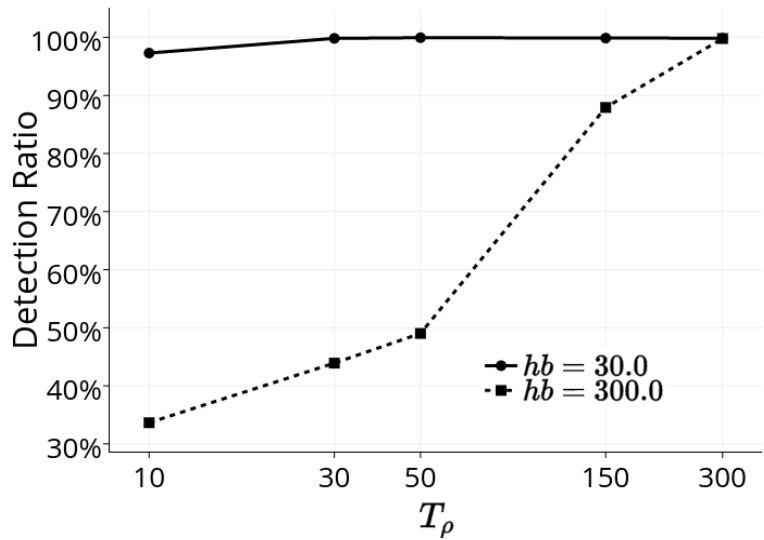

*Figure 29 Detection ratio for two different upper bound (hb) and congestion buildup times $T_\rho$ in seconds for $\zeta$=1.8.*

To conclude, a robust, scalable and efficient congestion detection mechanism based solely on throughput measurements has been implemented as a monitoring function that is well suited to integration into the UNIFY framework. The mechanism is distributed so that high rate operations are performed on or close to the monitored network element, and lower rate operations based on can be centralised, or distributed as required by the deployment scenario. The full details of the detection mechanism and the evaluation can be found in [59].

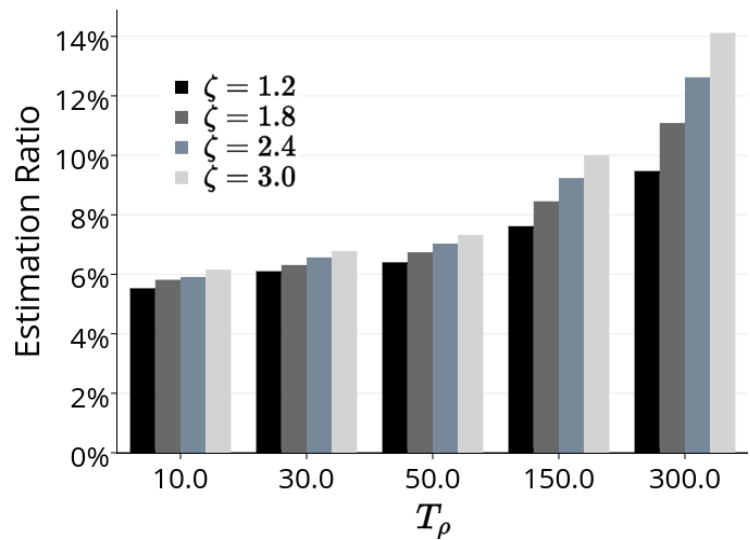

*Figure 30 Sampling cost relative to sampling interval corresponding to buffer depth for different slope mapping function slope parameter $\zeta$ and congestion buildup times $T_\rho$ in seconds. Upper bound 30 seconds.*

## 4.6 Probabilistic delay monitoring

### 4.6.1 Purpose and problem statement

Effective automation and autonomicity in large-scale networks require efficient ways of configuring the management functions at a higher abstraction level. Usually deterministic low-level and fixed thresholds are applied to control the network. For data-driven methods, applying thresholds directly on the data at hand requires more or less exact domain knowledge to be effective and is less robust to short-term variations in the data. A better alternative is to employ probabilistic approaches for data modelling as well as triggering of management functions and automated processes based on probabilistic requirements

Probabilistic approaches offer efficient ways to set detection thresholds and triggers by the means of configuration parameters normalized over the data. Increasing the configurability and controllability of the next generation networks is one of many challenges to address, where the employment of probabilistic approaches play an important role for various aspects of SP-DevOps and orchestration support, such as detection of service degradations and network faults, as well as resource management.

Based on our previous work reported in D4.2, we elaborate on the idea of probabilistically configurable monitoring functions for the purpose of conditional observability and troubleshooting support. Here, we report an extended probabilistic self-adaptive monitoring approach capable of varying its measurement intensity for modelling link delays and detecting changes given probabilistic precision and detection requirements. The benefits of both approaches encompass increased configurability, measurement efficiency and a normalized way to statistically profile changes that are of interest to detect based on probabilistic requirements.

For the purpose of obtaining a computationally efficient approximation of the underlying delay distribution, the approach is based on estimating the parameters of a parametric density function using a statistical moments estimation method. For both probabilistic precision and change detection, the work is focused on the case when the delay follows a Gamma distribution - however, the principles of the approach can be adapted to other delay distributions observed in the literature, such as log-normal or Weibull [61] [62] [63]. Parametric models, when applicable, often provide better and more complete models of the underlying statistical process and can in many cases also be estimated at a low computational cost, compared to non-parametric models.

### 4.6.2 Brief technical description

In the following, we describe the principles for probabilistic precision and change detection.

#### 4.6.2.1 Probabilistic precision

A specific subtask of network monitoring is the design of probing strategies. One of the aims of these strategies is to autonomously adjust the probing intensity or measurement session to meet requirements on precision of the measured values, without specifying the exact number of measurements. Applying such a strategy means that the probing intensity will vary in time and from one link to another to meet the given precision requirement. The advantage of conditional observability based on a statistical precision requirement is two-fold. First, the obtained probabilistic model is guaranteed to fulfil the specified precision requirement without the need of specifying the exact number of samples, which effectively eliminates the risk of oversampling or undersampling the link delay. Second, the approach allows for predicting the number of samples required to obtain a probabilistic model fulfilling the precision requirement, which allows for better planning of the deployment of monitoring functions and the resources needed for monitoring.

The principle of our probabilistic precision mechanism is described as follows: let $x$ be a random variable and θ be a measurement (E.g. mean or variance) of x. Let $\theta$ be the estimated measurement of a set of sample data D from x. We define the precision of an estimated measurement θ by using a pair $(p, c)$, where $p$ is the *precision probability* and where the real number $c$ is the *precision value*, such that, for some given method to construct a *precision interval* I based on a set of samples $D$ and the interval size $|I|$, we have:

$$\overline{\theta} \in I \quad (ii) \ \ P(\theta \in I) = p \quad (iii) \ \ I = \left[\overline{\theta} - \frac{c}{2}\overline{\theta}, \ \ \overline{\theta} + \frac{c}{2}\overline{\theta}\right] \tag{1}$$

The precision value $c$ is regarded as the relative interval size. The larger precision probability $p$ we have, for a constant precision value $c$, the higher precision. Similarly, the smaller precision value $c$ we have, for a constant precision probability $p$, the higher precision.

In order to guarantee the precision of an estimated measurement, we need to have enough samples. The number of required samples depends on random variable from which the samples are sampled as well as the precision requirement. In the case of link delay monitoring, link delay is often a Gamma distributed random variable. Thus, the needed number of samples $H$ for acquiring the estimated mean delay with a precision $(p, c)$ can be derived from the following formula:

$$H = \text{ceil}\left(2\left[\frac{2S}{Mc}\,\text{erf}^{-1}(p)\right]^2\right) + \delta \tag{2}$$

The δ is approximated to

$$\delta = -\text{round}\big(\log(2 - 2p)\big) \tag{3}$$

$M$ and $S$ are the sample mean and sample standard deviation of an initial set of samples $D_{\text{init}}$, respectively. We developed an adaptive on-line algorithm to decide the number of required samples $H$ and get the estimated measurement $\theta$ satisfying the precision requirement. Based on this principle, we designed an algorithm can find the appropriate number of samples to estimate a link self-adaptively. The algorithm first use an initial set of samples to calculate the required number of samples $H$. Then it takes $H$ samples to make the estimation.

### 4.6.2.2 Probabilistic change detection

For change detection, we propose a monitoring function that implements sophisticated detection mechanisms based on controllable requirements on the detection certainty for changes in the observed link delay. This allows for detecting changes with a certain profile (gradual or abrupt development) and magnitude specified by the operator and with probabilistic detection requirements. The advantages with the probabilistic approach of specifying desired changes to detect encompass alarm filtering and simplified configuration with less detailed knowledge needed about the data at hand. The idea is that the observed delays are generated by a statistical distribution which in periods is stable, i.e. it does not change, and in periods is unstable (during a change). Given this view of delay, the change detector detects changes in the delay distribution. The change detector designed by us is concerned with changes in the mean and in the variance of the delay distribution.

We assume that the delay data consist of a series of batches of delays. We let $D_1, D_2, ...$ denote the batches and $x_{i,j}$ the individual delay measurements, such that $D_i = \{x_{i,1}, ..., x_{i,n_i}\}$. We let $M_i$ and $S_i^2$ be the sample mean and sample variance of batch $D_i$, respectively. The concept of change presented here is governed by following two principles or requirements:

1. **False alarm rate** The false alarm rate is less than a given probability
2. **Minimum change** If a change of at least a given magnitude occurs in the delay distribution, then the chance of detecting that change must be larger than a given probability

The idea is to derive the number of sample needed to fulfil the first and second requirements above for the purpose of reporting a potential change with the specified certainty requirements. In the case of mean value change detection, by the false alarm rate requirement, the chance of detecting a change must be smaller than a given probability $1 - p_M$ if there is no change in the delay distribution. Therefore, we construct an interval $I_M$ around the mean $M$, if there is no change in the delay distribution from $D_i$ to $D_{i+1}$, then

$$P(M_{i+1} \in I_M) = p_M \qquad (1)$$

Let $\mu_i$, $\sigma_i^2$ be the mean and and variance of the delay batch data $D_i$. By a change of magnitude $|c_M - 1|$ in the mean value for $D_{i+1}$ with respect to $D_i$ is meant that $|u_i - u_{i+1}|/u_i = |c_M - 1|$. By the minimum change requirement, the chance of detecting a change must be at least equal to a given probability $(q_M)$, if a change larger than a certain magnitude $|c_M - 1|$ occurs. We assume that a change in the mean is detected whenever $M_{i+1} \notin I_M$, where $I_M$ is the same change detection interval for the false alarm rate requirement in equation (1). Then the mean value minimum change requirement, on the number of samples $n_i$ and $n_{i+1}$ is that, if the change magnitude is larger than $|c_M - 1|$

$$P(M_{i+1} \notin I_M) \geq q_M \qquad (2)$$

In the case that the delay data is Gamma distributed, we can derive that

$$I_M = [M_i - tS, M_i + tS] \qquad (3)$$

where $t$ is defined by

$$t = F_{n_i-1}^{-1}\left(\frac{1 + p_M}{2}\right)$$

and where $F_n^{-1}$ is the inverse cumulative distribution function for the t-distribution for $n$ degrees of freedom. The $n_i$ and $n_{i+1}$ is derived as the follows:

$$n_i \approx \text{ceil}\left[\frac{2(c_M z_q + z_p)^2}{\alpha(c_M - 1)^2}\right] \qquad (4)$$

for

$$\alpha = \frac{M_i'^2}{S_i'^2} \quad z_p = \Phi^{-1}\left(\frac{1 + p_M}{2}\right) \quad z_q = \Phi^{-1}(q_M)$$

where $\Phi^{-1}$ is the inverse cumulative function for the normal distribution.

Similarly, for variance change detection, we can also construct an interval $I_{S^2}$ around the variance $S_i^2$ and find the number of samples $n$ to satisfy both the false alarm rate requirement $P_{S^2}$, as well as the variance minimum change requirement ($q_{S^2}, c_{S^2}$).

### 4.6.3 Recent improvements concerning the tool implementation

The link monitoring function described in D4.2 [6] performs successive end-to-end measurements for the purpose of deriving intermediate link delay and loss, without the need of explicitly measuring each link individually. In [64] we evaluated the estimation accuracy of the method compared to the ground-truth link delay and loss in a simulated network environment in NS2. Moreover, we showed that the method, compared to a SDN-controller based link monitoring approach, can reduce the measurement overhead by the order of $O(n^2)$.

Since D4.2, we have focused on developing the statistical precision mechanism for controlling and configuring conditional observability points in a network with focus on delay measurements. Moreover, we developed a probabilistic approach for change detection in observed delays. Both cases have so far been evaluated in OMNET++ for single links. The approaches can currently be applied to cases where individual links are monitored (e.g. physical one-hop links or for monitoring service performance end-to-end), but is extendable to the successive end-to-end measurements as described in D4.2 [6]. . Details about the precision mechanism and the probabilistic change detection approach can be found in [65]and [66] respectively

### 4.6.4 Evaluation results and conclusion

The delay monitoring function is not part of any integration activity within the UNIFY project. The described functionalities of the monitoring function developed since D4.2 [6] have been evaluated in a stand-alone framework implemented in OMNET++.

#### 4.6.4.1 Probabilistic precision

We refer our probabilistic precision method as the *ad*, since it can adapt the number of required samples for an estimation to a precision requirement and sample distribution. We compare the results of using this method to reference methods using a fixed number samples per estimation, and refer to such methods as *ref n* for any given *n* number of samples.

In the comparison we use *hit rate* and *sampling cost* as metrics for our evaluation. The *sampling cost* is defined as the average number of samples used per estimation.  In our experiments, the delay of a link is a Gamma distributed random variable with two parameters: shape and scale.  For each random variable and under each precision requirement, we repeatedly use our precision mechanism to estimate the mean for 2000 times.  Hit rate is the number of the estimations that satisfies the precision requirement over the total estimations. Ideally, it should be equal to the $p$ of a precision requirement $(p, c)$.

Firstly, we test our method on link delay of different Gamma distributed random variables. We fixed the scale and vary shape parameter from 1 to 500. We compare the results with reference methods *ref* 10 and *ref* 100, respectively. From Figure 31, we observe that the hit rate of the *ad* method varies in a small range (0.89,0.92) when shape is below 100. This shows that the estimations made by our method can satisfy the required precision requirement $p = 0.9, c = 0.1$ with statistical significance. We also observe that with the ad method the sampling cost varies for each random variable. The method adapts the sampling rate to the link delay distribution that as the shape parameter increases, and the sampling cost decreases. For the reference methods, the sampling cost remains constant (10 for *ref 10* and 100 for *ref 100*). As a result, on one hand, too few samples are used when shape values are small, resulting in hit rates far below the desired precision. On the other hand, when the shape values are large, it does fulfil the precision requirement but often to a higher degree (and higher sampling cost) than required.

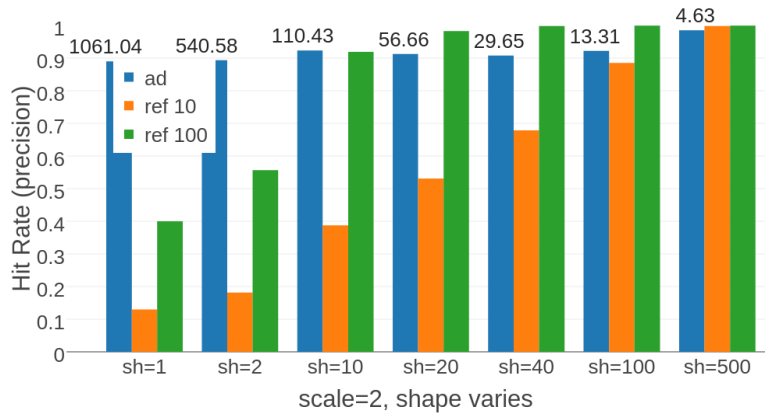

*Figure 31 Hit Rate of different models with different shape (sh) parameters for the delay distribution. Two reference models are used, ref 10 and ref 100. The annotations are the sampling costs of the ad model.*

Secondly, we apply our MF in variant networks of different sizes (from 26 links to 4090 links) generated by using Brite Barabasi model. As Figure 32 indicates, the total number of samples used to estimate all the links inside a network increases as the network size grows. For example, as Figure 32a shows, at a precision requirement (0.90,0.10), *ref 100* is the most costly one, while *ref 10* has the lowest cost. Our *ad* method falls between them, which means the sampling cost of our *ad* model is less than 100. In practice, 100 samples per estimation are very small cost. Probe tests are often much longer than that. However, our *ad* model can provide accurate estimations with less than 100 sampling cost.

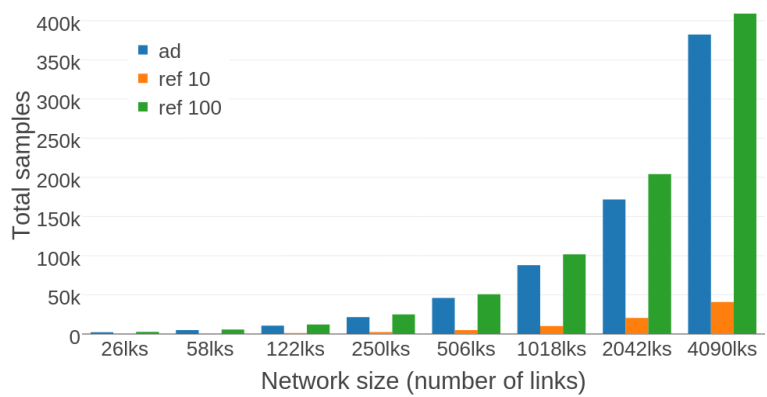

*Figure 32 Sampling costs in networks generated by Brite Barabasi model (p = 0.90,c = 0.10).*

In conclusion, our self-adaptive monitoring function is capable of estimating the parameters of Gamma distributed delays while fulfilling specified requirements on the model precision. This facilitates configuration as only the precision requirements are needed. The probabilistic design of the monitoring function provides means to control the monitoring result in a scalable and resource efficient manner across large-scale networks than can be done with most monitoring functions operating by a fixed number of measurements. The experiments illustrate that fixed monitoring functions deployed with identical configuration across a network will produce parameter estimates of highly varying precision, and with a higher total monitoring overhead (up to 22 times) compared to the suggested probabilistic monitoring function.

### 4.6.4.2 Probabilistic change detection

We implemented our probabilistic change detector in Omnet++ to monitor the delay of a link, which is a Gamma random variable. For evaluation, we randomly make changes on the shape and scale parameter of the Gamma random variable, in order to vary the mean between $0.02s$ and $1s$. The time interval between two neighboring changes is uniform randomly distributed with an average of $500s$. We setup our change detector to catch changes larger a required magnitude $c$ in every $50s$. We classify those random generated changes into three categories:

- *Valid changes* are changes that are larger than the magnitude $c$ and happen during the "ready" state of the change detector. For valid changes, our detector is supposed to catch them with a probability larger than $p_M$.
- *Small changes* happen during the "ready" state of the change detector, but with magnitudes smaller than the magnitude $c$. For those changes, the behavior of our detector is undefined. It may or may not detect those changes.
- *Invalid changes* are the changes apart from valid changes and small changes. They happen during the "training state" of our change detector. They are supposed to be ignored by our change detector.

Correspondingly, we also classify the detections of our detector into three categories.

- *Valid detections* are the detections for valid changes. A valid detection should occur within $100s$ after a valid change happen.
- *Small change detections* are the detections for small changes. A small change detection may occur after a small change within $100s$.
- *False detections* are detections made by our change detector when there is no change. False alarm rate is defined as the number of false alarms over the number of total estimations.

| Detection requirements | | | Test traffic properties | | | Detector performance | | | |
|---|---|---|---|---|---|---|---|---|---|
| $p_M$ | $q_M$ | $c$ | Total changes | Valid changes | Small changes | Valid change detection rate | Small change detection rate | False alarm rate | Avg. Samples per estimation |
| 0.95 | 0.99 | 0.05 | 1999 | 1676 | 85 | 100.0% | 71.8% | 2.3% | 3721.3 |
| 0.95 | 0.99 | 0.10 | 1999 | 1713 | 177 | 99.9% | 71.2% | 2.1% | 1262.6 |
| 0.95 | 0.99 | 0.50 | 1999 | 1129 | 182 | 99.8% | 66.1% | 3.8% | 66 |
| 0.95 | 0.99 | 1.00 | 1999 | 475 | 1447 | 99.8% | 75.6% | 8.6% | 25 |
| 0.99 | 0.99 | 0.05 | 1999 | 1614 | 55 | 100.0% | 68.8% | 0.4% | 4766.3 |
| 0.99 | 0.99 | 0.10 | 1999 | 1703 | 117 | 99.9% | 65.7% | 0.4% | 1610.2 |
| 0.99 | 0.99 | 0.50 | 1999 | 1119 | 462 | 99.4% | 57.2% | 0.6% | 96 |
| 0.99 | 0.99 | 1.00 | 1999 | 480 | 862 | 98.5% | 60.9% | 2.1% | 36 |

*Table 2 Test results of the change detector with different detections requirements.*

As Table 2 suggests that our change detector has a very good detection rate for valid changes. Besides this, the false alarm behaviour of our detector is as required. The average required samples per estimation vary with the detection requirement. The higher the false alarm rate, the smaller the magnitude the larger cost.

In summary, our change detector is capable of self-adaptively find out the appropriate number of samples for detecting changes. It can capture changes larger than a required magnitude with an expected probability while keeping the false alarm rate as low as required. The sampling cost varies with the detection requirement and false alarm requirement.

## 4.7 EPLE – efficient SDN loss monitoring

### 4.7.1 Purpose and problem statement

EPLE is a monitoring function that estimates packet loss for aggregated flows in SDN-based networks. For most practical applications of SDN in both transport and cloud domains, we expect extensive usage of aggregated (i.e. wild carded) flow entries in forwarding devices (in the current implementation OpenFlow switches), as opposed to fully specified microflows without wild-carded match fields. As a prominent example, traditional routing is based on flow-rules aggregated by destination addresses or address ranges (i.e. subnets). We thus identified proper performance measurements of aggregated flows in an SDN environment as an interesting and relevant research challenge [25]. Furthermore, our goal is to perform these measurements in a manner that is efficient from both data- and control plane perspectives.

### 4.7.2 Brief summary of concept and design choices

As outlined in D4.1 [25], our work was inspired by FlowSense [67] for low-cost monitoring and DevoFlow [68] in terms of using an automated devolving mechanism. In EPLE, we realize overhead efficiency of the measurement method by not requiring any additional probe packets and piggybacking signaling traffic and measurement results on OpenFlow messages. Furthermore, EPLE relieves the controller from analyzing each individual flow contained within an aggregated descriptor. Specifically, EPLE addresses the shortcomings with respect to network performance measurements (specifically packet loss as a first example) of existing methods by:

- supporting aggregated flows

- providing a policy-steered selection of flows to be devolved, as well as a timer and policy-based mechanism for selecting which flows should be communicated to the controller as candidate flows

- using a policy-steered timer that forces expiration of a flow from the table (through setting the hard_timeout parameter in the OpenFlow ofp_flow_mod structure associated with the devolved flow), thus forcing the transmission of a `OFPT_FLOW_REMOVED` message even while a long-lived flow is still being active in the network (note that this feature is optional is not supported in the current prototype)

EPLE calculates a simple packet loss estimate based on plain reading of packet counters at the expiration of a devolved flow. In the default setting, the counters are read at the ingress and egress points of the flow, although intermediary measurement points can also be configured in case more detailed investigations are desirable. The details of EPLE and more motivations for its design choices are described in D4.2 [6] and a concept paper [69].

### 4.7.3 Description of prototype implementation

The monitoring function has three major components:

- A monitoring application that installs measurement policies, estimates packet loss based on packet counter data received from the nodes and publishes loss results:

  The monitoring application is running as a feature on the chosen controller platform (OpenDayLight Lithium (ODL)). Two MD-SAL models have been defined: one for specifying the aggregated policies and the other for publishing loss results.

  The monitoring application implements three ODL listeners. As a DataChangeListener, it listens to the aggregated policy changes and calls the ODL REST API to add/delete aggregated flows to/from the switches. As an ODL PacketProcListener, it listens to the Packet-In message with reason 'devolve' from the ingress switch. In such a message, the selected microflow match is included, and the monitoring application uses the information to build/install a same-match microflow on the egress. An egress switch can be defined in the devolve policy by a user or if not defined, it can be calculated. The monitoring application uses the ODL topology service and calculates egress based on the microflow destination mac address.

  Finally, as an ODL SalFlowListener, the application listens to the flow expiration messages from which it gets the packet counts and calculates loss. Two flow expirations, one from ingress and the other from egress, are expected before a loss calculation. In order to match the expired microflows, the monitoring application implements a microflow repository that maintains the states and matches between the microflows. A flow is identified by a unique flow-id in ODL, but in OVS, there is no such identifier. We thus use cookie field for uniquely identifying the microflows from the OVS side. There are two parts of a cookie, devolve flow id and time stamp. The first part identify the devolve rule, and the second part identifies the microflow generated based on the devolve rule. The first part is set by the monitoring application when it installs the devolve rule on the ingress switch. The second part is provided by the ingress switch when the microflow is generated. The same microflow installed on ingress and egress shares the same cookie.

Once a result is calculated, the monitoring application will act as an MD-SAL provider that publishes the measurement results. Any consumers that subscribe to it get notifications when the results are available. Figure 33 shows the overall design of the monitoring application.

- Extended ODL implementation supporting additional messages:

  The ODL OpenFlowPlugin and OpenFlowJava packages were extended to support a *devolve* action and as well as *packet-in* with 'devolve' (OFPR_DEVOLVE) as a reason. The new devolve action is implemented as a vendor extension. Devolve policies (such as the microflow idle-timeout value) are encoded into the action body.

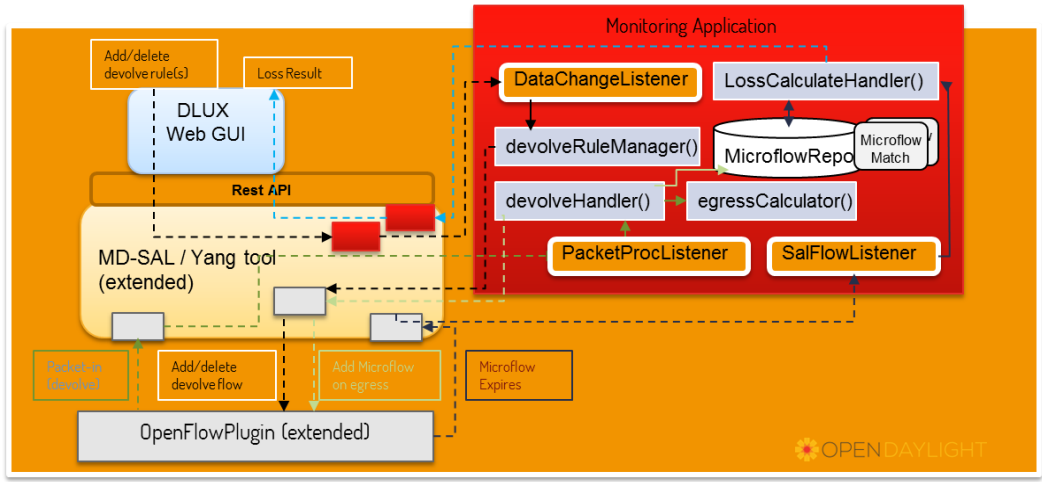

*Figure 33: EPLE control app and extensions to ODL.*

- Extended virtual switch supporting for specific functionality in the flow tables that allow automatic creation of individual packet flow rules:

    The chosen software switch (Open vSwitch 2.4 (OVS)) was extended to support a *devolve* action as well as *packet-in* with 'devolve' as a reason (see Figure 34). The *devolve* function generates an exact match microflow when a devolve rule applies, and consequently send *packet-in* to controller. The microflow generated is set to a priority higher than the devolve flow. This ensures that the next coming packets match the microflow instead of the devolve flow.

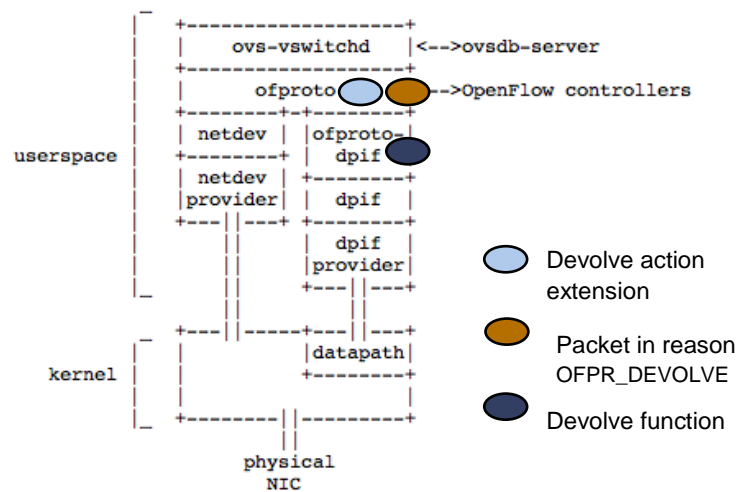

*Figure 34 EPLE extensions to OVS.*

### 4.7.4 Evaluation results and conclusions

In the following subsections (main results also published in [70]), we first validate the method with respect to the correctness of the results. We then assess the actual efficiency from both data- and control plane perspectives, which have been the original requirements leading to the design of EPLE. Specifically, we investigate how the method trades-off compared to competing active measurement methods with respect to dataplane and control plane traffic volumes as well as computational overhead.

### 4.7.4.1 Validation

The EPLE PoC has been verified on a setup with two VMs, each with 4 x 2094 MHZ CPUs and 8GB memory. Both VMs run Ubuntu 14.04. On one of the VMs, we installed the extended ODL controller and the monitoring application, and on the other, we run Mininet integrated with the extended OVS datapath elements.

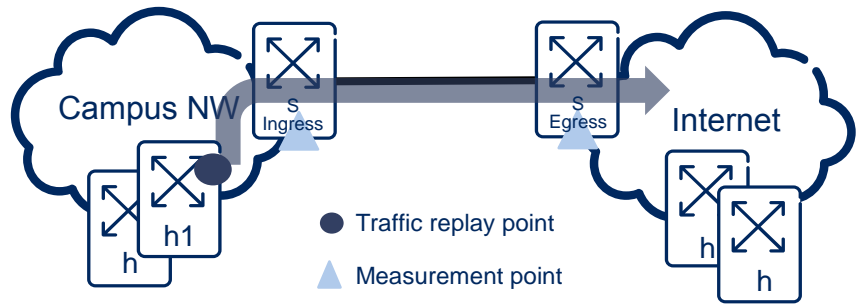

*Figure 35: Mininet test network setup.*

We have validated EPLE via various flow samples retrieved from real network traces that were captured on the border between a campus network and Internet [71]. We simulated the network with Mininet and play the traffic from a host connected to the ingress switch. The simulated network setup is shown in Figure 35. From the validation tests, we observed that EPLE was capable of estimate the loss for ICMP, TCP and UDP microflows. The loss estimation is 100% accurate if the interarrival time of the first two packets in a flow is larger than the microflow installation delay on the ingress switch, larger than the link transmission latency between the ingress and egress switch, and also larger than the egress microflow installation latency (which includes the sum of the ingress and egress switch control channels latencies and the processing latency at the controller). All of these latencies typically are in the range of few up to maybe a few hundreds of milliseconds. Selecting flows with sufficiently large initial packet inter-arrival time can thus be feasibly achieved by providing a policy for devolving rules to only auto-devolve microflows that typically show characteristics that meet these packet interval requirements, e.g. ICMP flows (often coming from ping operations with 1sec packet inter-arrival times) and new TCP connections, starting with slow start phases[3] (identifiable by the occurrence of SYN flags in the TPC header of the initial TCP packet).

In Figure 36, we can see an examples of loss rate estimates by single microflows generated by the EPLE monitoring application. It shows the frequency distribution for 1000 ICMP flow samples. The flow samples include on average with 195 packets (8190 bytes) per flow, with an average packet rate at 930 kbps. Using Linux *netem* [72], we emulated 10 % of random packet loss on the link between ingress and egress switches. In this case, the devolving rule configured on the ingress switch selected all ICMP microflows for loss estimations. We can see that loss rate calculated complies with the actual packet loss generated by the netem tool following a normal distribution centering around 10%, as expected. In fact, this particular set of samples has a mean loss estimate of 9,95% with a standard deviation of 2,08.

---

[3] Packets in one direction during a TCP slowstart phase (SYN or SYN/ACKs) are roughly spaced by 1 RTT, corresponding to approximately 2 times the one-way delay between the endpoints. Being only a sub-segment of the path between TCP endpoints, the EPLE ingress to egress latency is always less or equal to the TCP one-way delay, and has thus at least a safety margin of 1 one-way-delay compared to the RTT.

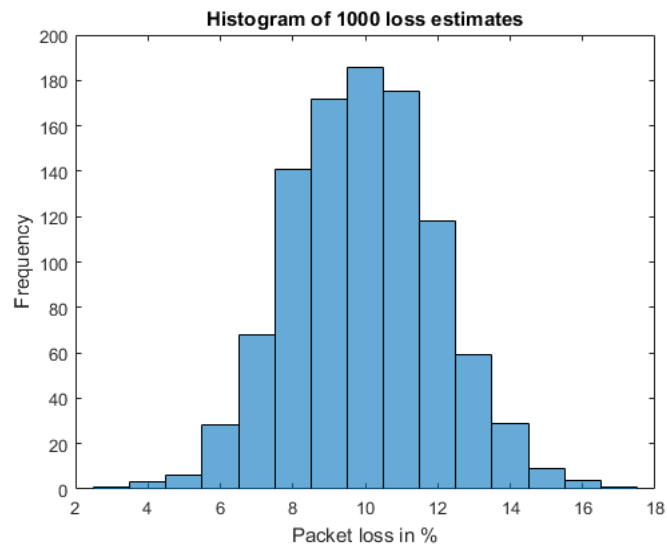

*Figure 36: Frequency distribution of loss rate estimates by 1000 flow samples.*

### 4.7.4.2 CPU and Memory overhead for devolving function

In this section, we present the measurement result for the devolvement processing overhead introduced on the ingress switch. The measurement method is described as follows: we create traffic traces with different numbers (10, 100, 300, 600 and 1000) of ICMP microflows and in each trace, with all the microflows having the exact same timestamp. This simulates concurrent packet flow arrivals.

The traces are run and compared with three configurations on the ingress switch: i) auto-devolving of all the microflows belonging to a wildcard flow match; ii) forwarding all the packets to next hop according to a wildcard flow match without auto-devolving of microflows; iii) and sending all the packets to the controller. This allow us to compare the extra switch resource overhead of EPLE (case i) compared to both a standard proactive OpenFlow usage (case ii) and a standard reactive OpenFlow usage (case iii), which could be used to setup microflows by the controller instead of an autonomous auto-devolve action on the switch. In each scenario, we measure the average and peak CPU and memory usages. In each measurement, we record the CPU and memory in a 10 second period and 0.1 second sampling interval.

Figure 37 and Figure 38 and illustrates the CPU related measurements, once as a metric averaged over 10 seocnds, and once as a metric averaging over only 0.1 seconds, which we use to estimated peak CPU utilization values. We can see that for devolving up to 100 concurrent microflows, EPLEs auto-devolve action causes no significant processing overhead on the ingress switch. However, there is an increase of processing overhead between 100 and 300 concurrent flows, consuming up to 40% of the CPU load at peak time. On the other hand, the number only increases by 3% on average in 10 second periods. From 300 to 1000 microflows cases, the CPU usage is levelling out at around 5% on 10 second average and 35% at peak.

We believe that the EPLE devolve processing overhead is reasonable because in real cases, it is unlikely to selecting more than 100 concurrent microflows on one specific ingress switch for loss measurement. And even if it would be the case, a 3% of CPU increment on average and the 35% CPU peak usage on a relatively weak hardware setup as in our test machine indicates that the additional resource consumptions for running EPLE on a soft-switch are within reasonable boundaries.

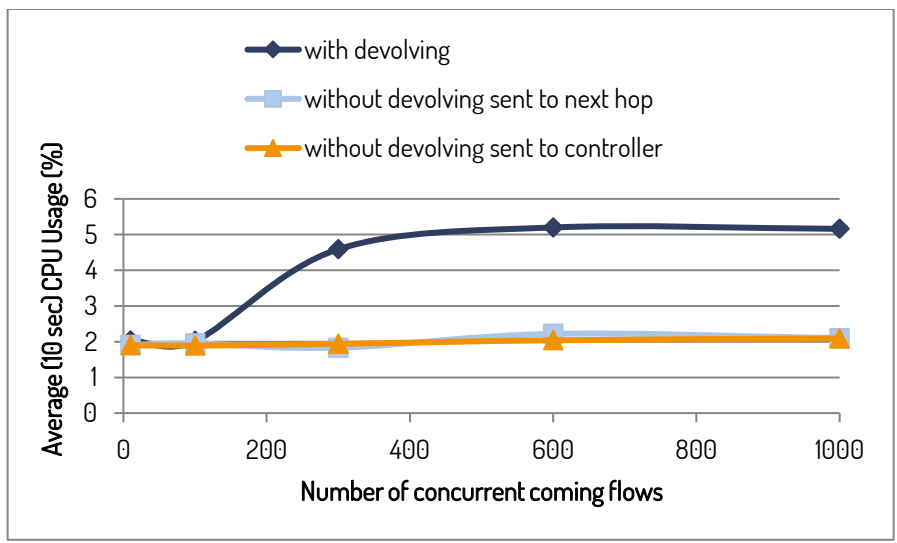

*Figure 37 Switch CPU overhead by devolving function - peak values.*

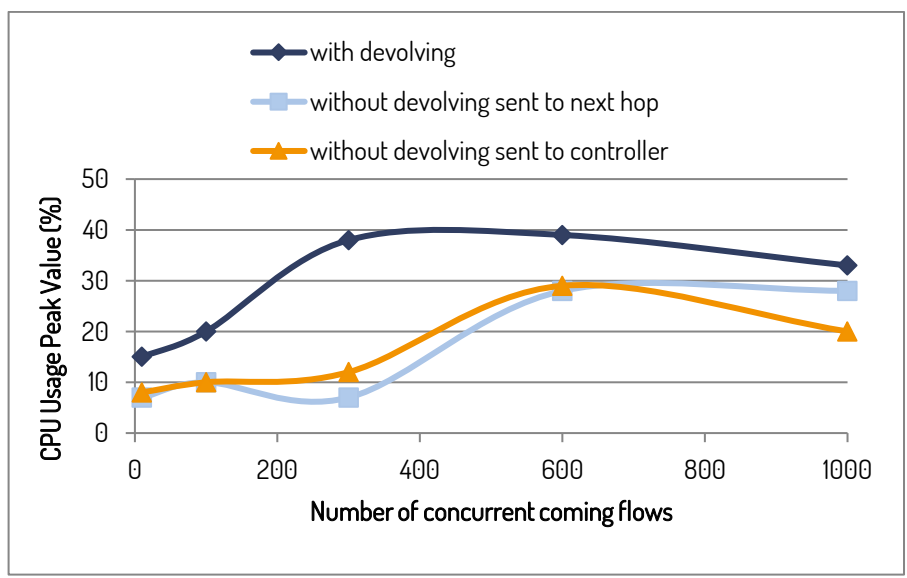

*Figure 38: Switch CPU overhead by devolving function - values averaged over 10 sec.*

In Figure 39 and Figure 40, we illustrate the resulting memory overhead. We can see that for devolving up to 600 concurrent microflows, there are no significant changes in memory usage compared to the average memory usage when the switch is not handling traffic at all. There is a small amount of memory usage increment for 1000 concurrent microflows. However, we observe no significant difference between the 'with' and 'without' devolving cases. We can conclude that the devolving function does not cause much memory overhead in a pure software implementation of a switch.

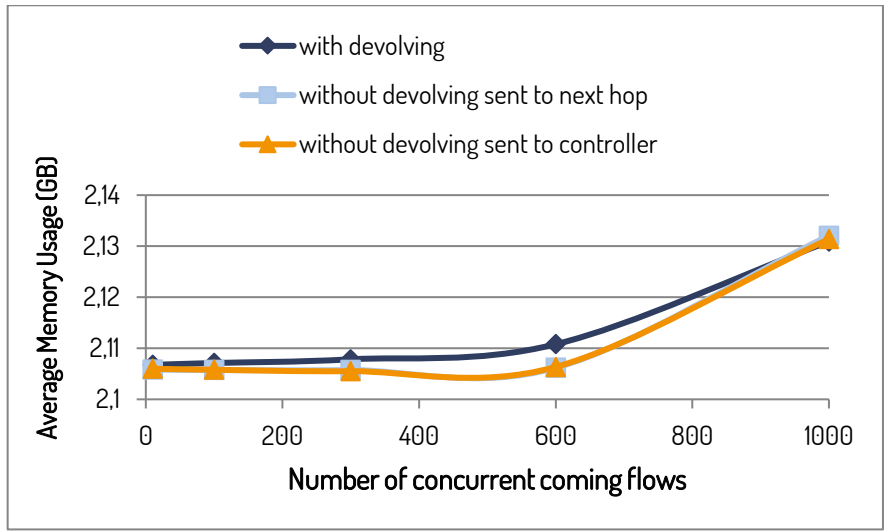

*Figure 39 Softswitch memory overhead by devolving function - values averaged over 10 sec.*

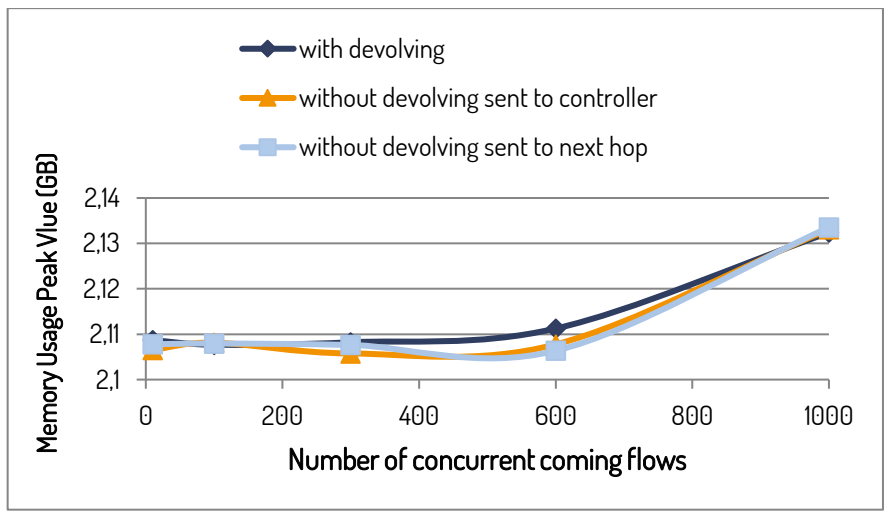

*Figure 40: Softswitch memory overhead by devolving function - peak values.*

### 4.7.4.3 Control Plane Overhead Evaluation

For evaluating the control plane overhead of EPLE, we consider two alternative ways of determining packet loss based on existing standard tools. While neither of these alternatives was designed particularly for SDN networks, they are widely used today from an operational perspective and comparing the control plane overheads allows us to determine whether a SDN solution such as EPLE may provide advantages beyond relieving the data plane overhead.

The first alternative is OWAMP [73], a one-way active measurement protocol that estimates both packet loss and delay at the IP layer. An implementation is publicly available in the toolset offered by Internet2 [74]. The second alternative is TWAMP Light [75], a version of the two-way active measurement protocol that estimates several metrics while removing the control plane part of the complete TWAMP protocol. We based our comparison on an existing internal Ericsson Research implementation of TWAMP Light.

For OWAMP, we used the following command lines for the server and client:

Server: `owampd -a O -P 21000-25000 -S 192.168.100.4:10002 -Z`

Client: `owping -c number_of_packets -A O -P 21000-25000 192.168.100.4:10002`

The command lines disable the authentication for each of the measurement sessions, because we did not implement authentication in EPLE. The server and client were executed each on its own virtual machine, on network interfaces that carried no other traffic. Since we intend to also evaluate the overhead of the control and management plane, this is a realistic environment that allowed us to easily capture and isolate the control and management traffic using the well-known tcpdump tool. The *number of packets* parameter on the command line of the Client represents the number of active measurement probes that are generated by the tool during one test session. This is equivalent with the length of the microflow selected by EPLE. We used the following values to represent both very short lived and longer-lived microflows: 10, 100, 1000, 10000.

The probes are sent at constant time intervals – the implementation would also allow specifying exponentially-distributed inter-probe times, but this is not relevant for determining the control and management plane overhead as such option would be included in the first packet that is sent anyway. In the calculation of the management overhead on the control channel, we ignore the TCP SYN/ACK and FIN/ACK packets that open and close the connection of the measurement session. This is because we assume that in a real deployment both EPLE and OWAMP would keep measurement sessions open when operated in a pro-active mode. Figure 41 shows the management and control overhead of OWAMP as reflected in the traffic traces, for a situation when one measurement session is active. The EPLE overhead is constant for one microflow, composed of:

- At the ingress node: one OpenFlow Packet_In message, one OpenFlow Flow_Mod message, one OpenFlow Barrier Request and one OpenFlow Barrier Reply, and one OpenFlow Flow_Removed message (in total 854 bytes for our OpenFlow 1.3-compliant implementation)
- At the egress node: one OpenFlow Flow_Mod message, one OpenFlow Barrier Request and one OpenFlow Barrier Reply, and one OpenFlow Flow_Removed (in total 576 bytes for our OpenFlow 1.3-compliant implementation)

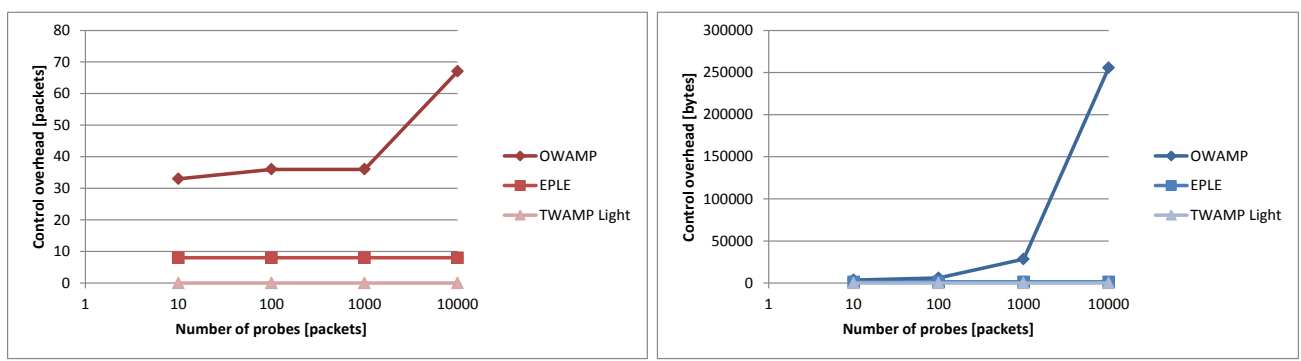

*Figure 41: OWAMP management/control overhead on control channel.*

The owamp tool, line with the specification of the standard, provided more information than EPLE. For example, results from one-way delay measurements are included. No packet loss was present in these measurements. The traffic comes in two bursts, one at the beginning of the measurement session that negotiates the parameters of the session and another burst at the end of the session for transferring the results from the server to the client. We repeated the OWAMP measurements when a 10% random packet loss was introduced on the measurement path using NetEm, similar to the scenario described for EPLE. We observed no significant change in the overhead of the control channel that would account for the reporting of the lost packets for the sessions length (up to 10000 probes) that we measured

For TWAMP Light, the implementation we used required no control plane. The following parameters were transmitted over the command line:

Server: reflIP reflPort

Client: sendIP sendPort reflIP reflPort sessionStart pktIntvl nPkts fwdTOS fwdTTL fwdPktSize

Probes were sent one packet a time with a pre-defined packet interval between them. Results are presented on the standard output of the terminal. Assuming that each command line parameter is represented by a managed object, and each managed object is configured and read using either the SNMP or Netconf protocols in line with the findings of Hedstrom et al [hws11], the overhead is shown in Table 3. The length of the requests and replies is indicative, and depends on how exactly the data model would be represented in an SNMP MIB or in a YANG description. We further assume that each parameter is read using one snmpget call, and written with one snmpset call using the Linux SNMP implementation (using snmpwalk would potentially reduce the read overhead).

*Table 3: Protocol overhead comparison of EPLE with SNMP and Netconf, based in [76].*

| Protocol | Get [packets] | Set [packets] | Get [bytes] | Set [bytes] |
|---|---|---|---|---|
| SNMP TWAMP Light | 10 server, 2 client | 20 server, 4 client | 90 | 2688 server, 13440 client |
| Netconf TWAMP Light | – | 49 server, 50 client | – | 8500 server, 10475 client |
| OpenFlow – EPLE | 1 ingress, 1 egress | 4 ingress, 3 egress | 202 ingress, 202 egress | 650 ingress, 374 egress |

In conclusion, the control and management plane overhead of EPLE is competitive compared to both OWAMP and TWAMP Light. While OWAMP provides additional information (such as delay measurements, which our current implementation of EPLE does not support), we note that adding one-way delay measurements would mean introducing a timestamp value in the Flow_Removed message which adds 4 additional bytes of overhead.

### 4.7.4.4 Data Plane Overhead Evaluation

One of the main advantages of EPLE is actually its efficiency with respect to data plane overhead. As the method is designed to use existing traffic as samples for performance estimates, the dataplane overhead introduced zero. To contrast this to the dataplane overhead introduced by active methods, Figure 42 illustrates the dataplane overhead of OWAMP and TWAMP light. In our configuration, OWAMP generated 60 byte sized probe packets, and TWAMP used 92 byte sized packets for the direction from the client towards the reflector. These were the default values for the packet size parameter. We decided to keep them in order to illustrate the simplest possible usage. For TWAMP Light, the graphs show the overhead for measurements in one direction in order to be able to compare with both OWAMP and EPLE which are uni-directional.

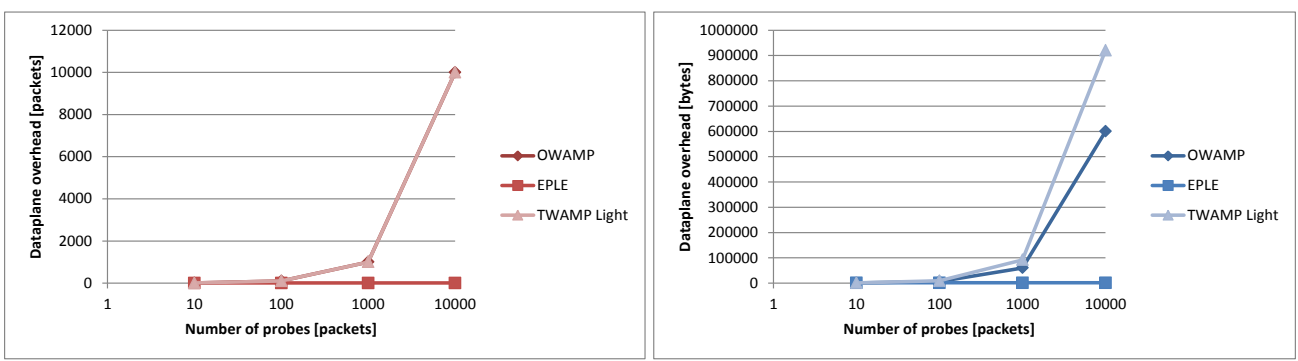

*Figure 42 Data plane overhead of OWAMP and TWAMP light*

### 4.7.4.5 Conclusions

EPLE was designed as a tool to meet a specific research challenge identified in D4.1, i.e. RC3 – "Low-overhead performance monitoring for SDN". With respect to the design goals specified, EPLE got a long way, but we still identify open issues which EAB plans to work on in internal project beyond the duration of UNIFY:

- RC3 design goal: Accurate determination of link and flow performance (i.e. loss, delay, throughput)

  - The current EPLE PoC can only estimate loss metrics for links and flows. However, the EPLE design allows for one-way delay measurements only by introducing a timestamp value in the Flow_Removed message. Furthermore, we are working on further ideas and concepts for additional fault and performance functions, but these are not published and implemented yet. As future work, Ericsson is working on expanding EPLE into a full-fletched assurance framework providing both performance and fault measurement functions for SDN.

- RC3 design goal: Scalability with respect to signalling and data-plane overhead

  - As shown in the evaluation section above, EPLE successfully reduces overhead on both control plane (signalling) and dataplane compared to competing active methods. This comes with a very modest cost in terms of additional compute requirements for the modified softswitch implementations.

- RC3 design goal: Generic, technology independent solution based on the existing SDN components

  - As described above, EPLE was implemented as add-ons to publicly available SDN components, i.e. OpenDayLight as controller, OvS as switch, and OpenFlow and control channel protocol. Utilizing the flat namespace of OpenFlow, it allow a generic way to assess loss metrics irrespective of the exact forwarding technology used.

- RC3 design goal: Effectiveness against granularity of flow definitions and flow-deployment mode

  - As shown in our evaluation, EPLE can estimate loss even in configurations with pre-deployed (proactive) aggregated flows. Obviously, using re-active mode of OpenFlow in the presence of microflows works well with native OpenFlow switches even without the auto-delolve action implemented.

- RC3 design goal: Programmable interfaces to facilitate automation (e.g. for troubleshooting purposes)

  - Right now, EPLE functionality is programmed (or rather configured) through the OpenDayLight REST interface. Testing this interface for usefulness in the scope of automated troubleshooting or on-demand monitoring purposes is subject of future work, to be done within Ericsson internal 5G integration activities outside the scope of UNIFY.

## 4.8 IPTV Quality Monitor

### 4.8.1 Purpose and problem statement

Today's Internet traffic is dominated by multimedia traffic which contributes more than 40% and 60% in Asia-pacific/European regions [77] and in America [78], respectively. A large fraction of this traffic is delivered over IPTV technologies and it shows a growing tendency since the Internet infrastructure bandwidth makes the delivery of multimedia content cheaper, clearer, and more reachable than ever before. It is undeniable that the IP-based infomediary will soon dominate the way of the content consumption if it did not happen yet.

However, the condition of a network may drastically change over time and thus the quality of content is highly dependent on it. Therefore, the measurement of the network performance is an important task not only for the network infrastructure owners, but also for those who provide service platforms and content delivery.

However, when it comes to the assessment of the audiovisual quality of multimedia content, it is no longer in a frame of the network performance evaluation since the value of the payload is different from one packet to another. Therefore, more application-specific network functions are essential to maintain the quality of the multimedia delivery within telecom service platforms. To this end, we introduce the IPTV Quality Monitor which is specifically designed network function in order to evaluate MoS (Mean Opinion Score) of IPTV Service Quality by taking packet-level Internet traffic as input.

To see the development history of the IPTV Quality Monitor, we refer to D4.2 [6] (Section 5.9)

### 4.8.2 Brief Technical description

The IPTV Quality Monitor is proprietary software that implements the standardized IPTV quality monitoring mechanism (ITU P.1201.2: Parametric non-intrusive assessment of audiovisual media streaming quality). DT has acquired a special license to use this software for the purpose of the research and demonstration. The software is originally intended to run in a separate hardware and to monitor the audiovisual quality of the IPTV playback from the viewpoint close to end-users (e.g., between a residential gateway and set top box). However, it can be used as an independent network function at any point of a network service when it is deployed in a virtualized environment (e.g., virtual machine).

ITU P.1201 series is a parametric quality assessment mechanism based on the various protocol headers (including IP, UDP, RTP, and MPEG-2 headers). Especially, ITU P.1201.2 model focuses on the analysis of audio and video codecs, e.g., H.264, AAC-LC, HE-AAC, MPEG1-LII, and AC3, which are typically used for the IPTV service. The model takes the quality impact of (audio/video) compression and transmission errors into consideration to assess the overall audiovisual quality. A detailed and mathematical description of the model can be found in the study of Garcia et.al. [79].

The IPTV Quality Monitor is intended to be used as a monitoring application (i.e., a VNF in the catalogue), but it can, in principle, be integrated in a Universal Node (UN) of the UNIFY framework as a Monitoring Function. In this case, the major role of the IPTV Quality Monitor would be to periodically inform the performance (e.g., performance degradation period, duration, MoS, and/or loss rate) of multimedia traffic to the higher layers.

A potential mapping scenario of the IPTV Quality Monitor to Universal Node (UN) is described in D4.2 (Section 5.9)

### 4.8.3  Recent improvement

The IPTV Quality Monitor is proprietary software and thus its modification is highly limited. Therefore, adapting this software for other service platforms was the major challenge from the beginning. Recently, the developers of the software have provided DT with the core of the software as a service-platform-independent library, thus it allowed us to integrate the tool into a stand-alone test-bed

### 4.8.4 Evaluation results and conclusions

The performance of the model (ITU P.1201.2) is thoroughly studied in Garcia et. al. [79]. Their experiment was conducted within standardized test rooms (ITU-R BT-500, ITU-T P.800/P.910) by using professional equipments. Performance degradation factors such as compression (audio/video codecs and bitrates) and transmission errors (random and bursty packet losses) were emulated in the experiment. The authors (the actual developers of the core software) report in their paper that the experiment results show 0.435 of RMSE (Root Mean Square Error) on 5-point scale and 0.911 of Pearson correlation based on 3190 test samples.

Besides, it is planned to perform an experiment using commercial IPTV service within a managed test-bed. Our goal in the experiment is to observe the correlation between the general network condition (e.g., throughput) and the audiovisual quality (e.g., in MoS) extracted by the IPTV Quality Monitor.

This tool will not be publically available due to its closed license, however virtualization of this network function will show a potential for an application-specific monitoring function within the scope of a modern service creation framework such as UNIFY.

## 4.9 AutoTPG – Verification of Flow Matching Functionality

### 4.9.1 Purpose and Problem Statement

AutoTPG is an OpenFlow based tool that performs verification of matching of incoming packets with the Flow-Match Header of a Flow Entry. These errors occur when matching of a packet with the Flow-Match Header gives an incorrect result (i.e., a packet gets matched with the Flow-Match Header which it should not, or a packet does not get a match although a matching Flow-Entry is present in the FlowTable). There can be two causes of incorrect matching of packets: (1) bugs in OpenFlow switch implementation and (2) errors in FlowTable configuration. Bugs in OpenFlow switch implementation may be caused by bugs in the hardware or software part of switches. The errors in FlowTable configuration may be caused by: (1) bugs in controller software for the addition of a Flow Entry and/or (2) presence of a high priority error-prone Flow Entry (added manually or by a controller) that gives a match with the incoming packets. The objective of verification is to find this incorrect matching and hence, to find the packet-headers that cannot be matched or are matched incorrectly with the Flow-Match part of a Flow Entry. Without such verification, it is difficult to find which packets cannot be delivered or are wrongly forwarded by a switch.

Most of the existing verification tools such as HSA (Header Space Analysis) [80], Anteater [81], and VeriFlow [82] detect matching issues by analysing configuration of switches. However, without sending real packets, it is difficult for these tools to find software or hardware issues in flow matching. Recently, the automatic test packet generation (ATPG) tool [83] is proposed to verify the network by transmitting test packets. ATPG verifies only one (or some) of the packet headers that gives a match with a wildcarded Flow Entry, whereas matching of all the other packet headers remains untested. Therefore, we implement AUTOmatic Test Pacekt Generation (AutoTPG) tool for verification of matching of all packet-headers with the wildcarded flow of a Flow Entry.

Our mechanism transmits test packets to verify that all the packet-headers match correctly with the Flow-Match Header or not. However, if test packets have to match with the Flow Entries of data packets, these would need additional bandwidth to be reserved for test packets on the links corresponding to the outgoing actions of the matched Flow Entries. To overcome the challenge (additional bandwidth requirement), we forward the test packets through duplicated Flow Entries, which either drop or forward the test packets to the controller (instead of forwarding these on the outgoing links). The details of the design choices of AutoTPG and types of bugs it can find have been described in D4.2 [6] and in the paper [84].

### 4.9.2 Brief Technical description

The tool performs three steps for verification: (1) flow duplication, (2) test packet generation, and (3) matching errors identification. In the flow duplication step, the tool duplicates the Flow Entries from a FlowTable to another FlowTable (the reason of duplicating Flow Entries is explained in the previous subsection). In the test packet generation step, the tool generates and transmits test packets that can match with the Flow-Match Header of duplicated Flow Entries. In the matching errors identification step, the tool calculates the matching errors either by reading the counters (statistics) of the duplicated Flow Entries or by comparing the sent and received test packets.

For flow duplication step, our tool uses a header field such as EtherType (or VLAN ID) for differentiating test packets from data packets and assumes that this header field (e.g., EtherType) is wildcarded in the Flow-Match Header part of Flow Entries. Therefore, test packets can be forwarded from a FlowTable (under verification) to another

FlowTable (containing duplicated Flow Entries) by adding an entry in the FlowTable (under verification) to redirect all the test packets to another table.

Our tool can be run from the controller or from a virtual machine (VM) establishing an OpenFlow session with OpenFlow switches. Therefore, test packets can be transmitted to OpenFlow switches through packet-out messages.

For matching error calculations, our tool either: (1) runs the binary search algorithm on the counters incremented in the statistics field of Flow Entries or (2) receives test packets (transmitted) back after matching Flow Entries. In the former case, the method is called binary search method, while in the latter case, the method is called packet-reception method. A detailed description of both the methods is provided in [84].

### 4.9.3 Recent improvements concerning the tool integration

Recently, a RestFul interface is provided to run our tool. The input and output of our tool is now received through RestFul commands. Currently, our tool is the part of the troubleshooting workflow. A detailed description about the troubleshooting workflow (included our tool) can be found in Section 5.4.

### 4.9.4  Evaluation results and conclusions

### 4.9.4.1 Emulation Methodology

We perform two different types of experiments - (1) (out-of-band) controller-induced verification, and (2) (in-band) VM-induced verification. Each of these will be handled in more detail in the following subsections. For our emulation, we implemented the proposed tool in the Floodlight controller that uses OpenFlow version 1.3 in its implementation. In addition, we used Open vSwitch for running OpenFlow in the switches of the emulated topology (Figure 43).  In the experiments, software matching errors are emulated by translating some packets of a flow to incorrect headers. In our emulations, we find these errors and find verification time. In all the experiments, multiple switches are verified at the same time.

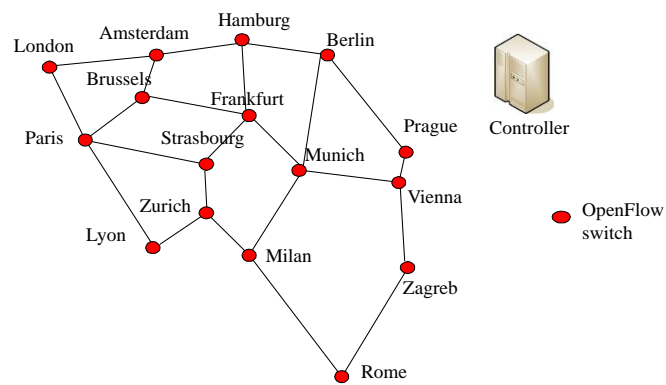

*Figure 43 Emulated Topology.*

### (Out-of-band) Controller-induced verification:

In the controller-induced experiment, the controller makes an out-of-band connection with switches in the emulated Mininet topology (Figure 43) and performs all the steps of verification. Flow-Match Header errors are detected either via the packet-reception or the binary-search method.

Figure 44 shows the emulation methodology using traffic captured on the controller when the binary-search or packet-reception method is used for verification. Traffic is shown from second -100 to 200, when 1 Mbps bandwidth is available between each switch and the controller for verification. In emulation, each switch in the emulated pan-European topology first establishes an OpenFlow session over the TCP session with the controller. The spikes in Figure 44 from -100 to -90 seconds are due to traffic exchanged between the controller and switches to establish OpenFlow sessions. At emulation time -50 seconds, the controller adds Flow Entries in each switch to forward data traffic. The Flow Entries contains the ingress port, source IP address and destination IP address as matching fields. The source and destination IP addresses are 24 bit addresses and therefore, the controller needs to transmit 65536 different test packets for verification. However, due to the generated bug, each Flow Entry is not able to match correctly 256 flows out of 65536 flows. At second 0, the controller starts verification of the Flow Entries. In Figure 44 traffic generated due the verification of one Flow Entry is shown.

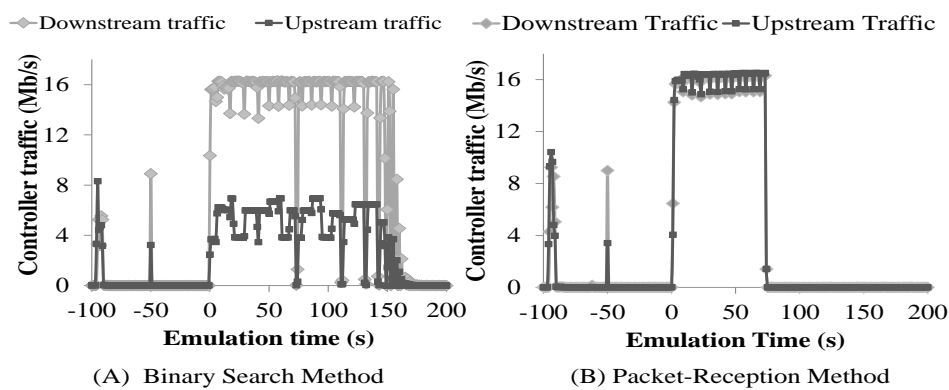

(A) Binary Search Method          (B) Packet-Reception Method

*Figure 44 Traffic on the controller link (introducing emulations scenarios for controller-induced verification).*

### (In-band) VM-induced verification:

We now explain the emulation methodology of the VM-induced verification experiment using the traffic captured on the controller and the VM link. Traffic is shown when the VM is created in the Hamburg switch of the emulated topology (Figure 43) and verification uses data plane links for transmitting or receiving test packets. The emulation methodology of this experiment is same as the controller verification experiment (Figure 45) in which switches establishes OpenFlow session paths with the controller from second -100 to -90 and at second -50, the controller adds Flow Entries in switches to forward data traffic. At second 0, the controller starts verification by copying Flow Entries to another table, creating a VM at the Hamburg switch, and establishing Flow Entries for making OpenFlow sessions (in-band) between the VM and switches [85] We see small spike in Figure 45 at second 0 due to this traffic.

In our emulation, the controller creates a VM using the RPC (Remote Procedural Call) commands. Figure 45B shows that the time to create a VM is about 19 seconds. After creating the VM, the VM establishes OpenFlow session paths with all the switches in the network and starts verification. For verification, we reserved 1 Mb/s bandwidth in each switch link. In our emulation, the controller controls the rate of test packets according to the bandwidth available in each link of its OpenFlow session. Additionally, the bandwidth between VM and the Hamburg switch is kept as 4Mb/s.

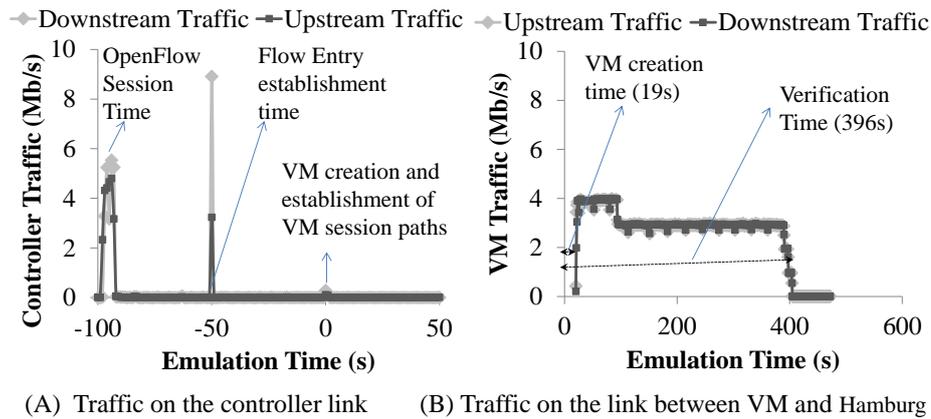

(A) Traffic on the controller link    (B) Traffic on the link between VM and Hamburg

*Figure 45 Traffic Intensity (Introducing VM-induced verification emulations scenarios).*

### 4.9.4.2 Emulation Results
**(Out-of-band) Controller-induced verification results:**

Figure 46 depicts the verification time when the bandwidth between the controller and switches is a value between 5 Mb/s to 700 Mb/s. For transmitting test packets, the controller sets the rate of the test packets according to the bandwidth available between the controller and switches for verification. In case of the binary-search method, the verification time is calculated by subtracting the time when the last counter read reply message is received by the controller with the time when the first counter read request message is sent by the controller. In case of the packet-reception method, the verification time is calculated by subtracting the time when the first packet-out containing a test packet is sent by the controller with the time when the last packet-in containing a test packet is received by the controller. All the results are taken 50 times and the average is shown in Figure 46.

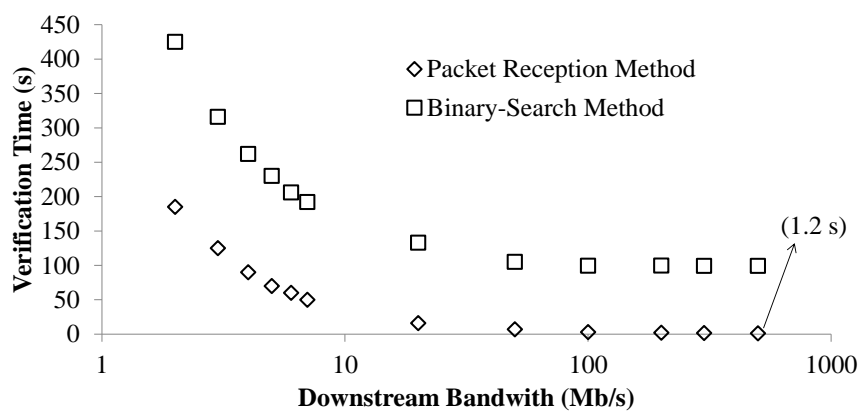

*Figure 46 Verification time using the packet-reception and binary-search method (controller out-of-band network scenario.*

Figure 44 shows the results when there are only 5 wildcarded Flow Entries (wildcards in the last 8 bits of source and destination IP address) in each switch for verification. In this case, verification is completed in limited time. However, when the number of Flow Entries is increased in switches, the verification time will increase significantly. Figure 47 shows the results when the number of Flow Entries is increased from 5 to 400. In this experiment, 5 Mb/s of bandwidth is available in between each switch and the controller for performing verification. Figure 45 illustrates that the verification time becomes significantly long as the number of Flow Entries increases. This is because if more Flow Entries are present for verification, more test packets is required to be sent to the switches. As the bandwidth is limited, the controller needs to wait long time to send all the test packets, increasing the verification time.

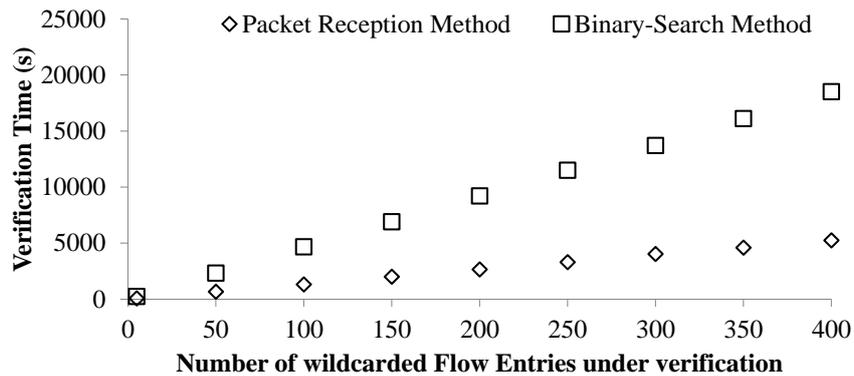

*Figure 47 Scalability experiment with respect to the number of Flow Entries.*

### (In-band) VM-induced verification results:

For in-band verification, we assume that different queues are installed in OpenFlow switches for control and data traffic. Configuration of these queues is described in [85]. . As traffic flows using different queues do not interfere with each other [in-band], we do not consider data traffic in our experiments.

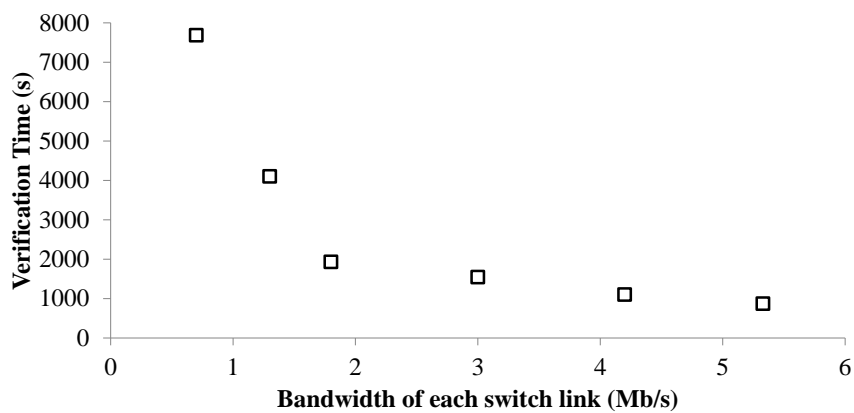

*Figure 48 VM induced in-band verification. The controller is connected to one of the switches in the network.*

Figure 48 shows the results of the experiments when the bandwidth of each switch links is limited between 0.5 to 5.5 Mb/s (value is shown in the figure) for verification. In these experiments, the VM makes an in-band connection with switches in the network and sets the rate of the test packets according to the verification bandwidth available in all the links along the OpenFlow session paths. As the paths for verification for some of switches are through other switches in the network, the switches may have to wait for verification until the intermediate switches in the path perform verification. This leads to increase in the verification time in the VM-induced verification experiment as compared to the controller verification experiment (Figure 46) in which the controller makes an out-of-band connection with the switches. However, an out-of-band control network is not always possible (for example, in widely distributed central offices in access networks). In this case, the controller may itself need to communicate with switches in the network using in-band control paths. Even if an out-of-band control network is present, there may not be enough bandwidth in this network to perform verification. Therefore, in this case verification traffic may need to send through the data plane links by reserving some bandwidth over the data network for verification.

### 4.9.5 Conclusions

We provided a brief overview of out our tool and extensive experimental study. The tool was evaluated in extensive emulation experiments. The results illustrated that the verification time depends on the bandwidth available in the network for verification. If bandwidth is unlimited, verification can be achieved in a very short time interval. However, if bandwidth limitations exist, the verification time might increase significantly.

## 4.10 DoubleDecker messaging system

### 4.10.1 Purpose and problem statement

In a rich SDN environment the variety of tools interacting together is more important than in a legacy network. Control messages flow toward and from the orchestration components. The monitoring functions transmit their results, sometimes to multiple recipients as time series databases, controllers or troubleshooting tools. A tool was needed to offer a simple way to transmit all kind of messages with performance constraints and heterogeneous domains. As a solution to this problem statement we propose the DoubleDecker messaging system.

To provide a scalable system for easily integrating different monitoring functions, and transporting their results, the DoubleDecker messaging system has been designed. The system consists of two protocols (client-broker and broker-broker) which allow point to point messaging as well as supporting a publish/subscribe mechanism. Messages are routed between connected clients through a local broker or a hierarchy of brokers if necessary, keeping messages as local as possible in order to avoid network overhead. The hierarchical messaging topology created by connecting brokers fits naturally to the hierarchical, recursive, UNIFY architecture.

### 4.10.2 Technical description

The main problem the DoubleDecker messaging system is trying to address is to provide a simple and easy to use transportation mechanism for control and monitoring data for the various monitoring functions developed in the project. As these monitoring functions can be quite different in their complexity and how they are implemented, the solution should be easy to apply in a range of scenarios. For example going from a set of counters in a switch that are read by a SDN Controller app, to daemons running on a UN, to dynamically instantiated VMs. Additionally, it should be easy to use in the dynamic environment envisioned in UNIFY, where functions may migrate and scale-in or out depending on various factors such as incoming traffic load. Finally, it should be easy to implement support for the messaging system in the clients, and to integrate with existing systems.

As it is built on top of the ZeroMQ messaging library it supports multiple underlying transport protocols, from shared memory within a process, to IPC, TCP, TIPC, and PGM multicast. Client endpoints are not identified by a network address but by a separate identifier. By separating network address and identity, components can more easily be migrated in the infrastructure without requiring for example network routes to be updated or tunnels moved between endpoints. Thanks to the publish/subscribe mechanism, it is possible to discover components in the system without manual configuration.

The DoubleDecker can be responsible for transporting very sensitive data from the orchestration layers down to the infrastructure layers. It is then critical to ensure the integrity of the data as well as authenticate the components. We have implemented authentication for both clients and brokers and point to point encryption. We use a standard public/private key pair mechanism and ECC algorithms for the encryption. This ensure that the transmission of the messages is secure but remains a marginal feature that VNF developers should take advantage of but not entirely rely on.

The client side library is implemented in the C, Python and Java languages to fit the requirements of the majority of the tools developed in Unify. The software is open source and distributed under a free license to allow any VNF developer to integrate the DoubleDecker with their tool. It also ensures a life for the tool beyond Unify.

### 4.10.3 Recent improvements concerning the tool implementation

Our prototype was functional for some time already, no major new feature was added since M4.3 [58] and no new feature is planned to be added in the scope of Unify. The integration work however has revealed several problems with the code. We have focused on fixing bugs and cleaning the code to make it easy to use for other VNFs developers. As the tool is integrated within the code of other prototypes we must ensure backward compatibility if any feature is added.

In the latest version we decided to maintain only the C version of the brokers. This choice was motivated for performance reasons. The broker is responsible for routing messages which can which is delay sensitive and heavier regarding computation if the messages are very heavy or if there is a huge number of clients registered as the routing tables become longer. The DoubleDecker is however not a dataplane tool but control plane only, it is a compromise between performance and flexibility plus robustness. As a first step during the development we offered the broker in Python which was unsatisfying regarding performances. We made some performance tests to evaluate the prototype; the results are available in the section 5.1.4 in D4.2 [6]

### 4.10.4 Recent improvements concerning the tool integration

We have worked a lot to integrate the DoubleDecker with other Unify prototypes and so far the results are:

- Integration with the Rate Monitoring (Section 4.5). The RateMon can publish its monitoring results through the DoubleDecker. This means that the monitoring data can be spread to several data sinks. In the Elastic Router use case two databases are collecting data. On one side OpenTSDB, which provides the row data to be used for troubleshooting, in particular the Recursive Querry Language exploits the data stored in OpenTSDB. On the other side, according to the MEASURE annotation, data are aggregated and stored in PipeDB. From the point of view of RateMon, the number of destinations is not even known as the pub/sub mechanism inside the DoubleDecker takes care of transporting the data to all the interested parties.

  Communication has also been able in the direction of the RateMon. RateMon features runtime configuration. In the case of a SDN, the exact position of the monitoring function is not always known. This makes runtime configuration via a direct IP address impossible. As the DoubleDecker provides an abstraction to the physical location of the clients, it is then trivial to send runtime parameters to any instance of the RateMon by simply knowing its identifier within the messaging system. It is also possible to use the pub/sub mechanism to notify all the instances to stop monitoring or resume monitoring for instance.

- The control application, Ryu, deployed in the Elastic Router scenario has also been connected to the DoubleDecker. This network function is responsible for managing the deployment and the elasticity property. As one of the tasks of WP4 is to provide automation, the monitoring framework has to be able to notify the control application when a scale up is needed. As of the properties of the DoubleDecker it made sense to transmit this notification through it. Again, the sender of the alert doesn't have to care about the exact location of the control application, by using pub/sub it is even possible to trigger the elastic functionality and log this action in a separate tool without affecting other components.

The current scenario describes an overload of the link which leads the router to be scaled. Adding functionalities to the controller to handle other situations would be a normal continuation. The DoubleDecker can then be used to transmit richer data, containing logs or runtime information for a more advanced decision. This system is totally flexible regarding the type of data passing through it, meaning that enriching the scenario would require work on the VNF and the controller but the link would remain untouched.

- With the important number of prototypes integrating an interface to connect the DoubleDecker, we have decided to integrate the Broker component as a part of the Universal Node (UN) prototype, developed in WP5. The Brokers are application independent and support multi-tenancy which makes them relatively architecture independent. The performance and deployment overhead of having a DoubleDecker deployed in every single UN is very small compared to its utility. The DoubleDecker interface is expected to be present as a standard component which eases even more the development of VNFs compatible with the UN.

The Brokers can handle several interfaces at the same time. One broker can be used to expose an IP address whether it is public or not, and at the same time use the IPC protocol to connect IP-less components.

- In addition to the integration done with monitoring functions such as the Rate Monitoring, the MEASURE framework, receiver of the monitoring data is also connected to the DoubleDecker. The generic code for aggregation points is connected to the messaging system to receive the raw data and transmit the transformed data. The genericity of this connection combined with the low latency and flexibility of messages formats and size made the Double Decker a natural solution. The MEASURE prototype described in section 4.2, has several independent components interconnected, without the DoubleDecker, the UN would have to allocate an important number of IP addresses only for the monitoring tools.

- Epoxide (Section 4.12) can use a DoubleDecker interface to receive troubleshooting data. When the controller is not able to solve a problem or the cause of the problem is unclear, manual troubleshooting tools need to be used. Epoxide is a generic tool aiming this goal. The more data are made available via the DoubleDecker the easier it will be for an operator to troubleshoot a service.

Apart from the per-tool integration we have improved the generic way to integrate the DoubleDecker with other components. If Network Functions follow standard configuration protocols (JSON RPC) and a standard way to format monitoring results (mPlane like) the integration is almost straight forward. This is not a feature added to our prototype but more a guideline on how to make the whole architecture more standard thus more extendable.

### 4.10.5 Evaluation results and conclusions

One of the major a benefit of using the DoubleDecker is the facility to integrate it with any component. We have experienced both integration within a completely new tool and with an already mature tool. Both lead to different conclusions.

Integrating the DoubleDecker with a new tool as it was done with the MMP was a trivial process. It influenced the choice of the language used for the tool but without putting too strict constraints. It is also easier to think in terms of interaction with the other components when the transport layer is abstracted and simplified. The developer can focus entirely on the format of the content instead of dealing with lower level problems. This characteristic fits perfectly in the DevOps concept described in the Section 3.1.

The integration with existing tools was a slightly more complex process. When the language was not compatible we had to find a workaround and spend more time with the developer of the tool to reach a good level of integration. This is a problem to be considered if we want to have the DoubleDecker being integrated in an already functioning environment. In addition to the original code, we have worked on a version of the DoubleDecker offering an even lighter coupling. By providing the DoubleDecker as an open source tool, and finding alternative solutions for integration with a large range of other tools we aim at providing a realistic solution for service providers. A SDN infrastructure has a lot to gain from easier messaging between components as much as each VFN does.

## 4.11 Recursive Query Language for monitoring results

### 4.11.1 Purpose and problem statement

In NFV environments, operators or developers sometimes need query the performance of virtualized Network Functions (VNF), for example, the delay between two VNFs. In existing systems, this is usually done by mapping the performance metrics of VNFs to primitive physical network functions or elements, statically and manually when the virtualized service is deployed. However, in UNIFY a multi-layer hierarchical architecture is adopted, and the VNF and associated resources, expressed NF-FGs, may be composed recursively in different layers of the architecture. This will put greater challenge on performance queries for a specific service, as the mapping of performance metrics from the service layer (highest layer) to the infrastructure (lowest layer) is more complex when compared to cloud infrastructure with single layer orchestration. We argue that it is important to have an automatic and dynamic way for decomposition of the performance query in a recursive way, following the different abstraction levels expressed in the NF-NGs at hierarchical architecture layers. Hence, we propose to use a declarative language to perform recursive queries based on input in form of the resource graph depicted as NF-FG. By reusing the NF-FG models and monitoring database already existing in the UNIFY architecture, the language can hide the complexity of the multilayer network architecture with limited extra resources and efforts. However, the expression of performance queries is significantly simplified for any potential receivers (i.e. operator and developer roles).

The proposed recursive query language allows both developer and operator roles to query monitoring metrics on high level of abstraction, corresponding to NF-FG and SG abstractions in higher layers of the UNIFY architecture. The queries are resolved by the query engine in a recursive way (following NF-FG definition throughout the hierarchical UNIFY architecture) until primitive infrastructure metrics are available in actual monitoring databases residing in infrastructure domains. The query language and the query engine have been described in D4.2 [6], an IRTF draft [86] and a paper [87]

### 4.11.2 Brief technical description

To address the challenge of automatic and flexible performance decomposition and abstraction in a recursive NFV architecture, we argue for a declarative logic-based language based on Datalog [88], which provides recursive querying capabilities. Query evaluation with Datalog is based on first order logic, and is thus sound (all provable statements are true in all models) and complete (all statements which are true in all models are provable). However, Datalog is not Turing complete, and is thus used as a domain-specific language that can take advantage of efficient algorithms developed for query resolution. In Datalog, rules can be expressed in terms of other rules, allowing a recursive definition of rules, together with reusability.

Like other Datalog based language, the recursive monitoring query program consists of a set of declarative Datalog rules and a query. A rule has the form:

$$h <= p_1; p_2; ...; p_n$$

which can be defined as "$p_1$ and $p_2$ and ... and $p_n$ implies $h$". "$h$" is the head of the rule, and "$p_1; p_2;... ; p_n$" is a list of literals that constitutes the body of the rule. Literals "$p(x_1; ; x_i;... ; x_n)$" are either predicates applied to arguments "$x_i$" (variables and constants), or function symbols applied to arguments.

The program is said to be recursive if a cycle exists through the predicates, i.e., predicate appearing both in the head and body of the same rule. The order in which the rules are presented in a program is semantically irrelevant. The commas separating the predicates in a rule are logical conjuncts (AND); the order in which predicates appear in a rule body has no semantic significance, i.e. no matter in what order rules been processed, the result is atomic, i.e. the same. The names of predicates, function symbols and constants begin with a lower-case letter, while variable names begin with an upper-case letter. A variable appearing in the head is called distinguished variable while a variable appearing in the body is called non-distinguished variable. The head is true for given values of the distinguished variables if there are values of the non-distinguished variables that make all sub goals of the body true. In every rule, each variable stands for the same value. Thus, variables can be considered as placeholders for values. Possible values are those that occur as constants in some rule/fact of the program itself. In the program, a query is of the form "$query(m, y_1, ..., y_n)$", in which "query" is a predicate contains arguments "m" and "$y_i$". "$m$" represents the monitoring metric to be queried, e.g., end to end delay, average CPU usage, and etc. "$y_i$" are the arguments for the query function. The meaning of a query given a set of Datalog rules and facts is the set of all facts of "query()" that are given or can be inferred using the rules in the program. The predicates can be divided into two categories: extensional database predicates (EDB predicates), which contains ground facts, meaning it only has constant arguments; and intentional database predicates (IDB predicates), which correspond to derived facts computed by Datalog rules.

In order to perform a recursive monitoring query, the resource graph described by an (V)NF-FG needs to be transformed so it is represented as a set of Datalog ground facts which are used by the rules in the program. The following keywords are defined to represent the NF-FG graph into Datalog facts, which are then used in the query scripts:

o *sub(x; y)* which represents 'y' is an element of the directly descend sub-layer of 'x';

o *link(x; y)* which represents that there is a direct link between elements 'x' and 'y';

o *node(z)* which represents an node in NF-FG.

It should be noted that more keywords can be defined in order to describe other properties of an NF-FG. The ground facts are usually generated by the query engine by analyzing the NFFGs on various level of abstraction (i.e. granularity). In addition to ground facts, some rules shall be defined by receivers (e.g., the network service operators) in order to describe how to translate the high-level performance query into primitive resource metric queries, and how to aggregate the primitive query results in order to be able to return a single high-level result. A set of function calls can be defined in order to support the decomposition of queries onto low level primitive resource metrics, for example, the CPU or memory usage of given VM. The function call will start with "fn " in the syntax and may include 'boolean' predicates, arithmetic computations and some other simple operation. The function calls can be added by either the provider of the query engine or the service developer.

The formal definition of the query language is documented in Annex 3 of D4.2 [6].

### 4.11.3 Recent improvements concerning the tool implementation

To perform recursive monitoring query, a query engine is also required in addition to the language. Since D4.2 [6], we have implemented a prototype (including query engine and GUI) which can perform the recursive query according to the defined language. The main functions of the query engine include:

o receiving a monitoring query from the receivers and sending back query results;

o parsing and compiling the query scripts;

o communicating with NF-FG repository and translating NF-FG graphs into Datalog facts;

o traversing NF-FG graphs according to query scripts;

o querying monitoring databases which contain the measurement results from monitoring functions.

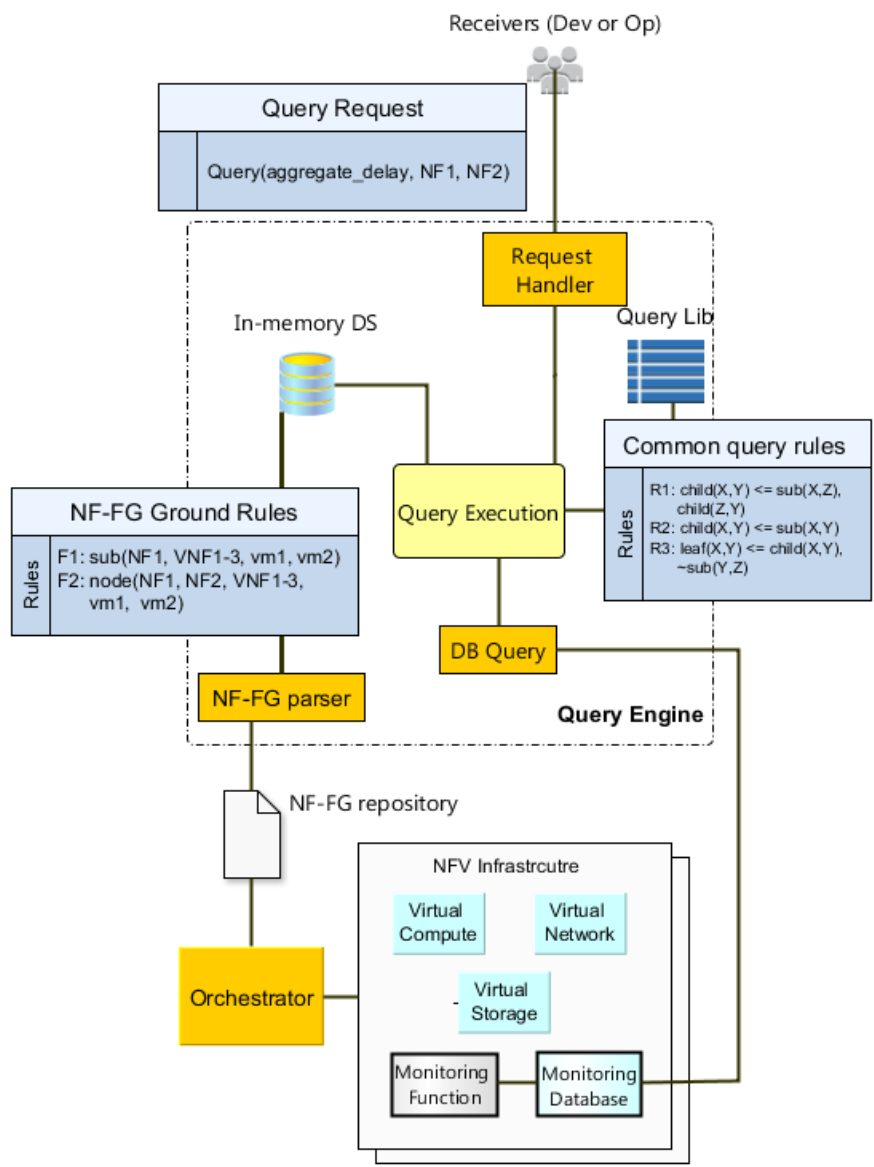

*Figure 49: The design of the RQL query engine.*

We have designed and implemented a query engine as illustrated in Figure 49. It consists of the following components: the Query Execution, the Request Handler, NF-FG parser, In-memory Data Store (DS), DB Query and Query Library. The Query Execution provides the compiler and running environment of the query language. The Query Execution is implemented using PyDatalog[4] (, an open source Datalog compiler . The Request Handler receives and parses the query request from developers or operators. The query request could be the scripts written with the proposed language or a simple query request command.

The NF-FG parser is used to check the NF-FG repository which is maintained by NFV orchestrators (i.e. ESCAPE in case of the UNIFY demo) and translates the graphs into Datalog ground facts automatically and stores them in an in-memory Data Store (DS). The query scripts are written in the language described in D4.2 [6]. Pre-defined Datalog

---

query rules or scripts can be stored in the Query Library, so that the receivers can just use simplified commands to perform the query. The Query Library contains the query scripts provided by developers or operators, and an API can be provided for each stored query script. The Datalog Execution will call the corresponding library according to received query command or scripts, and decompose the high level query request into primitive queries towards the monitoring data stored in the monitoring database. The Request Handler is also responsible for sending aggregated or abstracted query results to the receivers. In addition, the in-memory DS can also be used to store the intermediate results obtained by querying the monitoring database(s). The intermediate results can be used as cached results or to aggregate the data to be returned to the receivers. The DB Query retrieves low level monitoring data from the monitoring database and store intermediate results into the in-memory DS.

The query engine provides a tool for the receivers (e.g., developers and operators) to query the various VNF performance metrics from the lower NFV architecture layers (e.g. NVFI) at higher layers. The query could be input either by the GUI or via a RESTful interface. As a natural location, the query engine could be an application in a service layer. It is to be noted that the query engine does not collect monitoring data directly from the infrastructure layer. Instead it relies on the data collected by the monitoring functions developed in NFV infrastructure.

### 4.11.4 Recent improvements concerning the tool integration

We have verified the effectiveness of the language by integrating it with other UNIFY and 3rd party tools, including DoubleDecker (see Section 4.10), RateMon (see Section 4.5), cAdvisor and distributed OpenTSDB/InfluxDB monitoring databases. The two example usage cases (i.e., end to end delay and aggregated CPU usage of network services) described in D4.2 [6] have been implemented and verified in this testbed setup. The usage of the recursive monitoring language in the scope of the UNIFY wide integrate prototype (IntPro) for the purpose of supporting a troubleshooting process is described in Section 6.2.

### 4.11.5 Evaluation results and conclusions

In our implementation of the query engine, compiler and running environment is based on PyDatalog, an extension of python to provide Datalog support. The experimental environment of the NFV infrastructure is setup by using Docker container based VNFs. The NF-FG files are based on the format specified in UNIFY D3.2 [16]. Google cAdvisor is used to collect the primitive resource usage (e.g., CPU and memory) of individual containers, and for delay, simulated data are generated. OpenTSDB, an open source distributed and scalable time series database, is used as the monitoring database to store the collected performance metrics. We have used the HTTP RESTful API to query the performance metrics stored in OpenTSDB.

We have implemented the two example use cases: an end to end delay between network services; and aggregated CPU/memory usage of network services. We have proved the effectiveness of the recursive monitoring language and the query engine. Below we will discuss the scalability of the recursive query which is an important criteria considered in real system. We will use the query latency as the factor during the discussion. The query latency is defined as the time interval between a performance query is sent to the query engine and a response is received by the sender. It consists of the several parts and can be represented as:

$$l_q = l_s + l_d + l_{db} + l_r$$

in which $l_s$ denotes the latency from a performance query request is sent by a receiver and it is received by the query engine; $l_d$ is the time taken by the query engine to execute the Datalog rules and decompose the query request into primitive queries; $l_{db}$ is the time taken by the query engine to retrieve the primitive query results from the database; $l_r$ is the time taken to send back the aggregated or abstracted results to the receivers. For a single query from the receivers, $l_s$ and $l_r$ are independent from the scale of the NFV infrastructure and usually does not change much in the same setup. $l_d$ and $l_{db}$ will depends on the scale of the network infrastructure. $l_d$ will also vary given different Datalog rules in the same Datalog execution environment. $l_{db}$ will depend on the number of the primitive queries generated by the query engine, and the latency of each primitive database query in turn depends heavily on the monitoring database implementation. Below we measured the latency $l_d$ and $l_{db}$.

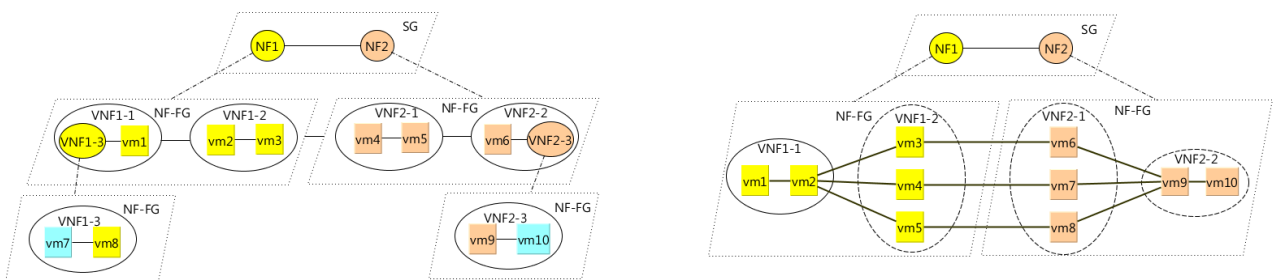

*Figure 50: Sample NF-FGs. (a) On the left, an example NF-FG with two layers of abstraction, i.e. VNF1-1 and VNF 2-2 are realized by finer grained sub-NF-FGs. (b) On the right, a sample NF-FG with parallel links between NFs.*

In the evaluation, two types of NF-FG are simulated. The first type is similar to the example shown in Figure 50a which has nested NF-FGs but there is no multiple paths between network functions and is refereed as *nffg1*. The second type of NF-FG is similar to the example shown in Figure 50b, and it has multiple paths between network functions but without nested NF-FG and is refereed as *nffg2*. In the emulation, we measured the query latency when the size of the NF-FG (i.e., the infrastructure size) is increased to 1000 for both *nffg1* and *nffg2*. When we increased the size of *nffg1* and *nffg2*, the shapes are kept the similar, and only the number of VNFs and nodes belonging to VNFs are changed. For *nffg1*, the number of layers is still two, and the number of nested VNFs and the number of nodes in each VNFs are generated randomly. For *nffg2*, we only change the number of nodes of VNF1-2 and VNF2-1 from 3 to around 500 respectively.

In Figure 51 we show the measured latency ($l_d$) for the query engine to execute the Datalog rules that decompose and aggregate the average CPU of a network service with both types of NF-FG. In Figure 52 we show the measured latency $l_d$ for the query engine to execute the Datalog rules for the end to end delay query between network functions. From both figures we can see that the Datalog engine execution latency ($l_d$) almost increase linearly when the size of the NF-FG increases, eventhough the increasing rates show minor difference. We think one main reason is that to get the aggregated CPU and delays the Datalog execution needs to traverse the NFFG definition(s) and decompose the performance query into primitive queries to individual nodes in the NF-FG. With the increase of the size of NF-FG, more nodes or links have to be traversed by Datalog execution environment. However, it has to be noted that this latency is only measured for Datalog engine PyDatalog. For other Datalog engines, the trend may be different.

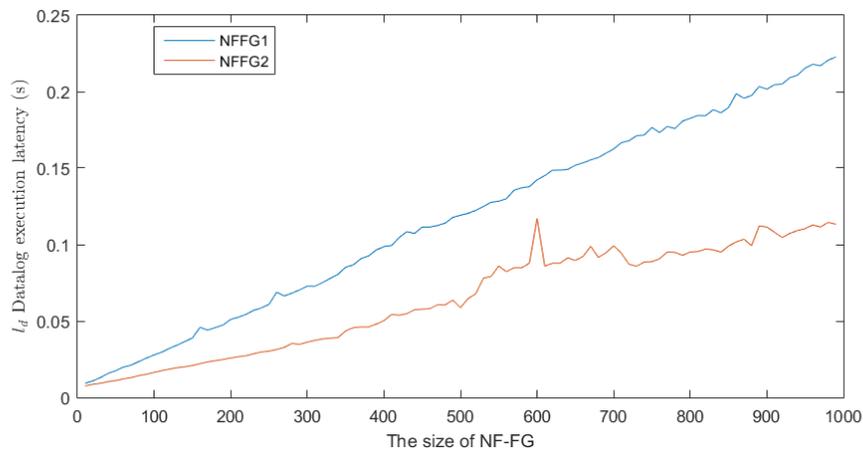

*Figure 51: The Datalog execution latency (ld) for querying average CPU of network service.*

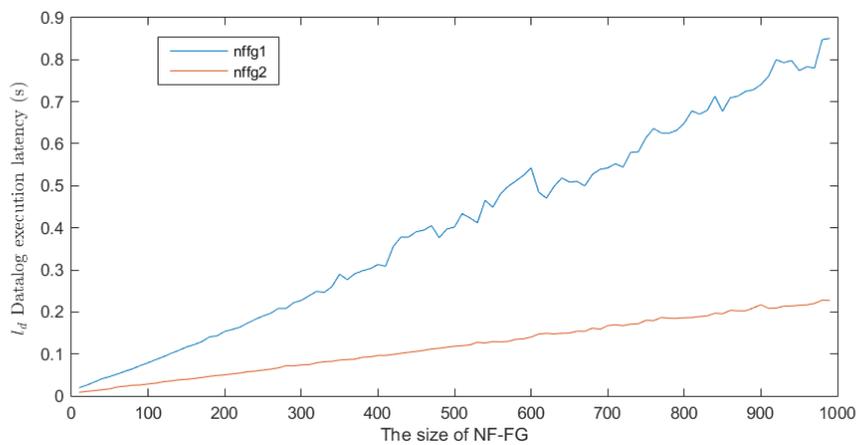

*Figure 52: The Datalog execution latency ($l_d$) for querying aggregated network delay between network services.*

It may cause high query latency when the size of the network service is extremely huge (e.g., greater than thousands). To reduce such potential high Datalog execution latency, one possible way is to cache the Datalog execution result into the in-memory Data Store because we expect the change of the NF-FG is not frequent.

In Figure 53 the latency ($l_{db}$) to query the primitive metrics from the monitoring database is measured. It shows that the latency also grows when the number of the nodes included in the query increases and the rate is greater than that of Datalog execution latency ($l_d$). As we mentioned above the latencies $l_s$ and $l_r$ are independent from the number of the nodes, therefore the order of complexity for query latency $O(l_q) = O(l_{db})$. It implies that the database query latency $l_{db}$ has more impact on the overall query latency in our experiment. Some monitoring databases (e.g., OpenTSDB) provide some aggregation functions though limited, which can be used to reduce the database query latency when the size of the network service is really huge. In addition, optimized placement and pre-aggregation methods of monitoring components can also be used to reduce the size of the collected monitoring data [32], hence the database query latency.

In summary, the experimental prototype verified that the proposed recursive monitoring query is practicable and effective. To tackle the potential high query latency for very large scale network service, the Datalog execution results can be cached in the query engine if the NF-FG doesn't change frequently.

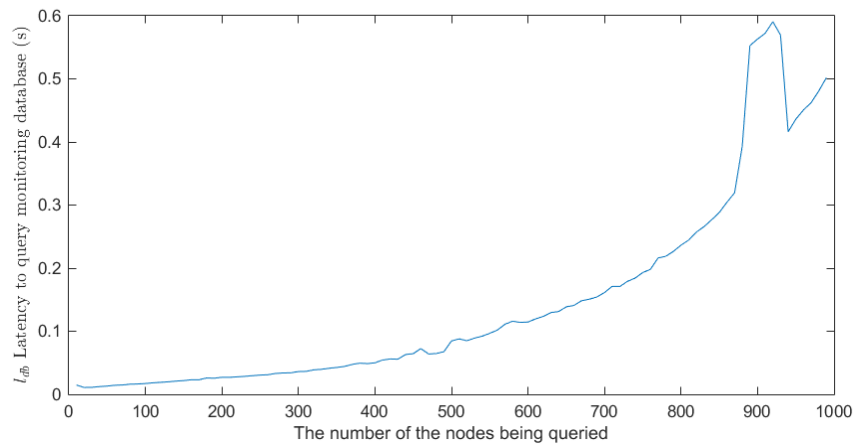

Figure 53 The latency ($l_{db}$) for querying primitive metrics in the monitoring database.

## 4.12 EPOXIDE – Multicomponent troubleshooting

### 4.12.1 Purpose and problem statement

Network troubleshooting is a complex task that requires accessing a plethora of troubleshooting tools and network nodes. Our tool's main goal is to help network troubleshooting by integrating pre-existing network troubleshooting tools and implementing a framework that provides a unified interface to interact with these tools. Instead of re-implementing functionalities available in existing, special purpose tools, EPOXIDE provides a simple way to create wrappers around these existing tools. Users then define high-level troubleshooting graphs (TSGs) written in a .tsg file that connect these special purpose tools to automate one hypothesis testing during a troubleshooting session.

### 4.12.2 Brief technical description

EPOXIDE is a multi-component debugging and troubleshooting tool written as an extension of the Emacs text editor [1]. The Epoxide framework interprets .tsg files and maps the nodes and links described by it to Emacs buffers. When execution of a .tsg is started, the framework creates an event queue and runs a scheduler that handles this queue. Events are generated when the content of a buffer is within EPOXIDE's scope changes. On such occasions, the nodes that connect to the updated buffer are looked up by the framework and notification events addressed to them are inserted into the frameworks event queue. The scheduler takes out these notifications one-by-one from the queue and calls the execution function of the appropriate node.

These node execution functions are responsible to call troubleshooting tools and handle the results returned by them. When they have new data, they should write their output links thus stimulating the frameworks scheduler to notify connected nodes.

Because of the modular structure of EPOXIDE, wrapper nodes can be created by anyone using the Emacs Lisp language. Node developers have to implement only a handful of functions requested by the framework (with the previously mentioned execution function among them). More information on the internal operation of EPOXIDE can be found in D4.4 [89].

### 4.12.3 Recent improvements concerning the tool implementation

To demonstrate the concept we had already written in the past wrapper nodes around simple troubleshooting tools like ping, iperf, etc. [64] [65] but since these basic wrappers can provide only the same functionality as the tools they wrap, more complex nodes became necessary in order to provide a higher level of utility for EPOXIDE. Recently we added four new types of nodes to EPOXIDE:

- nodes that wrap Emacs lisp functions,

- nodes that are able to channel new information from any Emacs buffer to the domain of EPOXIDE,

- Decision nodes that create conditional branches in a troubleshooting graph,

- Decision-summary nodes that summarize the results of different Decision nodes.

Using the first node type one is able to create simple manipulations on the text received on the inputs either by using one of the available Emacs lisp functions or writing a user-defined function. The nodes also provide the option to write the function in the .tsg file as a lambda function. This has the benefit that custom function definitions can be packed into the troubleshooting graph definition file itself in a self-contained way and no additional file is necessary for defining extra functionalities.

The second node type is useful when there is a process that updates an Emacs buffer that does not belong to a specific EPOXIDE session but we still want to access information from that buffer in an EPOXIDE session.

Decision and Decision-summary nodes, in their basic setup, filter their text-based input using a configurable filter expression then forward this filtered input to different output links. Although the logic of these nodes are relatively simple, they make possible to define TSGs that automate more than just one hypothesis testing of the troubleshooting session. The challenge here is to find an intuitive UI that can define useful decision logic easily and to create the node in such a way that it could support more automated troubleshooting processes by enhancing EPOXIDE with features of expert systems. All in all, the capabilities of EPOXIDE enhanced with the Decision/Decision-summary nodes are the following:

- Implementing simple tests and evaluating them. The decision node is able to interpret incoming textual data, evaluate it and decide whether it conforms to given criteria.

- Make conditional branches based on inputs. When a troubleshooting graph is built using different tests, the user is able to express that certain tests should run only when certain criteria are fulfilled.

- Provide input specific preprocessing functions. When a decision is based on multiple inputs, i.e., it is based on the results of multiple tools, then the TSG writer might need to use *preprocessing functions* tuned individually for each tool. These functions can analyze the text coming in on a certain input and return a value indicating a failure to comply with the criteria or some other value indicating success. For example, the Emacs lisp string-match function can be used as a preprocessing function as it looks up a substring in a string that matches a given regular expression. If it finds something, it returns the starting position of the match or it returns *nil* when there are no matches for the specified regular expression (the decision node considers this a failure).

- Combine inputs or select one from them using custom functions. When creating a hypothesis test the user can choose to have a simple success or failure message as an output of the decision node but sometimes it might be more appropriate to have more complex *selection logic* like, for example, to apply arbitrary Boolean logic on the preprocessed inputs.

- The output of the decision node is either the result of the selection function (i.e., which input is interesting) or the selected input itself. Consider the case when we want to check whether the input text contains a substring and if it does then use that exact text in a subsequent node of the TSG.

- Since different troubleshooting tools relay their results in different formats, it is necessary to provide a configuration option to the Decision node that defines how to process a given input. A certain tool's output may relay information on a line-by-line basis where each line ought to be interpreted individually (e.g. *ping*) or in bigger chunks (e.g. *ifconfig*).

- The result of an individual tool arrives asynchronously to EPOXIDE, in some cases users want to wait for all the inputs of a decision node to draw a decision, in other cases it makes more sense to conclude a decision when the first input arrives, or right after a given timeout expires.

- For convenient usage, the decision node provides three kinds of output links. The first one outputs the result of the selection function or the selected input. The second outputs a timestamp when the selection function does not select any of the inputs. And finally, the third kind of output is a status output that contains the results of the decision in all cases. These kinds of outputs help writing conditional branching and overview tables implemented by Decision-summary nodes with ease.

- The preprocessing function normally processes the data of one input link. However, it is also possible to assign multiple input links to one preprocessing function to work on, which in some cases simplifies the definition of the decision logic.

Figure 32 illustrates the extended flowchart of the Decision node that summarizes the main functionalities listed above. For a more detailed description of the Decision node's in/output and configuration parameters see D4.4 [89].

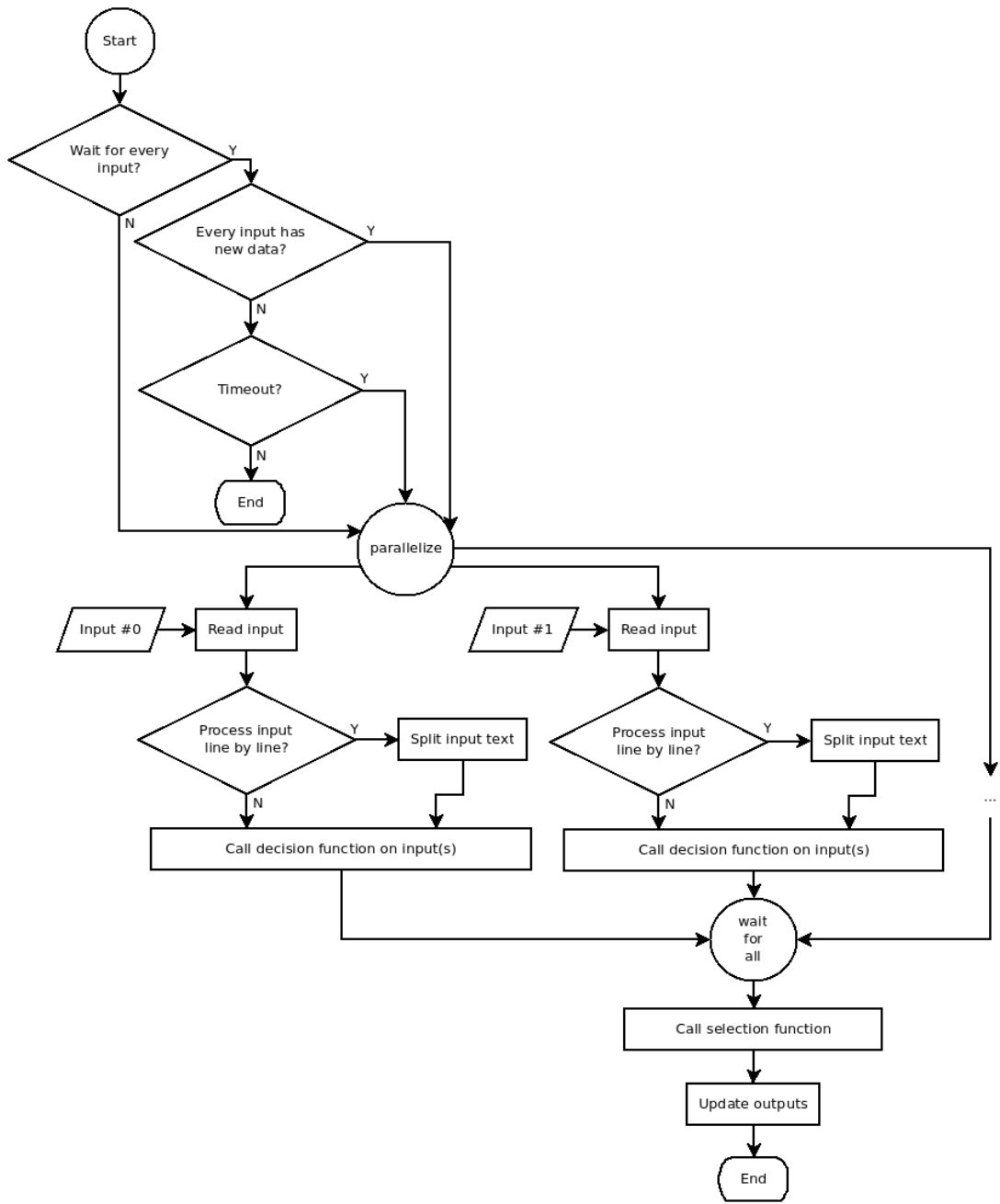

*Figure 32 Extended flowchart of the Decision node.*

### 4.12.4 Recent improvements concerning the tool integration

We have identified workflows that allow integrating EPOXIDE with the ESCAPE framework. To achieve this we implemented a new wrapper node that is able to connect to the ESCAPE framework and query topology information from it by accessing its REST API. The node performs processing on the received data and is able to forward it to a Graph node that displays it using COGRE.

We also have identified the DoubleDecker, Rate monitoring, RQE and the AutoTPG tools as possible candidates for UNIFY monitoring/troubleshooting tools that can be integrated with EPOXIDE. We created a wrapper node for DoubleDecker and possible troubleshooting graphs for RQE and AutoTPG.

The DoubleDecker node wraps around a special DoubleDecker client that is able to display and receive messages in JSON format. It has no input links; when it is instantiated it creates a sub-process that continuously updates its only output link. The node has one mandatory and four optional configuration arguments. The mandatory first argument specifies the host where the DoubleDecker client should run at. The second argument specifies the topic to which the node should subscribe in the form of <topic>/<scope> (or a semi-colon separated list of those in case of more than one topic should be used). The following arguments have default values: the third argument defines the location of the private key file for authentication, the fourth the tenant and the fifth specifies the DoubleDecker name of the node.

The single output of the node relays every message received from the DoubleDecker client unaltered. These messages are usually JSON formatted but in case of the startup they are unformatted strings.

The RQE and the AutoTPG tools provide REST APIs to interact with or control them. We created an EPOXIDE node that is able to establish communication with REST APIs and relay information received from them. These two tools communicate with JSON formatted messages so we implemented another EPOXDE node that is able to dissect a JSON message and return the value belonging to a certain JSON key. Using these nodes and the previously mentioned node that is able to run the *format* Emacs lisp function we came up with the following two troubleshooting graphs as sample scenarios to interact with the RQE and AutoTPG tools.

Figure 33 shows a simple scenario for the AutoTPG tool. In this scenario we want to check switch 00:...:01 in every minute for errors then filter the error part of the result and show it in an EPOXDE *view*.

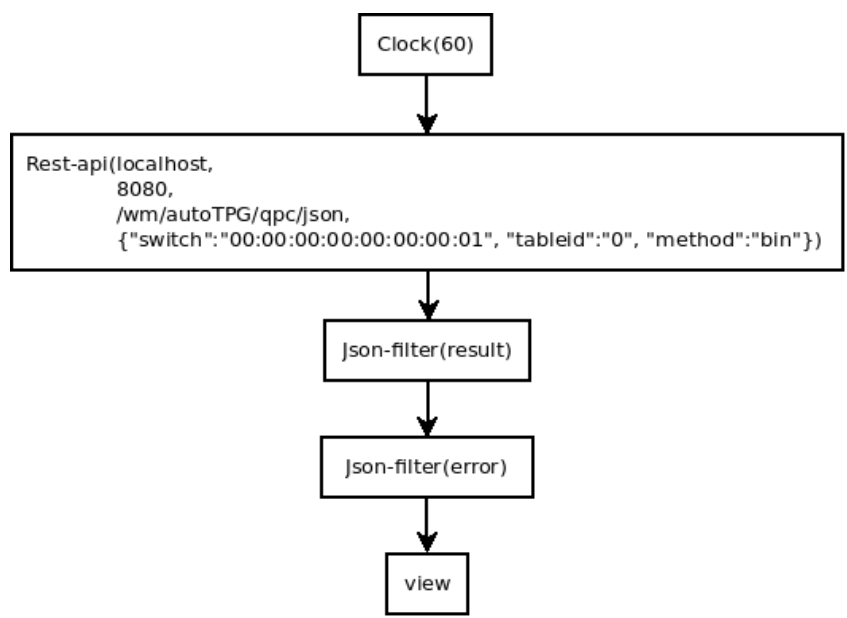

*Figure 33 Sample TSG using AutoTPG.*

For the RQE tool we came up with the sample TSG shown in Figure 34. First we compile a string that is used for accessing the tool's REST API – the left Format node is responsible for this. To make the TGS more versatile, we take the three arguments the REST API requires (*resource*, *nfid*, and *type*) from their respective Emacs buffers. The RQE tool is accessible at the localhost on port 8000. The query is performed every 10 seconds, as indicated by the Clock node. When a response arrives from the RQL tool's REST API, the requested parameter gets selected by the Json-filter node. Since the parameter's key can change depending on the queried resource and type, the right side Format node creates this key dynamically and it passes the resulting key to the Json-filter node as its first configuration argument. The result of the query is showed via a *view*. Notes on Figure 34 show the links' contents in a sample execution. In the presented scenario we want to query the mean CPU utilization (right and left side Emacs-buffer nodes). The RQE tool responds that it is 80% (visible on the output of the Rest-api and the Json-filter nodes).

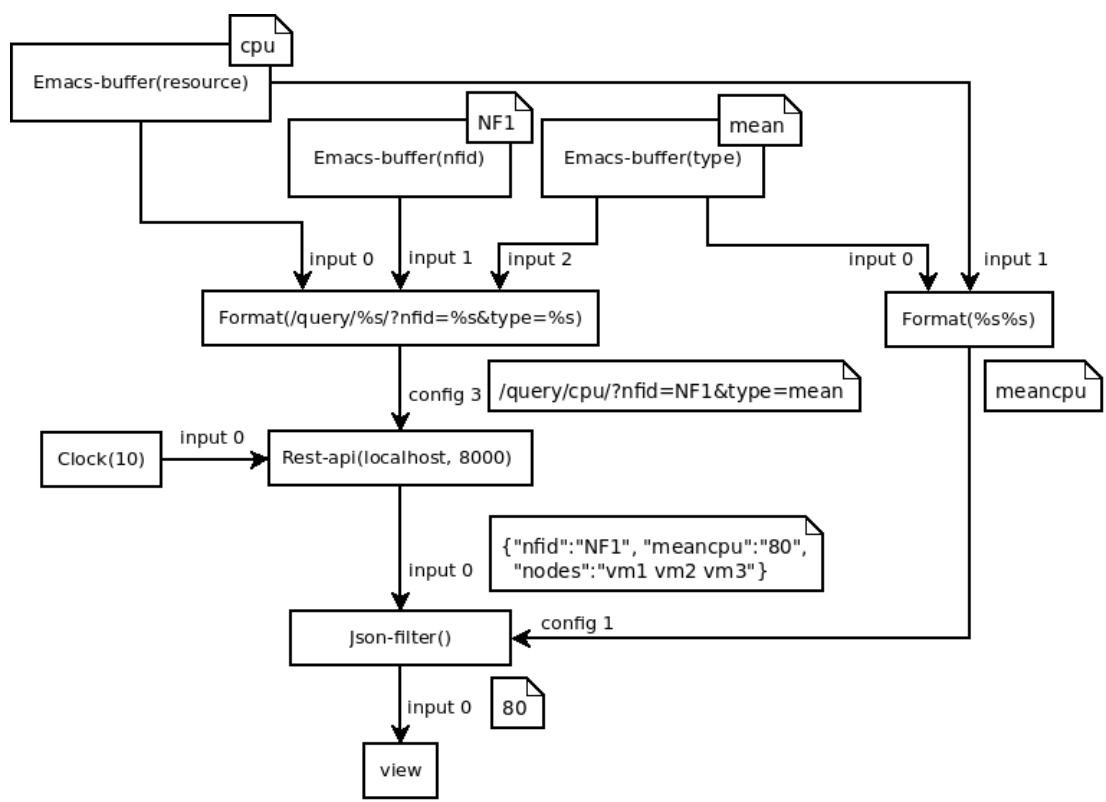

*Figure 34 Sample TSG using RQE.*

### 4.12.5 Evaluation results and conclusions

Taking the Decision node approach further would take us closer to our goal to have a more automated troubleshooting process. This concept of creating tests by evaluating data returned by troubleshooting tool greatly resembles to the methods used in recent years for performing automated inferences in network troubleshooting [66].

If EPOXIDE links forward probabilities expressed as real numbers then EPOXIDE with Decision nodes would also be able to run a Bayes network based decision support system in the background and provide help for which tools to run in the troubleshooting process or to build a troubleshooting graph without supervision.

## 4.13 Final toolkit composition and status

The SP-DevOps Toolkit is a set of tools developed in as part of the UNIFY SP-DevOps activities, and are collectively published under open source licenses via UNIFY dissemination channels. The tools solve particular problems on network performance monitoring, troubleshooting and service graph verification, and various combinations of these tools have already been demonstrated (e.g. DoubleDecker and Ramon in [27]) or are integrated to demonstrators as part of SPDevOpsPro or IntPro in the scope of UNIFY (see D4.4 [89]. and D2.4 [14], respectively).

The following table shows the tools for the final SP-DevOps toolkit released in two batches. While the four tools comprising the first release are already publicly available since November 2015, the additional tools of the second release have been published in June 2016. The final toolkit release was announced via the UNIFY webpage[5] with pointers to the separate repositories. Furthermore, the toolkit was advertised via presentations at IETF 94 in the NFVRG [90] and at the OPNFV summit 2016 [91]. Note that additionally to the tools listed in Table 4, an experimental version of MEASURE is also publicly available, but did not yet reach the maturity to be officially included to the SP-DevOps toolkit. Documentation and links of all the public tools, including MEASURE, can be found in D4.4 [89].

*Table 4 SP-DevOps toolkit components.*

| SP-DevOps Tool or Function | Problem solved | 1st Toolkit Release | 2nd Toolkit Release | License |
|---|---|---|---|---|
| DoubleDecker | Flexible and scalable messaging | x | | LGPLv2 |
| Ramon | Scalable congestion detection | x | | Apache 2 |
| AutoTPG | OpenFlow forwarding verification | x | | iMinds License |
| EPOXIDE | Multicomponent troubleshooting | x | | GPLv3+ |
| VeriGraph | Service graph and configuration verification | | x | AGPLv3 |
| TWAMP Data Model | Programmable active network measurements with a legacy OAM tool | | x | Simplified BSD License |

- DoubleDecker, a scalable messaging system based on ZeroMQ that features built-in message filtering and aggregation and sending opaque data between virtual network and monitoring functions, as well as towards higher layer orchestration modules

  o Tool release date: 2015-11-01

  o Repository: https://github.com/acreo/doubledecker

  o License: LGPLv2

---

[5] https://www.fp7-unify.eu/index.php/results.html#sp-devops-toolkit

- Ramon, a tool for determining a precise statistical estimate of link utilization in a distributed manner and offer congestion indications

    o Tool release date: 2015-10-31

    o Repository: https://github.com/nigsics/ramon.git

    o License: Apache 2

- AutoTPG (Auto Test Packet Generator), a tool able to verify/test the matching header part of the OpenFlow forwarding entries, even in the presence of aggregated (i.e. wildcarded) flow rules

    o Tool release date: 2015-11-01

    o Repository: http://users.intec.ugent.be/unify/autoTPG/

    o License: iMinds License

- EPOXIDE, an Emacs-based framework capable of assembling general and SDN specific debugging and troubleshooting tools in a single platform and make it easy for a developer or operator to combine them

    o Tool release date: 2015-08-14

    o Repository: http://github.com/nemethf/epoxide

    o License: GPLv3+

- VeriGraph, a formal verification tool for complex networks containing active NFs that is able to automatically verify NF-FGs and their configuration against some desired network policies

    o Tool release date: 2016-04-30

    o Repository: https://github.com/netgroup-polito/verigraph

    o License: AGPLv3

- TWAMP Data Model, a data model for client and server implementations of the Two-Way Active Measurement Protocol (TWAMP), defined through Unified Modeling Language (UML) class diagrams and formally specified using YANG.

    o Model release date: 2016-03-21

    o Repository: https://datatracker.ietf.org/doc/draft-ietf-ippm-twamp-yang/

    o License: Simplified BSD License

# 5 SP-DevOps workflows and processes

The expected operation of SP-DevOps is specified in this section through the definition of a number of workflows, each one capturing a particular aspect or viewpoint of SP-DevOps: verification, monitoring (or observability), and troubleshooting. For each workflow, a number of different processes is defined, each one corresponding to different conditions or phases. In particular, this work concentrates on the tools described in Section 4 and represents the processes combining these tools. The next sub-section presents the criteria we used for distinguishing different processes that belong to each workflow. Then, the next sections deal each one with a different workflow. The notation chosen for specifying processes is BPMN (Business Process Model and Notation) [92], which is a standard graphical intuitive language commonly used for the specification of business processes.

## 5.1 Scopes and Roles of processes

When defining SP-DevOps processes for observability, troubleshooting or verification, it helps to categorize them by its purpose (i.e. WHY the workflow exists), the scopes where the process operates (i.e. WHAT the process does), and the roles involved in the process (i.e. WHO does what). Based on these dimensions, various different processes can be classified, pertaining to different kinds of scopes and roles.

In order to break down various scopes and purposes of SP-DevOps processes, we devised a simple WHY/WHAT/WHO model:

- WHY: Purpose of the operational (SP-DevOps) processes (monitoring/verification/troubleshooting)

    - Status updates (e.g. for orchestration, optimization of embedding, infrastructure status …)

    - SLA follow up (e.g. for SLA reports)

    - Accounting

    - Troubleshooting (including RCA and debugging)

    - Quality assurance (Verification, Validation/Activation)

- WHAT: Technical scope of the process (monitoring/verification/troubleshooting)

    - Physical Infrastructure Resources (including physical network functions – PNF)

    - PNF (i.e., middlebox) functionalities

    - PNF performance

    - Service functionalities (entire service chain or NF-FG)

    - Service performance

    - VNF functionalities and performance (seen in isolation)

- WHO: Initiator of monitoring/verification/troubleshooting processes and receiver/user of the results produced by such processes

- Service Developer (in SL space, Operator internal or external)

- VNF Developer (in SL space, Operator internal or 3[rd] party)

- Operator (in SL space, e.g. for "internal" SLA follow up and service status)

- Operator (in OL space, e.g. for service optimization, embedding, scaling)

- Customer/Tenant (in SL space)

Note that the WHO category may also include modules that act on behalf of the operator/developer/user role, e.g., a scaling module in the OL on behalf of the operator, or a MF on the controller on behalf of the VNF developer.

In the following table, we show a mapping according to this model. The list contains the SP-DevOps processes defined within each UNIFY workflow.

| Workflow | SP-DevOps Process (Ref. to relevant section) | WHY (Purpose) | WHAT (Techn. Scope) | WHO (Initiator/Receiver) |
|---|---|---|---|---|
| Verification | Pre-deployment verification (Sec. 5.2.1) | Correctness assurance (Verification) | Service (functionality) | Customer, Operator (SL), Service Developer (SL) |
| Verification | Post-deployment verification (Sec. 5.2.2) | Quality assurance (Validation) | Service (functionality) | Operator (OL) |
| Observability | Service monitoring (Sec. 5.3.1) | SLA reporting | Service (func.& perf.) | Customer, Operator (SL), Service Developer (SL) |
| Observability | Orchestration support (Sec. 5.3.2) | Status updates | Resources (per VNF) | Operator (OL) |
| Troubleshooting | VNF development (Sec. 5.4.1) | Troubleshooting (debugging) | VNF (func. & perf.) | VNF Developer (SL) |
| Troubleshooting | Service troubleshooting (Sec. 5.4.2) | Troubleshooting (RCA) | Service (func. & perf.) | Operator (OL) |

## 5.2 Verification workflows

As previously discussed, for verification we can distinguish two main kinds of verification: pre-deployment verification, which works on models and is performed before deployment in order to prevent that bad graphs or bad configurations are deployed, and post-deployment verification, which works on the infrastructure after deployment and aims at detecting possible errors that show up in the deployed graph. The next two sub-sections deal with these two different kinds of verification. For pre-deployment verification a single process is defined while for post-deployment verification several processes are defined according to the different tools that have been experimented.

## 5.2.1 Pre-deployment verification

The pre-deployment verification process in BPMN notation is shown in Figure 54. It represents how the pre-deployment verification tools operate within the UNIFY architecture, focusing on the key components, and which verification tasks they perform, according to the main roles involved in the UNIFY architecture.

The whole verification process starts with a request of service graph deployment coming from the tenant or operator. The request includes: (i) a description of the Service Graph to be instantiated; (ii) the set of verification policies that have to be verified in the graph (in other words, the properties, as already defined in Section 4.1). These inputs are managed by the Service Graph Adaptation Sublayer of the UNIFY architecture, which triggers the first verification step, i.e. Service Graph verification.

In detail, during SG verification, the system interacts with the Neo4JManager service (an internal component of the pre-deployment verification module, see Section 4.1) through its REST API, specifying the service graph, which is stored in the Neo4J database, and the type of properties to check on that graph. At this level, topology verification of the Service Graph is done by Neo4JManager ("SG verification" activity in Figure 54). The verification tool returns a verification result, which is processed by the system in order to either alert the user of failing in deploying the service request or continue with the request process.

In case of successful verification, the Orchestration level of the UNIFY architecture comes into play. In particular, the system operates several transformations on the NF-FG, related to its translation, decomposition, and embedding into available resources. In order to check the coherence of at least some of these operations, the orchestration layer also triggers the second stage of verification, namely the NF-FG verification.

The NF-FG verification is actually split into two steps, compliantly with the orchestration process described in D3.3 [7], the "Topology Verification of Decomposed NF-FG" and the "Flow-level Verification" (Figure 54).

After the NF-FG decomposition phase, the orchestrator triggers the first NF-FG verification step, which is a topology-level verification ("Topology Verification of Decomposed NF-FG" in Figure 54). Similarly to what happens in SG verification, the Neo4JManager tool performs the NF-FG topology verification process and needs both the (decomposed) NF-FG description and the verification properties as inputs. The Neo4JManager tool returns a verification result, which is further processed by the system. Part of this processing includes a service for fixing possible topology errors detected by the Neo4JManager in NF-FGs ("NF-FGBugFixing" in Figure 54). After the fix, the modified NF-FG must be verified again, in order to be sure about its correctness. A loop is thus created, which can be repeated for a finite number of times. If the process for fixing NF-FG errors is not able to correct the graph within a fixed number of iterations, the system must alert the tenants of the verification failure. In the case of a successful fix, the system will continue in deploying the NF-FG.

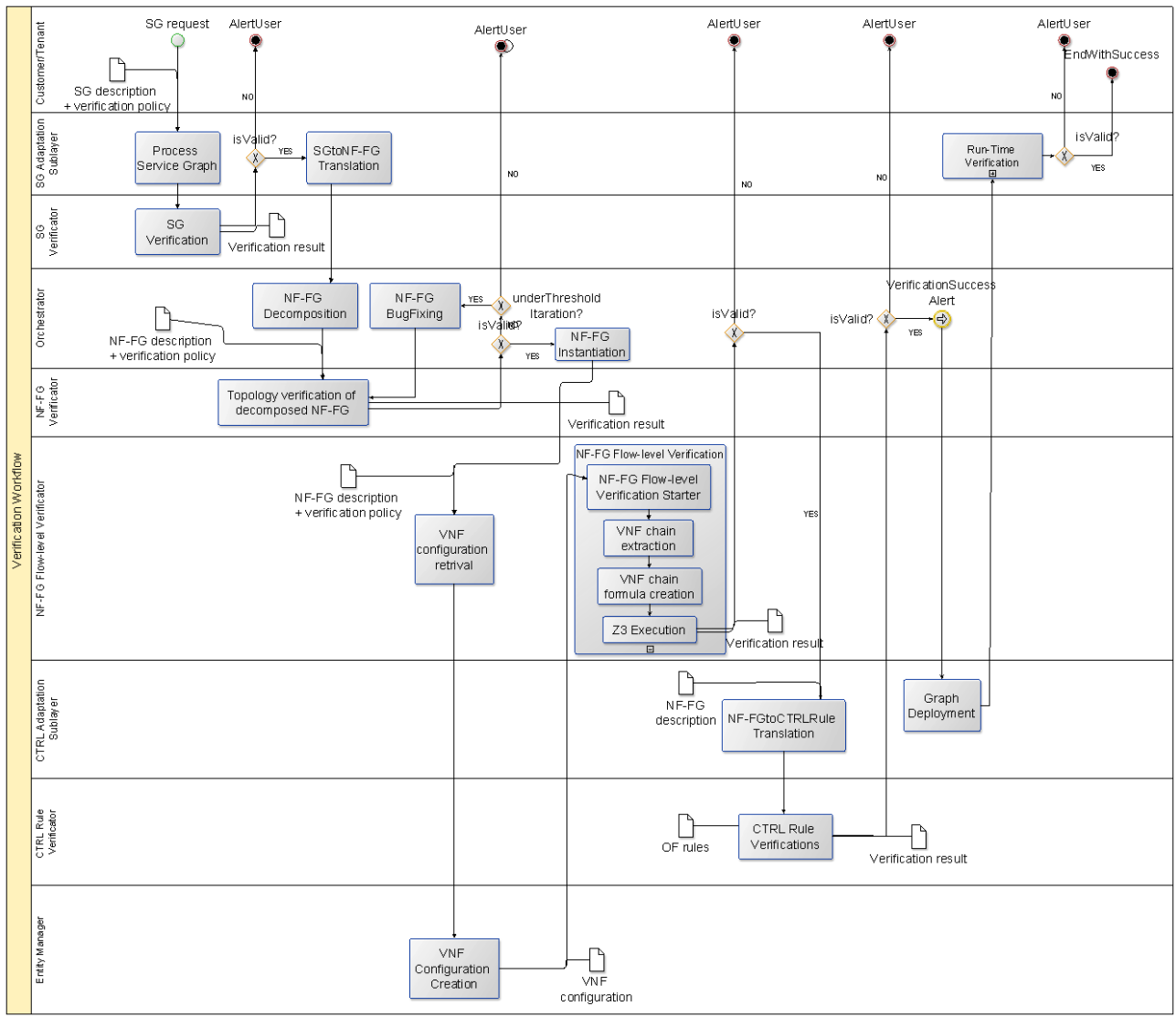

*Figure 54 Pre-deployment verification workflow.*

At this point of the process, and in essence after the embedding phase of the orchestration process (see the extended view of orchestration process in D3.3 [7]), the second phase of NF-FG verification, i.e. flow-level verification, is performed by the means of VeriGraph ("NF-FG Flow-level Verification" activity in Figure 54). Actually, before performing this kind of verification, further steps must be done by the system. The details of such steps have not been reported in the diagram as they do not concern the verification workflow. However an example of these preliminary steps is the retrieval of VNF configurations. This step is done in order to deploy the NF-FG correctly. In this process representation, we are assuming that the VNF configurations are already available. The system could retrieve such configurations from an external Entity Manager. Another solution could be that tenants configure directly their own graph, but these are details not essential from a verification perspective. Hence we can assume that the verification tool (namely the activity "Flow-Level Verification" in the workflow figure) receives the NF-FG description, the configuration of its VNFs and which properties have to be verified.

"Flow-level verification is composed of a set of sub-steps that are run in an automated fashion. As it is depicted in Figure 54, the verification tool has to do some preliminary tasks including:

- Retrieval of all the paths that connect the source and destination nodes in the NF-FG, thanks to Neo4JManager, which exposes an API for doing this task ("VNF chain extraction" activity);

- for each of these paths, VeriGraph must create a model though First Order Logic formulas and initialize the Z3 environment. In fact Z3 is the core engine of the main pre-deployment verification tool proposed by UNIFY, as it was already mentioned in D4.2. [6]

In case of failure in verifying flow-level properties in the NF-FG, the system must alert tenants, otherwise it can continue with the NF-FG deployment. Once again, here the diagram does not detail the next steps performed by the system to deploy the graph, like the creation of Controller Rules (e.g., OpenFlow Rules). Moreover, we can image that in this phase a pre-deployment verification of the correctness of such rules can be performed by another verification tool (shown in figure as "CTRL Rule Verification" task). However, as this verification problem has already been addressed and solved by many existing works in relevant literature (many tools are available like VeriFlow [93], NetPlumber [94] and others), this task will not be part of the demonstrated UNIFY pre-deployment verification tool functionality. Instead, the tool functionality will be limited to verifying the correctness of graphs in the higher levels of the UNIFY architecture (namely Service and Orchestration levels), which was the main open challenge.

### 5.2.2 Post-deployment verification

When all the pre-deployment verification processes have been completed successfully, the tenant graph is deployed, and run-time verification can be performed, possibly alerting the tenant if any error is detected. In UNIFY, two different run-time verification techniques can be activated: verification of path functionality by tagging and verification of match flow functionality by automatic test packet generation. Consequently, we present here two instances of the post-deployment verification workflow, each one using a different verification technique. The two instances share the same purpose, scope and roles. These workflow instances aim to give a clear overview of how the post-deployment verification tools operate within the UNIFY architecture.

Regarding the first post-deployment verification process, the operator would like to ensure the in-order policy updates in the data plane as there is an asynchronous medium between OpenFlow switches and controller so the consistent update verification tool takes care of it. The NF-FG converted from the service graph along with corresponding verification policies is passed as an input to the resource orchestration layer through the Sl-Or interface. The verification policies come in the form of requests from the user (operator in this case). The operators receive requests in the form of policies to determine the exact behaviour on the data plane and whether the behaviour is in sync with the control logic. As explained in D4.2 [6], there can be many reasons that could lead to an anomaly such as: a) the asynchronous medium between control and data plane could lead to out of order updates in policies, b) switches from different vendors have different reaction time and c) h/w or s/w bugs in switches, middleboxes and d) configuration errors.

The resource orchestration layer maps the verification policies and the graph received from the service layer to the substrate resources and therefore, it allows the reservation and configuration of resources on the physical substrate. This step is automatic. The path verification tool works in the infrastructure layer and performs verification of data plane paths. This tool can have one of the two possible flavours: a) as an application running on the SDN controller for the domain or b) as a data plane feature.

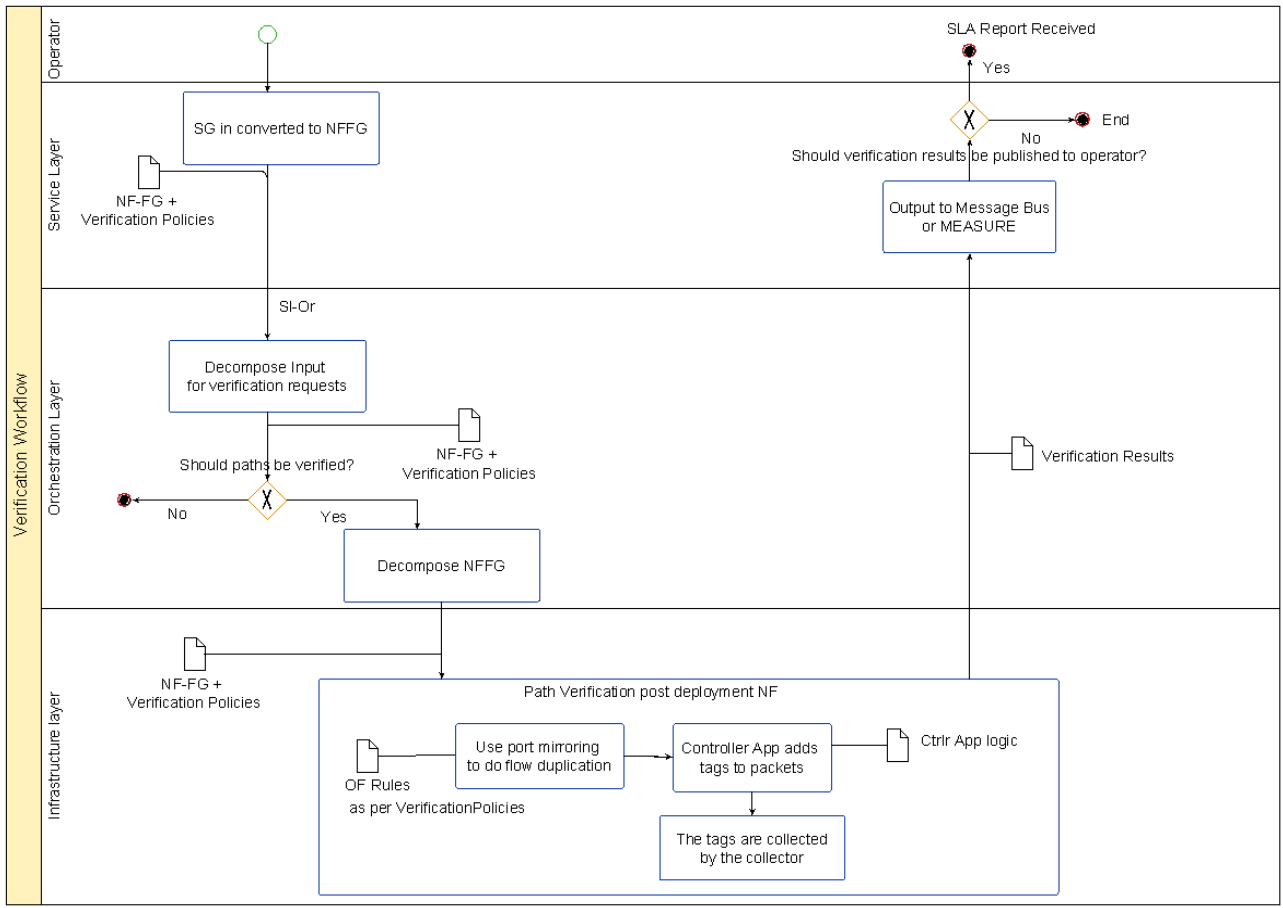

*Figure 55 Controller application-based approach for path verification.*

The controller-based application (Figure 55) works in three steps as shown in Figure 55: a) port mirroring b) tagging c) collection of tags. The flow rules are inserted with a timeout to mirror the traffic so that the line rate of production traffic is not disturbed. The tagging of the mirrored traffic is done by the controller application (RYU controller is used, it can be used with other controllers like ODL, Flowvisor having corresponding controller applications) with the corresponding logic or it can be done by the switch itself (data plane approach explained in the next paragraph). The controller application resides on the SDN controller in the infrastructure layer and adds unique tags per links to packets which eventually are gathered by the collector for analysis The collector sends the data to the parser and the gathered statistics are sent out through the output interface. The results are published to the user (operator). If there is an anomaly, the users (operators) are informed by the service layer.

The data plane-based approach (Figure 56) tags packets on the line rate directly on the datapath element (OpenFlow switch) by pushing VLAN or MPLS tags. VLAN or MPLS tags are used as free space on packet header is already a scarce commodity. VLAN ID tags should be used in a way that forwarding behaviour is not affected by the tags inserted for verification. Moreover, there has to be a lot of consideration to make sure that rules inserted for tagging do not interfere with the forwarding rules. There is a configurable soft timeout of this tagging rule so that not all packets get tagged and later the collector gathers the packets in the same way as in previous controller application approach and analyses the tags for determining paths. This is where our tool terminates the proceeding steps such as the representation of the verification results for higher layer metrics, which are considered as automatic or taken care by the UNIFY architecture.

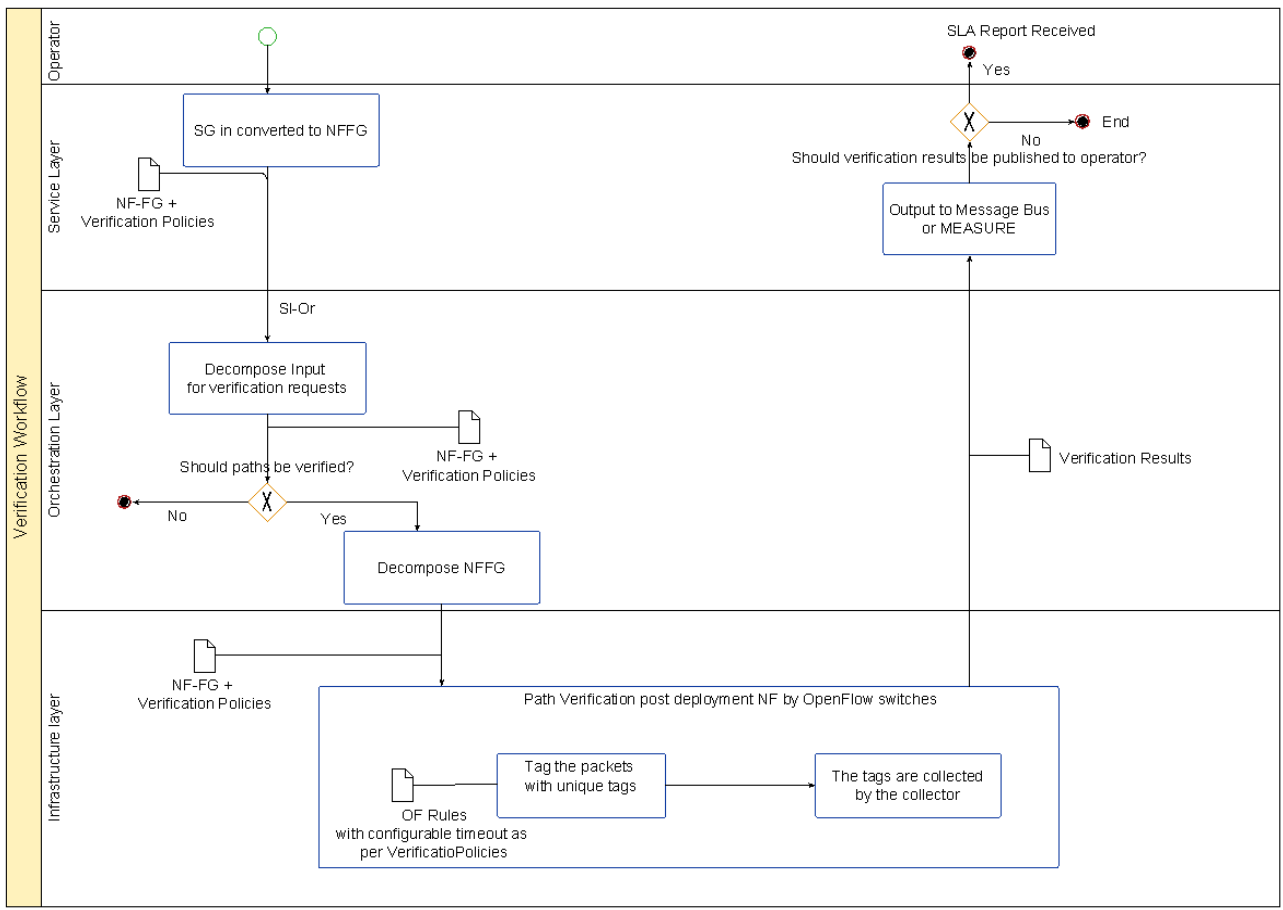

*Figure 56 Data plane-based path verification.*

The second instance of the post-deployment verification workflow is based on the AutoTPG tool and is shown in Figure 57. It shows how AutoTPG operates within the UNIFY architecture.

In the UNIFY architecture, the service graph adaptation layer communicates with users (such as operators). For the post deployment verification, users provide the service graph together with verification requests in a set of policies. In this set of policies, the users request to verify the deployed graph for finding flow-matching errors. The causes of such errors are: (1) configuration errors such as presence of an error-prone high priority forwarding entry and (2) software/hardware bugs.

In the service graph adaptation layer (shown in Figure 57), the service graph is converted into the NF-FG graph, and the NF-FG graph description together with verification policies translated into NF-FG-related policies is passed to the resource orchestration layer through the SL-OR interface. The resource orchestration layer is the core of the UNIFY architectures. This layer maps the graph (including polices) received from the service layer to the physical resources. Based on this mapping, the layer automatically reserves and configures resources and management functions (including Verification Network functions for post deployment verification) through its southbound interface towards the controller adaption layer. The controller adaptation layer creates a domain wide joint abstraction of all the underlying resources which then creates virtualized resource view per user. Using this layer, the AutoTPG control network functions are deployed over the infrastructure layer.

Through the NFs deployed (shown in Figure 57), the post deployment verification tool AutoTPG performs verification of the data plane network function. As shown in Figure 57, it performs three steps for verification: (1) flow duplication, (2) test packet-generation, and (3) matching error calculations. In the flow duplication step, the controller in the orchestration layer copies the Flow Entries in the infrastructure layer from a FlowTable to another FlowTable. The entries are duplicated to maintain the same state between the FlowTable (under verification) and the duplicated FlowTable.

In the test packet generation step, the controller generates and transmits test packets that can match with the Flow-Match Header of duplicated Flow Entries. In the matching errors identification step, the mechanism calculates the matching errors either by reading the counters (statistics) of the duplicated Flow Entries or by comparing the sent and received test packets.

The output given by our tool identifies which deployed NFs (found through errors in Flow Entry) have errors. If there is no error, the output will be that there is no Flow Entry specific issue in the networks.

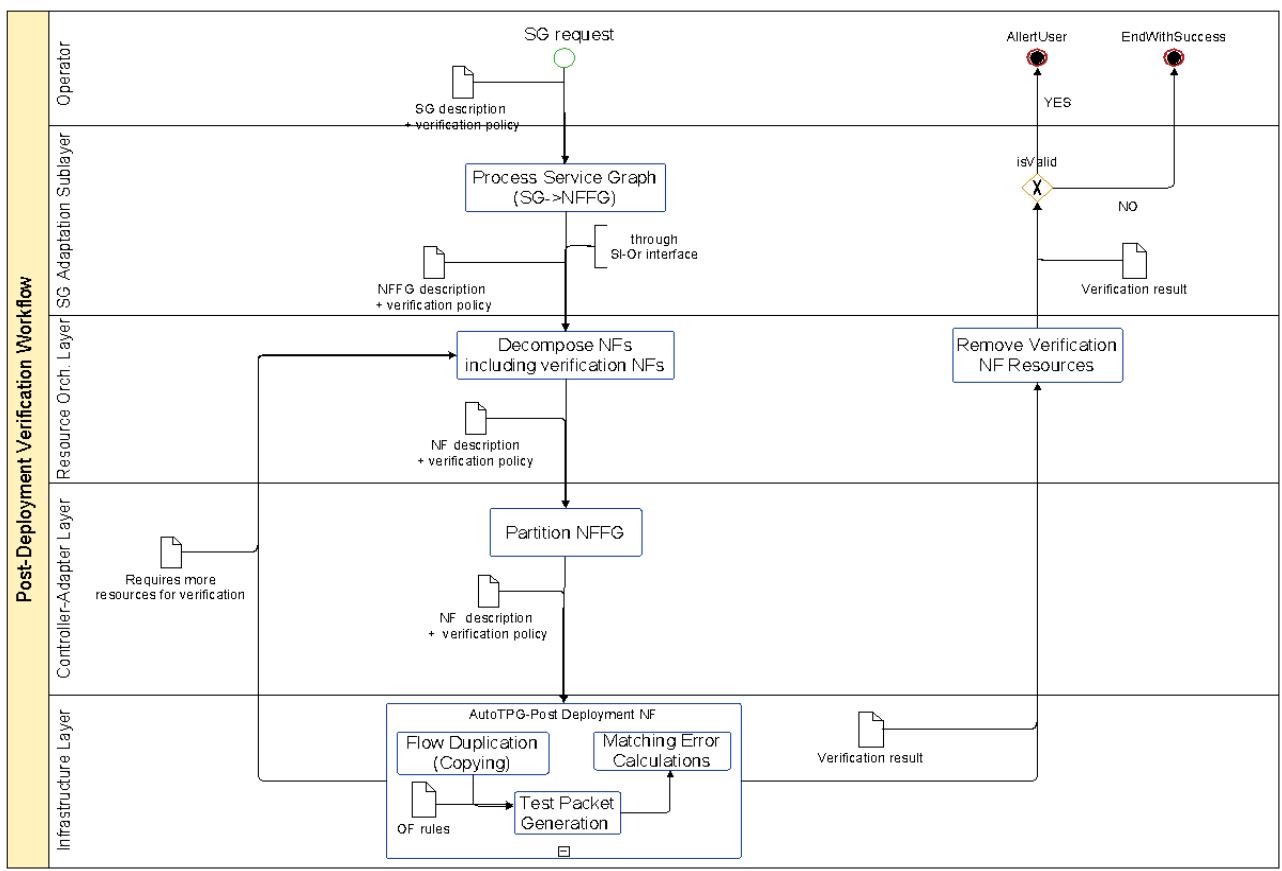

*Figure 57 Post-deployment verification workflow (using automatic test packet generation).*

## 5.3 Monitoring workflow

In this section, we describe two monitoring processes part of the SP-DevOps observability workflow. They outline concrete processes evolving the generic observability process definition documented in [25] (D4.1, Section 4.2.2.1).

The first process corresponds to the SP-DevOps support for the service deployment and orchestration processes as described in D3.3 [7], relating to Service SLA monitoring of an elastic router function, ranging from the service layer all the way into the infrastructure and back again. The purpose (WHY) of this process is SLA follow up and reporting, the scope (WHAT) is service functionality and performance; and the initiator and receiver of this process (WHO) is the operator in the service layer

The second process supports dynamicity aspects of the UNIFY orchestration layers and relates to the service elasticity example enabled by the Cf-Or interface as described in D3.3 (sections on elastic router and state migration). The purpose (WHY) of this process is real-time status updates, the scope (WHAT) is VNF resource usage and performance; and the initiator and receiver of this process (WHO) is the resource orchestration layer in charge of the operators services.

### 5.3.1 Service SLA monitoring process

In this section, we will outline a service SLA monitoring process in alignment with the UNIFY architecture and the WP3 deployment process. It is important to note that the process depicted below is focusing on monitoring related activities and interactions. Instead SG/NF-FG and general service deployment related activities and interactions are described in D3.3 [7] (section on orchestration process).

The observability process for service SLA monitoring (shown in Figure 58) is supported by SP-DevOps tools and functions developed as part of SP-DevOps by means of MEASURE annotations to NF-FGs, DoubleDecker messaging for dissemination of configuration and results, various monitoring components (including RateMon, E2E Delay and Loss Monitoring, EPLE, TWAMP, IPTV quality, AutoTPG and path-verification) treated as one generic monitoring function in the depicted process. As an alternative to the monitoring results dissemination part of the process (right part of Figure 58), the recursive query language and engine (RQL) could be used to support on-demand querying and processing of specific results from the monitoring DBs in the infrastructure, bypassing many of the data merging and composition steps of MEASURE.. The RQL is not shown in Figure 58, but described in Section 4.11 and visualized in Figure 3.

The process starts by a tenant (customer or user) requesting an abstract service graph (SG) together with a set of requirements expressed in an SLA. The process flow through service (SL), resource orchestration (RO), and controller adaptation layers (CA) is as follows: first, the SLA is translated into specific observable KPIs; second, the KPIs are expressed in MEASURE annotations to the derived NF-FG, defining measurement definitions (metric, locations, configuration), zone definitions (aggregation and threshold configuration), and reactions (how to disseminate the results).

In the RO sublayer, MEASURE annotations need to be decomposed following the decomposition chosen for the NF-FG. Next, the MEASURE annotations need to be scoped according to the scoping of the NF-FG onto available infrastructure domains. At this stage, the NF-FG together with MEASURE annotations reach the domain specific monitoring controller, e.g. Local orchestrator including a Monitoring Management Plugin (MMP) within a Universal

Node (UN). The MMP interprets the MEASURE measurement definitions and selects a suitable monitoring tool available in this domain, and if applicable, it instantiates a MF control app (this would be the case for many SP-DevOps tools like RateMon, AutoTPG, EPLE or TWAMP). It also retrieves the Zone and Reaction definitions of MEASURE to configure domain wide aggregation points (which might be implemented within as part of MF control app or as a generic DoubleDecker client). In order to compile the final configuration of actual observability points (OPs), the MMP will receive relevant parameter from network and compute controllers (e.g. through the LO within an UN). Based on the MEASURE annotations and the low-level details gathered locally, the MF control app can configure one or several OPs within the domain.

During the operational phase, the OPs perform their continuous monitoring tasks according to configuration. In regular intervals or on events (like threshold breaches) the monitoring results are collected by the MF control app via DoubleDecker. Optionally, a DoubleDecker aggregation client might reduce the amount of data based on the earlier configuration. The MF control app can further pre-process and aggregate results from several OPs and push the results into a monitoring DB in the MMP. The SLA reporting reads from this DB in periodic polling intervals (order of seconds to minutes) and uses the scoping and composition rules of CA and OR to merge and re-compose metrics corresponding to the higher abstraction levels of the NF-FG of the observed service. In the SL, the KPIs are stored in an SLA database for logging and reporting purposes. Additionally, the periodic results can be compared to the requirements of the SLA and trigger notifications to the tenant on immediate SLA breaches. In a periodic reporting interval (order of weeks or months), SLA reports are created out of the timeseries of metrics stored in the SLA DB and provided to the tenant.

The above described process is probably the most generic observability process and was chosen as a descriptive example due to its completeness through all architectural layers. It serves mainly the purpose to show how UNIFY SP-DevOps tools together can be used to create a process in concertation with UNIFY orchestration modules and on UNIFY infrastructure in the form of the UN. SP-DevOps tools will support the lower part of the process (with focus on controller and infrastructure layers), as well as rudimentary versions of monitoring orchestration modules in the orchestration layers CA and RO, which is developed primarily for the sake of integration with dynamicity use-cases such as elasticity and state migration as documented in [7] and [18]. The specific process to support orchestration dynamicity is outlined in the next section.

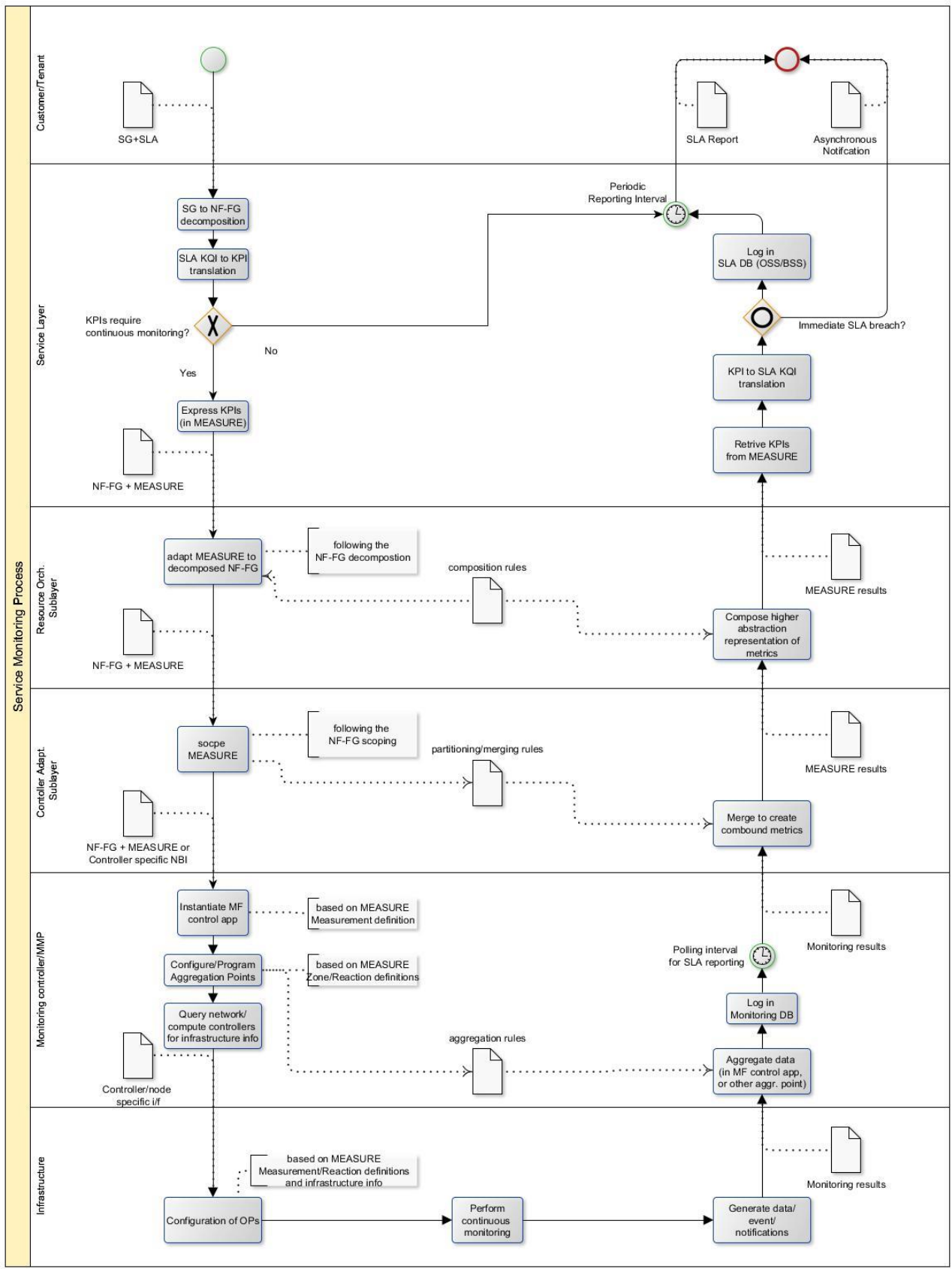

*Figure 58 Process diagram for generic service SLA monitoring.*

### 5.3.2 Orchestration dynamicity support process

In this section, we describe the process supporting orchestration dynamicity in alignment with the UNIFY architecture and the service deployment and orchestration process, relating specifically to the service elasticity examples enabled by the Cf-Or interface as described in D3.3 (sections on elastic router and state migration).

This monitoring orchestration support process (shown in Figure 59) is supported by SP-DevOps tools and functions by means of MEASURE annotations to NF-FGs, DoubleDecker messaging for dissemination of configuration and results, and selected monitoring components (RateMon and cAdvisor as a third party tool for computer resource monitoring are considered) treated as one generic monitoring function in the depicted process.

Like the more generic service SLA monitoring process, this process starts by a tenant (customer or user) requesting an abstract service graph (SG) together with a set of requirements in form of an SLA. In the SL, the SLA is translated into specific observable KPIs. These KPIs are then expressed in MEASURE annotations to the derived NF-FG. In this example, the RO sublayer chooses some network connections to be realized through an elastic router, i.e. a CtrlApp based (black-box) decomposition [7], due to some strict network requirements on a certain ports (e.g. capacity or availability). The elastic router is decomposed into VNFs realizing not only data plane (DP components), but also a special control network function (control NF) component, which inherently knows the service logic of the specific network function (in this case the router elasticity logic), corresponding to the declarative option [7]. In cases when the procedural option is followed, the elasticity logic is not inherent part of the VNF instantiation representing the control NF, and it can be programmed dynamically based on a procedure template residing with composition rules.

The remaining activities in the RO and CA sublayers are to decompose MEASURE annotations following the complete decomposition chosen for the NF-FG, and to finally scope the MEASURE annotations according to the scoping of the NF-FG onto available infrastructure domains. The next steps until the operational phase are similar to the generic Service SLA monitoring deployment phase: the NF-FG together with MEASURE annotations reach the domain specific monitoring controller, e.g. Local orchestrator including a Monitoring Management Plugin (MMP) within a Universal Node (UN). While network and compute controllers deploy and connect the service VNFs (including the control NF), the MMP interprets the MEASURE measurement definitions and selects a suitable monitoring tool available in this domain, and if applicable, it instantiates a MF control app (in this example for RateMon). It also retrieves the Zone and Reaction definitions of MEASURE to configure domain wide aggregation points (which might be implemented within as part of MF control app or as a generic DoubleDecker client). In order to compile the final configuration of actual observability points (OPs), the MMP will receive relevant parameters from network and compute controllers (e.g. through the LO within an UN). Based on the MEASURE annotations and the low-level details gathered locally, the MF control app can configure one or several OPs within the domain.

Basically, the operational phase within the infrastructure is similar to the generic monitoring example: the OPs perform their continuous monitoring tasks according to configuration, which in the elastic router example means monitoring of the VNFs representing elastic router DPs. In regular intervals or on events (like threshold breaches) the monitoring results are collected by the MF control app via DoubleDecker. Optionally, a DoubleDecker aggregation client might reduce the amount of data based on the earlier configuration. The MF control app can further pre-process and aggregate results from several OPs, but besides pushing the results into a monitoring DB, the relevant metrics (traffic rates, resource usage, etc…) are made available to the elastic router control NF though DoubleDecker. This can be either done via direct notifications or the publish/subscribe interface (which is the preferred way). The elastic router control NF can decide on reception of the monitoring results to initiate a scaling operation via the elastic control loop as described in the process flow in Section 6.2, and in more detail in D3.5 [18]. In short, the elastic router control NF instruments the Cf-Or interface to first request the NF-FG infrastructure context from the RO, and then locally performs a CtrlApp based decomposition which might result in an updated NF-FG with MEASURE annotation, for instance including further DP instances. This information is then the input to another deployment cycle, starting with an NF-FG + MEASURE at the Resource Orchestration.

The above described process specifically supports the elastic router use case. However, it is general enough to support orchestration also for other dynamicity features such as state-migration. The main parts of this process have been implemented within the project-wide IntPro, supporting the WP3 elastic router use-case and with the Universal node as the infrastructure execution environment. The implementation is described in Section 6.2, with a focus on the parts involving SP-DevOps tools and functions.

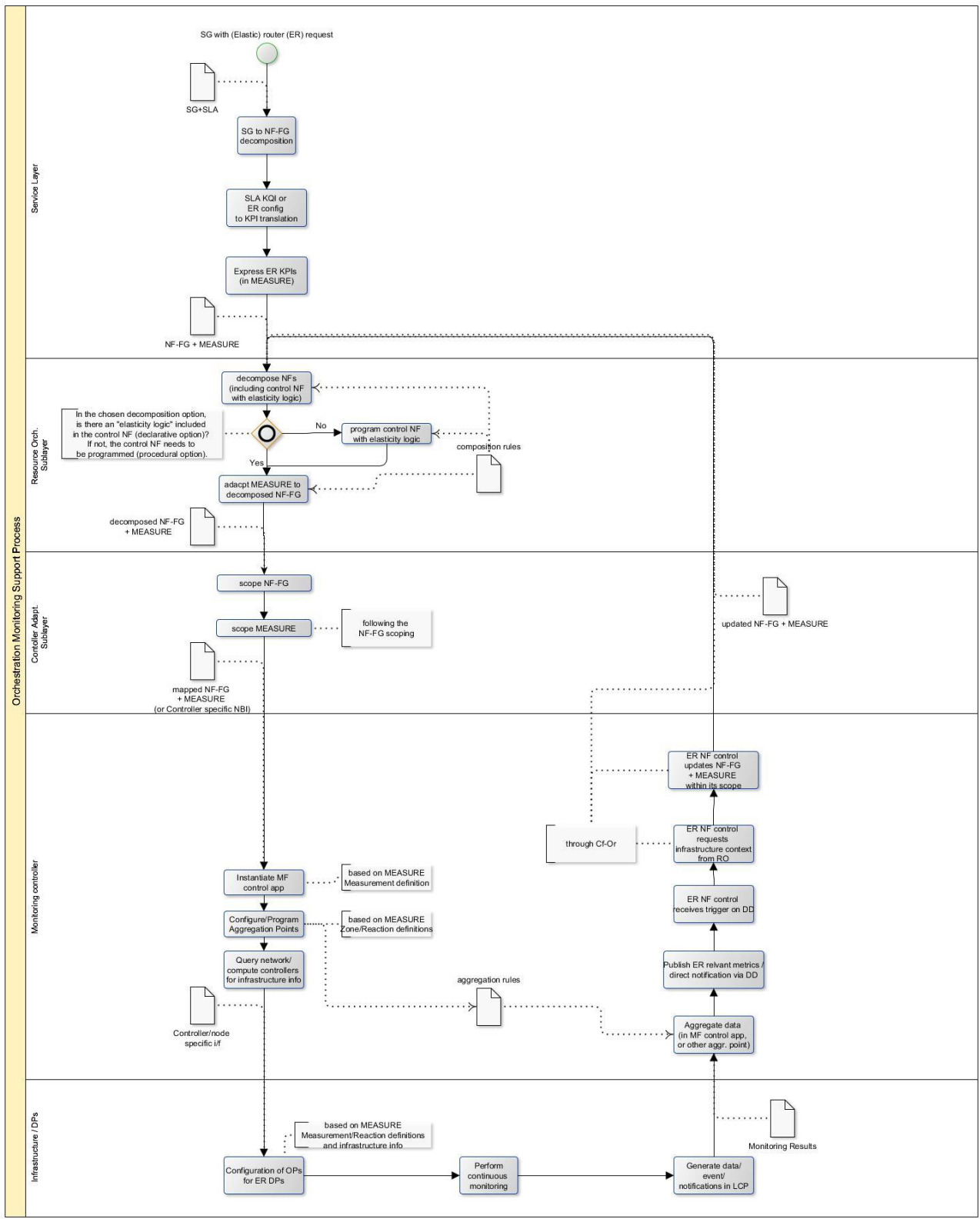

Figure 59 Process diagram for monitoring in support of orchestration dynamicity on the example of the elastic router.

## 5.4 Troubleshooting workflows

A troubleshooting process is typically triggered either by a user request (Operator/VNF developer/Service developer), or by a running monitoring process and its functions used for monitoring KPIs, run-time verification, or conditional observability (e.g. breakpoints) in a deployed service graph. The information level of the output depends on the role of the requester and associated permissions [26].

A troubleshooting process gathers data from three sources - MFs, VNFs and DBs (i.e. logs), for the purpose of analysis (e.g. event correlation) and reporting. For run-time troubleshooting (triggered by a monitoring process), the report or notification can be used for triggering actions to automatically and seamlessly mitigate the problem by re-deploying affected parts of the NF-FG. Moreover, troubleshooting can be based on either of the following or both:

- the creation of a troubleshooting NF-FG, which is an extended NF-FG with additional MFs and VNFs (e.g. functions for duplication and comparisons) used for testing and observing the components of the service,

- queries to DBs (such as logs) and direct requests to MFs and VNFs for the purpose of gathering information about the network state and behavior.

In many cases, troubleshooting based on gathering information from the system is sufficient for identifying the cause of a detected problem. In other cases, a modification of the NF-FG into a troubleshooting NF-FG may be necessary in order to gather additional information not already available to the troubleshooting process, or for performing specific tests or actions (e.g. debugging) on a VNF or service graph, for instance during VNF or service development phases.

A Troubleshooting Function performs three tasks: Translation of a Troubleshooting Specification into Troubleshooting Actions; execution of Troubleshooting Actions; analysis and reporting. Troubleshooting Functions can be implemented in three different ways: as a function invoked by a user at User Layer (UL) (Section 5.4.1); as a functional block in the Resource Orchestrator Sublayer; or as a dedicated Troubleshooting Controller App (TCA) (Section 5.4.2). Troubleshooting Functions invoked at the UL are suitable for VNF developer support or for additional troubleshooting support of an SG. Troubleshooting implemented as a Resource Orchestrator functional block or as a controller is more suitable for fault management of a SG for seamless resolution of detected problems.

The Troubleshooting Specification defines the steps and behavior of an automated troubleshooting process. Essentially, the Troubleshooting Specification defines which components of the SG should be tested or observed and how (e.g. in which order), including data sources (MF, VNF, DB) to be used and corresponding configurations. The Troubleshooting Function implemented at the UL, Orchestrator Layer (OL) or in a TCA, processes and translates each step defined in the Troubleshooting Specification into a detailed description of Troubleshooting Actions to be executed in the system for the existing SG. Note, that the formal language used for creating a Troubleshooting Specification is Troubleshooting Function specific, and will not be addressed here. If the design of the Troubleshooting Function and the Troubleshooting Specification language allows it, a Troubleshooting Function may call another one for the purpose of nested troubleshooting.

A Troubleshooting Action is a description that includes: a Troubleshooting Graph (TG) with the required MFs and VNFs; KQI/KPI to be observed by the MFs; VNF configuration; query expressions for accessing DB information (e.g. logs, historic monitoring information); and, instructions for pulling information directly from MFs or VNFs.

The TG includes the components (MFs and VNFs) that should be added or inserted to an already existing NF-FG to form a troubleshooting NF-FG. The definition of the TG may require access to MF/VNF information bases. The deployment of a TG requires functionality in the OL to extend an already existing NF-FG with additional VNFs and MFs. An empty TG corresponds to only performing troubleshooting based on gathering information from existing NF-FG via queries and direct requests to existing MFs and VNFs. The translation from Troubleshooting Specification to TG is done in a similar fashion for troubleshooting functions in OL and TCA. The extension from a TG to a troubleshooting NF-FG is handled by the Resource Orchestration and Controller Adaptation Sublayers in the OL.

Depending on the scope and purpose there are several ways of defining a troubleshooting process. We focus here on presenting two cases of troubleshooting workflows (see Section 5.1 for a summary of scope and role overview). The first workflow relates to VNF development, with the purpose (WHY) of troubleshooting and debugging of undesired behavior and performance at the service layer (WHAT), initiated by a VNF developer (WHO) (see Section 5.4.1). The second workflow relates to troubleshooting and root cause analysis (WHY) of service performance and functionality (WHAT), initiated (automatically) at Operator level in the OL (see Section 5.4.2).

### 5.4.1 Troubleshooting process for VNF development support

The process description in Figure 60 is in general applicable to any type of user (Operator/VNF developer/Service developer) and can be employed for debugging, testing and troubleshooting one or several VNFs under development. In this case, a VNF (or a chain of VNFs) is observed and debugged using arbitrary troubleshooting tools invoked at the UL. As an example, a SP-DevOps troubleshooting function could be implemented by a combination of EPOXIDE and the Recursive Query Language (RQL).

The requesting user provides a Troubleshooting Specification as input to the Troubleshooting Function, which is used for producing a TG, a query, or both for the purpose of gathering data. In some cases, the creation of the TG requires access to MF/VNF information bases. The resulting TG or query is sent to the Orchestration layer for processing. In the case the troubleshooting requires deployment of MFs and VNFs, the TG is decomposed (along with the specification and annotations) as a normal service graph. Information queries or requests may also be executed as a single or parallel task to TG deployment. If any query (or other type of information request) depends on the output of a TG, it can simply be executed as an additional Troubleshooting Action. Data (measurements, test results, notifications, etc.) gathered from the infrastructure layer are aggregated and forwarded to the specified aggregation point of the troubleshooting function in the user layer. The Troubleshooting Function processes and analyses the data and may, if necessary, carry out further queries or perform further measurements in line with the Troubleshooting Specification until a result can be reported to the requesting user or network function.

The outlined workflow supports automated troubleshooting support for different types of tools. As an example based on EPOXIDE, a Troubleshooting Specification would include the definition of steps for setting up a management network, followed by carrying out a set of tasks (manual or automated) for a certain VNF of the TG. A Troubleshooting Function using EPOXIDE would thus take the Troubleshooting Specification and translate the steps into Troubleshooting Actions, out of which the first would be to create the TG (i.e. the management network). The remaining steps would be to use the management network to obtain status information of the VNF under testing and carry out specified tests (manual or automated), which corresponds to processes related to the "Aggregate logs from distributed DBs/perform direct information requests" block in Figure 60.

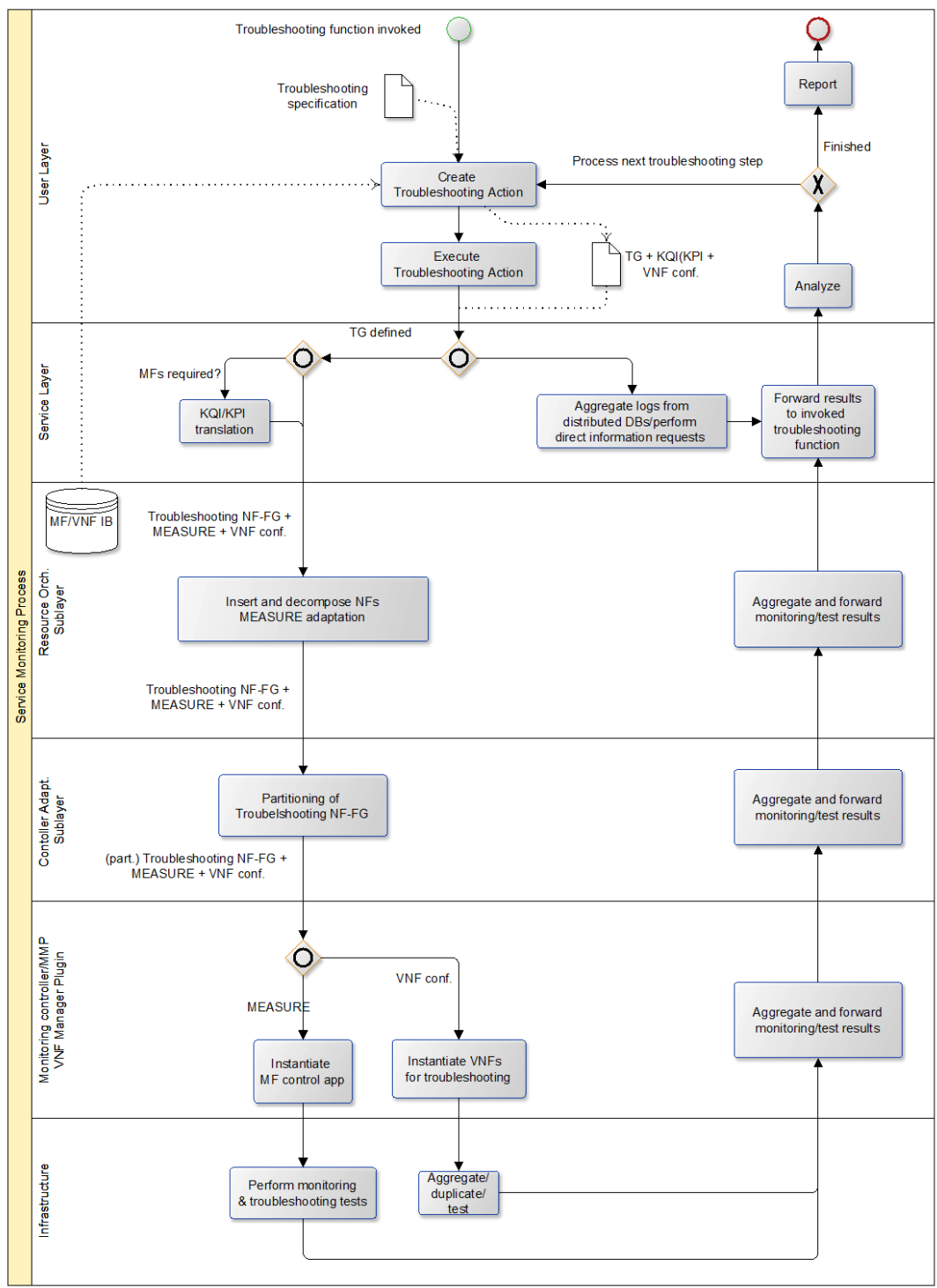

*Figure 60 Principles of the requested VNF troubleshooting process.*

## 5.4.2 Service troubleshooting process

Automated Service Troubleshooting is mainly aimed as part of service operation (Section 5.4.1). This process Figure 61 is triggered by a detection event notified by a monitoring function (e.g. continuous monitoring through RateMon, EPLE, or AutoTPG [6]). The notification (along with other relevant information) is logged and forwarded towards the Resource Orchestrator Sublayer where a Troubleshooting Function is invoked. Based on a Troubleshooting

Specification, assumed to have been provided along with the SLA definition at service graph deployment, the NF-FG is modified and updated to include specified troubleshooting support functions (MF, VNFs). The modified NF-FG together with MF and VNF specific annotations is decomposed, scoped and deployed on the infrastructure. Additionally, DB queries and information requests are performed as a parallel task when needed. Data from the infrastructure and DB queries are gathered at a data aggregation point and forwarded for further processing and troubleshooting analysis (e.g. event correlation). Data from lower layers of the architecture may be recomposed and presented at varying level of abstraction depending on role and permission. The results of the analysis may require further troubleshooting steps according to the Troubleshooting Specification. When the analysis is done, the result of the troubleshooting is logged and reported to other parts of the management system. Although the actual troubleshooting process can be regarded as finished already at this stage, Figure 61 also includes the description of invoking a problem-resolving management function and reporting on, for example, the location of identified trouble and whether the problem was resolved or not.

The alternative to implementing troubleshooting functionality in the Resource Orchestration Sublayer would be to implement a Troubleshooting Controller App (TCA) associated to the SG. In this case, a notification to the TCA from e.g. a continuous MF or WatchPoint instance, would trigger a request to the Resource Orchestration Sublayer to update the deployment of the NF-FG in accordance with a decomposed troubleshooting NF-FG (done dynamically in the TCA upon trigger or ready at deploy-time) given a Troubleshooting Specification (somewhat similar to the Elastic Router described in D3.3 [7]). The data from deployed MFs and VNFs of the troubleshooting NF-FG, as well as the results from DB queries, are gathered, processed, and analyzed by the TCA. In the end of the process, the TCA reports to the specified recipient, i.e. the user (Operator/VNF developer/Service developer) and/or a management controller function for resolving the identified problem.

Implementation of a troubleshooting logic as a functional block in the resource orchestration layer relates more to the early architecture defined in D2.1 [10] and D4.1 [25] whereas the TCA approach appeals more to the narrow-waist architecture as described in D2.2 [11]. In practice, which variant to implement depends on the capabilities of the Orchestration layer, performance requirements and technical features of the infrastructure.

*Figure 61 Principles of a service troubleshooting process.*

# 6 SP-DevOps use-cases and demo implementations

The UNIFY project has considered three process categories to realize the main technical challenges of SP-DevOps: verification, monitoring (or observability), and troubleshooting. This section presents possible use-cases for SP-DevOps processes, each consisting a subset of SP-DevOps tools.

In particular, the state migration scenario related to the UNIFY FlowNAC prototype serves as a first use-case for SP-DevOps. FlowNAC is supported by programmable and efficient observablity through DoubleDecker messaging and MEASURE. Furthermore, the FlowNAC demonstrator allowed us to verify the applicability of our definition of monitoring functions (MF) as discussed in Section 3.2.3. The ongoing evaluation of the FlowNAC demo shows that the SP-DevOps MF concept can be nicely generalized to include also VNF-internal monitoring. However, we learned that our current definition of MEASURE for automatic programmability of MFs still has some shortcomings, as highlighted in Section 4.3.4. As the FlowNAC state migration demo has primarily been carried out in a parallel UNIFY workpackage, we want to refer for detailed discussion to D3.5 [18].

A second use-case involves a selected NF-FG as a scenario to demonstrate the pre-deployment verification process as integrated with the ESCAPE orchestration framework. Finally, as the third and main use-case, an Elastic Router allows us to demonstrate the implementation of a more advanced monitoring process together with automated troubleshooting. The next subsections discuss the latter two use-cases and their implementations in more detail.

## 6.1 Integrated verification of NF-FGs

The integration of VeriGraph into ESCAPE aims at enabling the verification of SGs which involves the potentialities and functionalities offered by the UNIFY architecture. Nevertheless, a verification process must have a reasonable cost in order to be actually feasible in the UNIFY context.

In this section, we show the impact of the verification process on the overall service deployment. A possible strategy to evaluate this impact is to compare the percentage of time spent by ESCAPE for the verification of a requested service (i.e., SGs, NF-FG) with the total deployment time. In order to evaluate this verification time, we consider as use case the NF-FG shown in Figure 62. This graph has been designed mainly to highlight the usefulness and potentialities of a verification process into the UNIFY architecture: thus, it is a simple graph that can be deployed into ESCAPE, for example, by means of Click elements. In particular, this NF-FG (Figure 62) is composed by three middleboxes, which forward and modify packets based on their internal state, and some end-nodes to verify the reachability between each other.

Note also that this kind of NF-FG that contains several types of nodes can be suited to apply the verification process to the different functionalities of the UNIFY architecture, like the deployment of a graph (i.e., SG or NF-FG) over several physical domains (i.e., multi-domain deployment). A multi-domain graph can be simply the NF-FG proposed in Figure 62, where the two users are deployed onto Mininet, for example, the middleboxes on the Universal Node, while the web server on OpenStack.

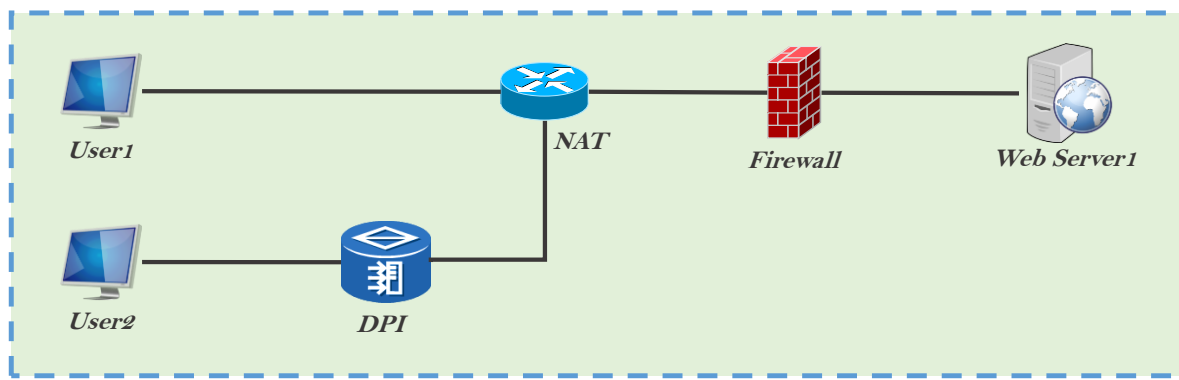

*Figure 62 Verification demo scenario.*

To briefly recap the purpose of the pre-deployment verification and how it is performed (see Section 4.1), VeriGraph is able to check the correctness of flow-level properties (expressed by means of verification policies) on a given NF-FG, by modelling how each function involved in the network and the network itself forwards and processes packets. The network properties (or verification policies) considered to verify the correctness of the proposed NF-FG, are mainly related to the existence of reachability between two nodes in the graph. In particular, the NF-FG shown in Figure 62 has been verified in terms of reachability between two pairs of web client (User1 and User2 in picture) and web server (WebServer1).

The satisfiability of a reachability property between two nodes is strictly intertwined not only with the network topology (if no path between the two nodes exists, certainly there will not be connection), but also with the configuration of each function in the network path that connects the two nodes. In order to check the tool with configurations that lead to different results, the following two scenarios have been defined.

***Scenario 1:*** Figure 62 contains a DPI function that has been configured to discard all packets containing some words in their payloads (e.g., packet contains the "sex" word). On the other hand, the firewall function considered in the proposed NF-FG is modelled to drop packets based on an Access Control List, whose content is a list of pairs of IP addresses not allowed to send packets between each other. In particular, in this first scenario, the ACL firewall has been configured to drop all packets that contain the IP addresses of the NAT and Web Server, as source or destination addresses.

In particular, we verified reachability from User1 to WebServer and reachability from User2 to Webserver. We run the verification five times in order to measure the average times taken by the tools. In this scenario, VeriGraph takes an average time of 0,971 seconds to conclude that the desired policies are not verified in the chosen NF-FG. This means that the verification module completes its task in about 11% of the overall deployment time employed by ESCAPE to handle the SG request, which is of 8,812 seconds. In Figure 63, we have also reported the partial times of each sub-process triggered in the verification module, whose description is presented in Section 4.1.4.

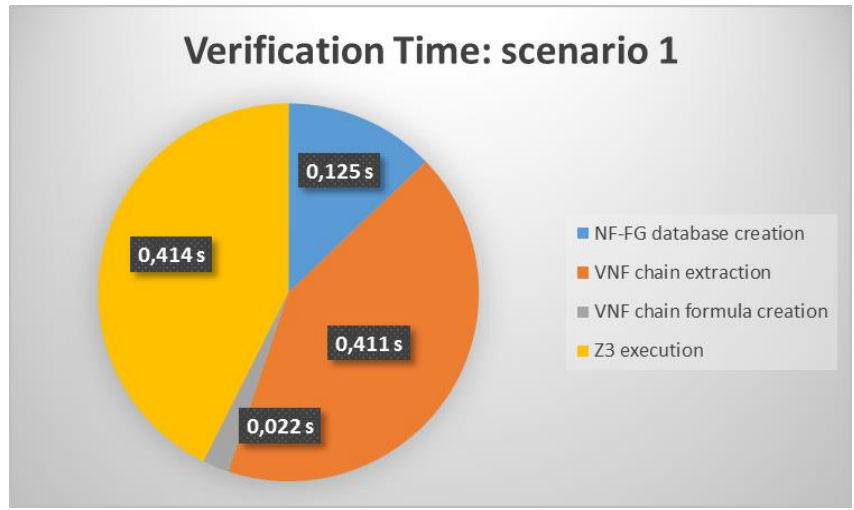

*Figure 63 Verification time in Scenario 1.*

*Scenario 2:* the second scenario shows a possible configuration of the involved functions, which allows the connection between the end-host. In particular, the DPI function has the same configuration of the previous case (i.e., drop packet containing the "sex" word), while the firewall function forwards packets with source or destination IP addresses belonging to the NAT or the Web Server. Hence, the reachability between <User1, WebServer1> and <User2, WebServer1> is satisfied when the two clients are configured such that they send packets that do not contain the word denied by the DPI.

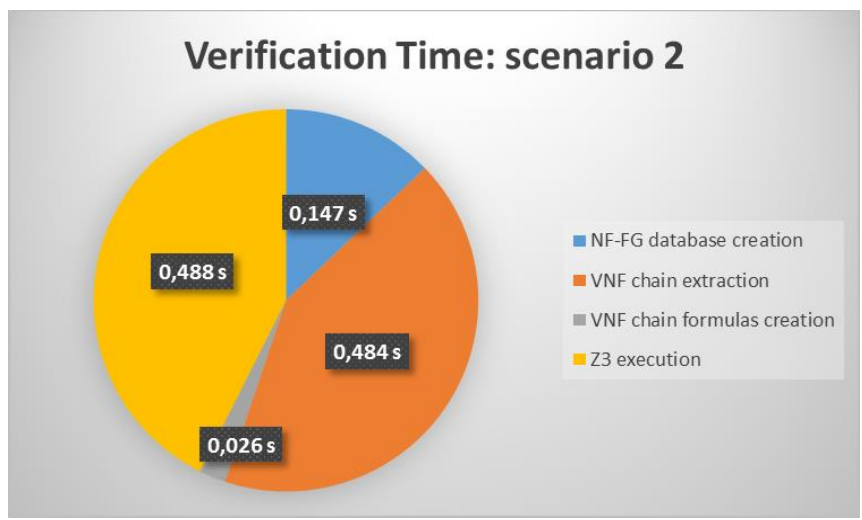

*Figure 64 Verification time in Scenario 2.*

We use the same metric as in the previous scenario to report the overall time required by VeriGraph to provide a verification output to ESCAPE. The total time employed by ESCAPE to handle the SG request in each case is 10,669 seconds for each test run (five times), while VeriGraph takes 1,144 seconds in average. The division of the total verification time into the several sub-processes (presented in Section 4.1.4) is depicted in Figure 64.

## 6.2 Elastic router monitoring and troubleshooting support

In this section, we describe the use case and demonstrator implementation for monitoring and troubleshooting processes support for dynamic network functions. Specifically, a SP-DevOps monitoring process supports dynamic scaling of network functions, corresponding to the generic process depicted in Section 5.3.2 and Figure 59, whereas a SP-DevOps troubleshooting process extents the monitoring process by considering the handling of a troubleshooting condition which is identified by the monitoring support, largely similarly to the generic process depicted in Section 5.4.2 and Figure 61. The demo descriptions in this section focuses on the SP-DevOps related components, and thus complements the demo description of D3.5 [18] (Section 4.1), which focuses on the Elastic Router logic and the orchestration functions and interfaces required for a dynamic network function.

### 6.2.1 Elastic Router use-case

An overview of the Elastic Router use case in depicted in Figure 65, which shows an overview of the UNIFY architecture on the left, the data model of the UNIFY reference points (i.e. NF-FG and MEASURE) in the middle, and the resulting deployment of the VNFs on the right. For a detailed description of the elastic router function itself, the reader is referred to [7], section 2.5.1, where it is more elaborately described. A brief introduction, however, is provided below.

The initial deployment of the elastic router starts with a Service Graph (SG). In this high-level description, the elastic router is a service, routing traffic between four ports (SAP 1-4). In the Service Layer, the SG is translated to an NF-FG, a more practical description of the service, describing the VNFs, SAPs and links to be deployed. Specifically, in the demonstrator, the mapped NF-FGs (visualized in the figure as NF-FG 1 and 2) represent the software router as decomposed into:

- A control plane, implemented by a dedicated Control NF using an OpenFlow controller (Ryu) with an extra layer on top of it, which holds the scaling logic. This logic will generate the updated NF-FG to be sent via the Cf-Or interface to the global orchestrator (realized through ESCAPE).

- A data plane, implemented by one or more Open vSwitch instances (OVS VNF), whose flow entries are set by the Control NF. The SAPs are connected here and all traffic is routed through these OVS VNFs. Every OVS VNF also has a control interface that connects the OVS to the OpenFlow Controller in the Control NF. This control network and the SAP connections are routed via the LSI (logical switch instance) of this NF-FG in the UN.

In addition to the SG, a high-level service description includes a set of requirements in the form of a Service Level Agreement (SLA). In the Service Layer, the SLA is translated into specific observable KPIs, which are expressed as MEASURE annotations to the derived initial NF-FG (e.g. requiring to retrieve metrics like throughput and CPU utilization from specific ports of VNFs). The global orchestrator now contains a NF-FG including definitions of the VNFs, SAPs and links, and additionally a set of monitoring intents in terms of required metrics and aggregation rules (i.e. MEASURE), which can be used to trigger actions like scaling of the elastic router or automated troubleshooting.

In our integrated prototype, both service layer and global orchestrator are implemented in the ESCAPE orchestration framework [17]. A Universal Nodes (UN) serves as the infrastructure domain, as described in [22]. Each UN has a local orchestrator, which deploys the actual network functions and their interconnections as specified in the NF-FG. The global orchestrator (ESCAPE) could even map to different infrastructure nodes, also in other domains than the UN, as described in D3.5 [18], section 4.4.

The deployment, operation and troubleshooting of the elastic router in the UNIFY framework consists out of five distinct stages, roughly illustrated in Figure 65:

1) At initial service deployment, a Service Graph (SG) is sent to the Service Layer. This SG is a high-level description of the elastic router, describing the external interfaces connected to the service, and a description of SLAs.

2) The Service Layer generates the NF-FG from the SG. The SLA is translated into monitoring requirements expressed as MEASURE indents.

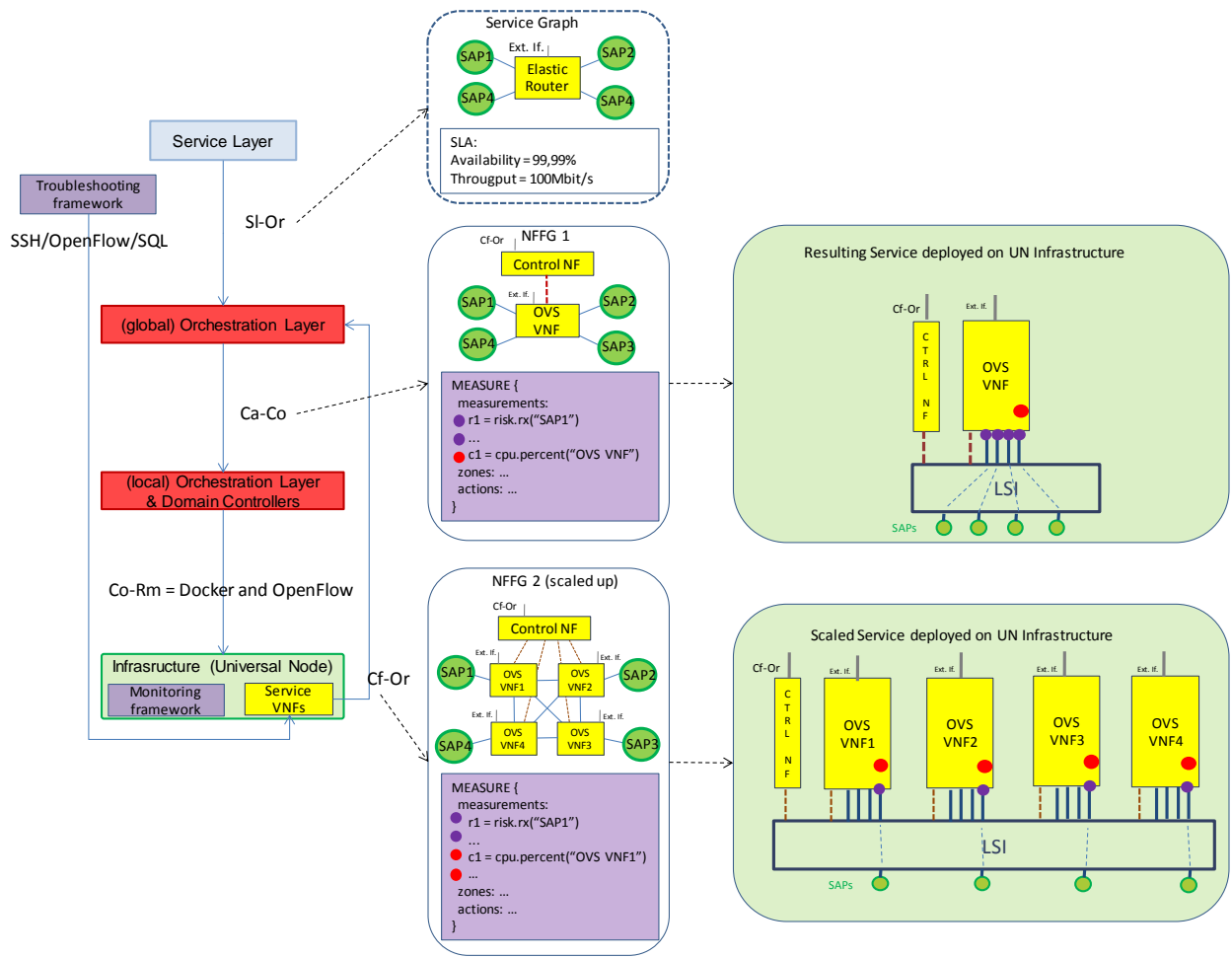

*Figure 65 The elastic router use case. The UNIFY architecture (left), with data model of the UNIFY reference points (middle), and resulting service deployment in the infrastructure (right).*

3) The NF-FG is passed to a (global) orchestrator, which determines the best choice of infrastructure mapping, deciding where the VNFs will be deployed. The local orchestrator on the infrastructure node will deploy the VNFs and trigger the installation of the monitoring functions defined in the MEASURE part of the NF-FG. The monitoring controller (called MMP in the UN), that installs the required monitoring functions, is further detailed later in this section.

4) At the service runtime, elastic scaling is possible. A condition derived from the MEASURE annotations triggers the generation of a new, scaled version of the NF-FG and MEASURE annotations in the elastic router Control NF. Via the Cf-Or interface, the updated NF-FG is then sent to the (global) orchestrator to be deployed.

On a different condition derived from the MEASURE annotations, a troubleshooting request is sent to an external troubleshooting node. This node will facilitate a semi-automatic troubleshooting process involving a range of SP-DevOps and 3[rd] party tools to identify and solve the problem.

## 6.2.2 Integration of monitoring components to an Elastic Router service

The elastic router use-case is used to demonstrate the deployment process of a NF-FG in the UNIFY framework on a Universal Node, together with monitoring and troubleshooting features. During the service lifetime, monitoring tools will trigger the Elastic Router Control App (i.e. the Control NF) to scale-out/in more or less Elastic Router OVS instances (i.e. OVS VNFs). Additionally, troubleshooting tools enable efficient debugging of the elastic-router. This illustrates how the UNIFY framework integrates SP-DevOps processes like automated monitoring and troubleshooting Figure 66 depicts this integrated prototype.

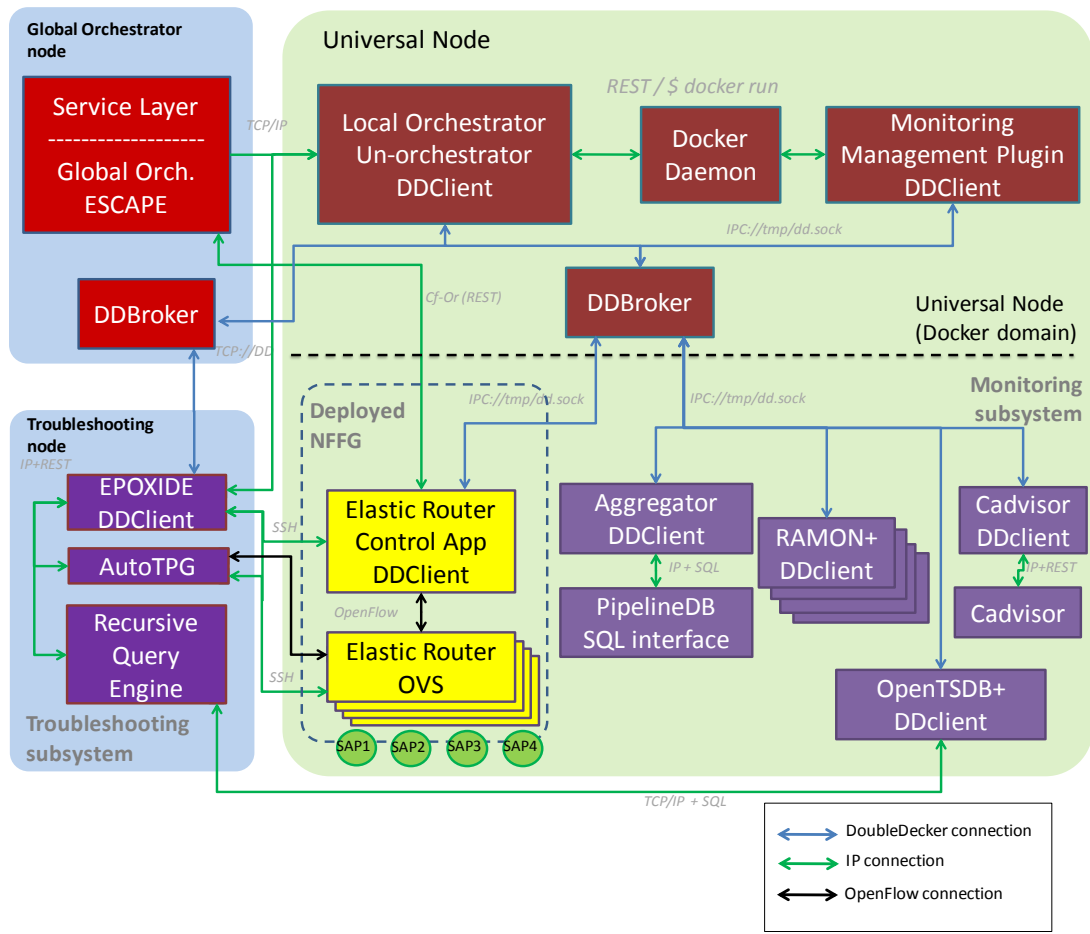

*Figure 66 System view of the prototyped elastic router, depicting integrated functional blocks and their interfaces.*

The system components are deployed over three physical separate nodes:

- The Global Orchestrator Node is primarily running ESCAPE, i.e. realizing the service layer and (global) orchestration layer. This is where the initial NF-FG is generated. Additionally, the global orchestrator is connected to the DoubleDecker messaging system, and is thus hosting a local DoubleDecker Broker connected to the Broker in the UN.

- The Troubleshooting Node executes the semi-automatic debugging and troubleshooting tasks triggered by the monitoring subsystem. In this use-case, troubleshooting is performed by EPOXIDE, instrumenting RQL (Recursive Query Engine), AutoTPG and 3[rd] party OpenFlow swtich debugging tools. More details on the integration and usage of troubleshooting tools will follow in Section 6.2.3.

- The Universal Node deploys the NF-FG, i.e. realizing local orchestration and infrastructure resources. The local orchestration and management functions are depicted in dark red in Figure 66. They include the UN local orchestrator, a node-local DoubleDecker Broker, a Docker Daemon, and an infrastructure monitoring controller (in the case of an UN as infrastructure domain, we call it MMP) that instantiates and configures monitoring related functions according to the definitions in the MEASURE annotations to the NF-FG. In terms of infrastructure resources, the Universal node is configured to use OVS for the virtual switching environment (VSE), realizing the LSI. Furthermore, the VNF execution environment (VEE) is realized through Docker (thus the existence of the Docker Daemon), which means that the elastic router VNFs are actually Docker containers (depicted in yellow in Figure 66). Also, the monitoring subsystem on an UN is realized as Docker containers (depicted in purple).

In the following, we will focus on detailing the infrastructure components realizing the observability process supporting the elastic router use-case:

- **DoubleDecker** (DD) Brokers and clients: As documented in Section 4.10, DD is a scalable, carrier-grade messaging system based on ZeroMQ, sending opaque data between connected clients. For this use case, DD is realized by a two-layer broker hierarchy, one in the infrastructure (i.e. on the UN), and another one serving higher layers including the service and global orchestration layers as well as the troubleshooting node. Locally on the UN, DDclients use IPC transport methods, whereas the DDbrokers and EPOXIDE use TCP. DD is used by the local orchestrator to trigger the MMP on reception of a new NF-FG including MEASURE annotations. The MMP (Monitoring Management Plugin) uses DD messaging for starting, stopping and re-configuration of monitoring tools. The monitoring functions (RAMON and cAdvisor) are publishing their results on DD, with the Aggregation Point and the OpenTSDB database as their subscribers. Finally, also the Aggregator is publishing filtered and aggregated results on DD, where scaling requests are subscribed to by the Elastic Router Control App, whereas troubleshooting requests are received by EPOXIDE. The preferred message format used in the demonstrator is JSON, but the details on interfaces and messages are detailed in [89].

- **Monitoring Management Plugin** (MMP): This entity represents a monitoring controller inside a UN, and is one of the two main components realizing MEASURE (as described in Section 4.3). It acts as an MF control app for the observability points (OPs) in this demo (i.e. RAMON instances and cAdvisor). Via DD, the MMP exposes one JSON-RPC command for receiving all the information required for monitoring from the Local Orchestrator. Specifically, the required information includes the MEASURE descriptions from the update NF-FG, as well as the infrastructure mapping of the abstract VNFs and interfaces specified in the NF-FG (e.g. VNF ids to Docker names; or port ids to interfaces on the LSI). This is called whenever an NF-FG is deployed, modified, or removed. This information allows the MMP to either instantiate, update, configure, or remove the associated monitoring tools (i.e. OPs). Details on the implementation of the MMP and its helper functions like a MEASUREparser and an mPlane [95] inspired MF-IB can be found in D4.4.

- **Aggregator**: This component is the second main component of MEASURE, realizing a generic local aggregation function for monitoring results. The Aggregator aggregates raw metrics streamed from monitoring functions like RAMON and cAdvisor via DD. To do so, it uses PipelineDB [96] as a backend container. PipelineDB is a Streaming SQL database, able to process streams of incoming data and discard the data once it has been used for a calculation. It is used to process incoming measurement results and to calculate which zones are currently active. The Aggregator receives configuration from the MMP, derived from the "Zone" and "Actions" definitions in the MEASURE annotations to the NF-FG (see Section 4.2.3.1). More implementation details on the Aggregator can be found in D4.4 [89]. In this use-case, the aggregator collects monitoring metrics provided by RAMON for the purpose of processing the observed link utilization and congestion risk over multiple ports. Note that the aggregator may also gather other metrics such as CPU load provided by e.g. cAdvisor. Specifically, the monitoring information provided per individual SAP facing port is inspected for the purpose of classifying three cases: whether (i) the total (combined) load of the Elastic Router is below a threshold, (ii) the total load is above a threshold, and the throughput per port is roughly equal, which triggers a scaling request, or (iii) the total load is above a threshold, but the throughput per port is very imbalanced, indicating that scale-out cannot solve this overload issue and a troubleshooting process should be triggered. Depending on the outcome of the inspection, a trigger for either scaling or troubleshooting is published - the former is subscribed to by the Elastic router control app, whereas the latter is subscribed to by the EPOXIDE troubleshooting framework.

- **RAMON**: RAMON is a rate monitoring tool acting as OP for determining a precise statistical estimate of link utilization and other congestion indications in a distributed manner (see Section 4.5). In the current RAMON version, one instance is required per monitored interface. In this demo, four RAMON instances are configured to monitor the elastic router ports representing the external visible SAPs (cf Figure 66). The RAMON instances are instantiated and configured by the MMP according to the MEASURE annotations requesting transmission rate and overload risk metrics for the SAPs. The results of RAMON are via DD streamed to the Aggregator, resulting in triggers for either scaling or troubleshooting operations.

- **cAdvisor**: This is a resource monitor for Docker containers [51]. cAdvisor realizes an OP that provides various resource usage and performance characteristics of the running VNF Docker containers. In this demo, cAdvisor results are not used by the Aggregator, but could easily be considered to be combined with RAMON results for more advanced, compound triggers. Instead, we use cAdvisor to stream CPU and memory usage into a time series database, which is used by RQL as part of the troubleshooting process.

- **OpenTSDB**: This is a scalable, distributed time series database [97]. In this demo, it stores the raw-metrics generate by monitoring functions (i.e., RAMON and cAdvisor). It is used as a local logging facility allowing to perform off-line Root Cause Analysis (RCA) based on the maximum available data granularity. While monitoring data is streamed in the DB via DD, it is also accessible via TCP from external nodes through the port-mapping feature of the Universal node. The external connection is used by RQL to perform SQL queries during the troubleshooting process.

### 6.2.3 Integration of troubleshooting components to an Elastic Router service

The integrated prototype includes an example of a troubleshooting scenario of the ER use case as triggered by the observability process. The troubleshooting process aims at demonstrating the interaction between SP-DevOps tools operating across the architectural layers and the UN. We assume a troubleshooting process to isolate the cause of observed load imbalance, detected by the Aggregator based on a MEASURE definition using the measurements of RAMON. The aim is to detect increased overload, trigger the troubleshooting process, and perform a set of troubleshooting actions via EPOXIDE using RQL and AutoTPG for the purpose of isolating buggy or faulty flow rules as the root cause for the observed behaviour.

The components realizing the troubleshooting process supporting the elastic router use-case are listed below:

- **RAMON, Aggregator/MEASURE, and DoubleDecker:** These components jointly generate and publish the troubleshooting trigger according to the observability process described above. Specifically, the Aggregator processes monitoring information for the purpose of identifying an unexpected load imbalance after scale out operations, mainly based on link utilization and congestion risk metrics provided by RAMON. Detected imbalance triggers publication of a troubleshooting request.

- **EPOXIDE**: EPOXIDE, a multicomponent troubleshooting tool (Section 4.12 has subscribed on troubleshooting related topics via the DDBroker in the orchestration layer. On reception of a troubleshooting request, EPOXIDE performs automatic or manual troubleshooting steps for RCA instrumenting both RQL and AutoTPG, in our prototype running on the same troubleshooting machine.

- **RQL:** The RQL (Recursive Query Engine) provides a prototype for the recursive query language (RQL, Section 4.11). In this demo, it is used for efficient retrieval of all available measurement results stored in the distributed time-series database in the infrastructure (OpenTSDB). The results are used to verify the validity of the troubleshooting request, and to localize and identify further troubleshooting steps.

- **AutoTPG**: As described in Section 4.9, AutoTPG is able to verify data-plane functionality of OpenFlow-based forwarding elements.It acts as a secondary OpenFlow controller that temporarily connects to the OVS VNFs. In this use-case, we use it to verify the possibility that the imbalance of the elastic router is caused by faulty configuration of one or more flow-tables in the OVS VNFs. If such misconfigured flow-tables are found, the problem could be solved by manual manipulation of the OVS instance (adding or removing flow entries). Furthermore, since the Elastic Router control app is responsible for setting up the OpenFlow tables in the OVSs, a debugged connection (e.g. via SSH) to the control app could be issued to pinpoint the code resulting in the faulty *flowmod* instructions.

We assume that SAP links of the ER are monitored for the purpose of load balancing. In this case, the estimated risks output from each instance of RAMON are inspected for the purpose of classifying three cases:

Whether (i) the total (aggregated) load of the Elastic Router is below a threshold, (ii) the total load is high on all SAP ports, which triggers a scaling request, or (iii) the total load is high but very imbalanced between SAP ports, indicating that scale-out cannot solve this overload issues and a troubleshooting process should be triggered. In other words, we assume that scaling should be triggered when the aggregated risk estimate is above a certain probability. Troubleshooting is triggered when one rate monitoring instance signals that the probability of overload on the monitored SAP link is above a certain threshold. Once a troubleshooting request has been published by the Aggregator, EPOXIDE (which is subscribed to troubleshooting related messages) receives the request via the DoubleDecker message bus, and takes over to perform the remaining steps towards root-cause analysis.

The first step involves the RQL, which here is used to gather information about the estimated overload risk in other instances of the ER deployed on this and other UNs. The information gathered is used for deciding over which set of nodes the next troubleshooting actions should be executed and is part of fault localization and root cause analysis. As a configuration step of RQL, EPOXIDE queries MMP for the port mappings of the deployed NF-FG and relays this information to the Recursive Query Engine.

In this example, the output from RQL indicates that there is only one SAP link (the triggering one) at one UN that needs to be further investigated by the use of AutoTPG. One instance of the AutoTPG controller will be used to send test packets over the selected UN in order to identify any faulty policies or flow rules deployed in the switches under testing. To configure AutoTPG, EPOXIDE once again queries the MMP via DoubleDecker for the management addresses of the OVS containers (i.e., IP address and ssh port pairs) and logs in to set the secondary controllers of the OVS switches to AutoTPG. Then it waits for the switches to automatically connect to AutoTPG before it instruments AutoTPG to perform its tests.

In the demo, one of the four OVS OpenFlow switch instances of the Elastic Router contains one erroneous high priority flow-entry. The AutoTPG tool identifies this faulty flow entry that caused the increased load on the SAP link. EPOXIDE reports the outcome of the testing process, which shows the destination IP addresses of those packets that are lost or misrouted due the erroneous flow-entry. At this point, the trouble-shooter has to switch to manual troubleshooting, but EPOXIDE queries and shows the flow tables of the OVS switch in question to help starting this manual detective work. In this use-case, the erroneous flow-entry is a result of the misconfiguration of the control application of the elastic router, but the further process of finding this misconfiguration is not part of the demo.

The steps are summarized below:

1) RateMon monitors each switch (RX on SAP links)

2) The Aggregator (through MEASURE) detects increased overload risk, and notifies EPOXIDE via DoubleDecker

3) EPOXIDE triggers RQL to inspect overload risks in other parts of the service to identify which parts should be part of the remaining troubleshooting process:

   a) Assuming that there is not only one UN but several infrastructure domains under control of the global orchestrator.

   b) Assuming that a symptom of a fault can have a root cause somewhere else.

4) RQL analysis indicates that the switch monitored by the triggering instance of the rate monitoring is the switch to be further inspected.

5) EPOXIDE starts and configures AutoTPG with the necessary addresses to test the flow rules (switch by switch). It also configures the switches to connect to AutoTPG as well.

6) Based on the AutoTPG tests, one rule is identified as faulty.

Further analysis regarding when/who/how the rule was installed may be performed. Examples of such a process can be found here [98].

# 7 Conclusions

This deliverable reported on the final results of the UNIFY SP-DevOps work. After a short summary of the UNIFY project as a whole and activities in other WPs, we described and assessed the UNIFY SP-DevOps concept. We continued in Section 4 with details of the SP-DevOps tools, also including experimental evaluations of each tool in isolation. In Section 5, we showed how the SP-DevOps tools can be combined into powerful workflows supporting verification, observability and troubleshooting processes accessible for both operator and developer personnel and thus integrating them into one SP-DevOps framework. Finally, evaluation of the concept as a whole was addressed in Section 6. We did this by designing particular instances of the processes as examples associated to specific use case scenarios shared with the programmability and orchestration results as well as the Universal Node prototype.

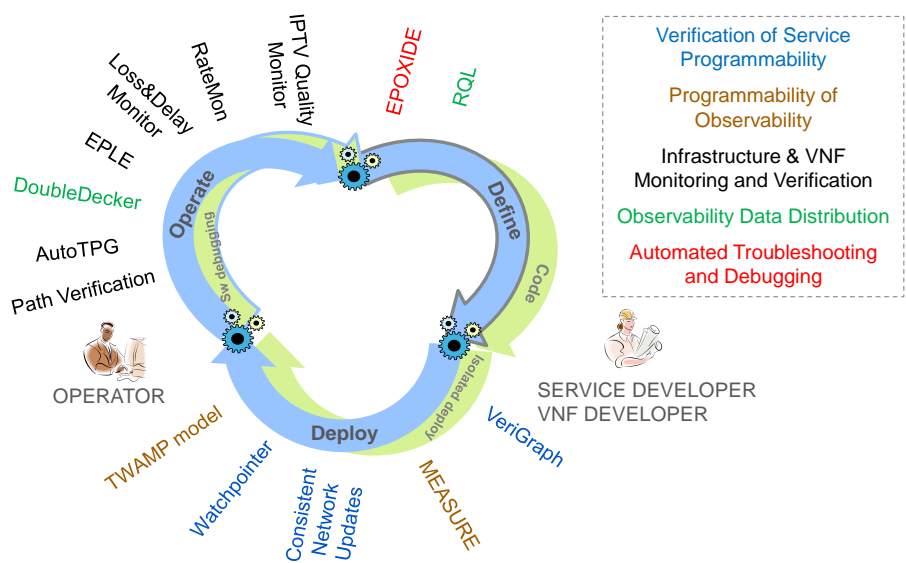

*Figure 67 SP-DevOps tools mapped on the UNIFY Service Lifecycle.*

In Figure 67, we depict how the SP-DevOps tools developed in UNIFY map onto a service lifecycle as introduced in Section 3. Our tools support operators and developers across all stages, with a focus on the deployment, operations and debugging/troubleshooting phases. A summary of all the tools and their main value compared to the respective state of the art as studied in D4.1 [25] was provided in Section 4 – Table 1. Note that several of these tools reached a maturity level that allowed them to be published as part of the open-source SP-DevOps toolkit (see Section 4.13).

Constant verification of service graph functionality, both at design time and runtime, is a key requirement in software-defined infrastructure that was addressed through the VeriGraph tool as well as infrastructure and VNF verification tools such as AutoTPG and Path Verification. The integration with the orchestration ensures that policies can be consistently deployed to the infrastructure and potential deviations are identified at early stages. Furthermore, software-defined monitoring approaches were developed to take advantage of programmable infrastructure platforms (e.g., RateMon, EPLE, Delay&Loss Monitoring) to provide key quality indicators such as congestion notifications and packet loss estimates with a fraction of the overhead of typical management tools.

Furthermore, these tools may be deployed programmatically using MEASURE and provide data through the scalable DoubleDecker messaging system that eases integration with third-party software addressing both Developer and Operator roles. Finally, a way to automate debugging processes was realized through a developer-friendly troubleshooting tool (EPOXIDE).

Each SP-DevOps tool itself advances the state-of-the-art and solves a specific research challenge (see Table 5 in Annex 1). However, in order to meet the ambitious objectives of UNIFY SP-DevOps, we have combined them together into powerful, automated processes supporting verification, observability and troubleshooting. We demonstrated selected SP-DevOps processes integrated with the UNIFY global orchestration platform and the Universal Node. The technical integration of SP-DevOps tools with the orchestration and infrastructure layer realization of UNIFY was enabled by a few key tools and functions: DoubleDecker and MEASURE facilitated efficient integration of observability features with entities in all UNIFY layers; EPOXIDE proved to be a powerful framework to instrument many types of SP-DevOps monitoring and verification tools as well as third-party debugging tools; and the modular design of VeriGraph allowed straight-forward integration of verification features into the global orchestrator (i.e. ESCAPE).

The process definition and integration work proves the feasibility of combining individual SP-DevOps components to bring extra value to automated processes. Applying processes for verification, observability and troubleshooting as one SP-DevOps solution to software defined infrastructure, we calculated a possible OPEX reduction in terms of time of investment by 32%-77% through purely technical means.

With this deliverable, we could successfully report results with respect to all related initial project objectives. Being a research project, UNIFY SP-DevOps results have also been extensively shared with the research community through numerous journal, conference and demo publications [6]. Additionally, we have continuously influenced the standardization community (IETF, IRTF, ETSI) with drafts [86] [52] [99], presentations and demos. Finally, the SP-DevOps toolkit allows us to directly address the open-source community with SP-DevOps results. In this context, our focus is on the industrial open source NFV platform OPNFV. As a first step, we have given a talk on an OPNFV summit advocating Software-Defined Operations with help of our SP-DevOps Toolkit [91]. By mapping SP-DevOps contributions to relevant OPNFV projects, we created significant interest in the industry and started discussions in how to engage in several OPNFV activities (most notable with the SP-DevOps tools DoubleDecker, MEASURE, and Verigraph). As next steps beyond the duration of UNIFY, we will continue to support the ongoing IETF drafts and concretize and realize the adaptation of open-source SP-DevOps toolkit components within the scope of the quasi-industry-standardization forum OPNFV.

To sum up, we could show how DevOps principles can be applied in the control and management plane of a future "softwarized" network infrastructure, also highlighting the benefits of such approach. In fact, current legacy OSS/BSS systems (mostly still based on human made manual configurations) will not be able cope with the dramatic growth in complexity brought by network Softwarization, which makes it necessary to manage and control, instead of thousands of physical nodes, millions of software processes, up to the user equipment.

---

[6] https://www.fp7-unify.eu/index.php/results.html#publications

Starting from the project results, future work in this area should include the definition of:

- heuristics, autonomic/cognitive algorithms, data analytics methods for automating and optimising the processes of Operations;
- Control Orchestration Management Policy (COMP) solutions for infrastructure slicing;
- Security-by-design techniques;

This technical transformation will have education and cultural implications, with particular reference to the new skills, which will be eventually required by the Softwarization. The future of creativity may depend on younger generations being taught mastering the software skills just as they are taught foreign languages, mathematics, and science.

# List of abbreviations and acronyms

| Abbreviation | Meaning |
|---|---|
| NF | Network Function |
| DP | Data Plane |
| VNF | Virtual Network Function |
| SDN | Software Defined Network |
| UL | User Layer |
| OL | Orchestration Layer |
| SL | Service Layer |
| RO | Resource Orchestrator |
| CA | Controller Adaptation Layer |
| LO | Local Orchestrator |
| PNF | Physical Infrastructure Resource |
| SLA | Service Level Agreement |
| KPI | Key Performance Indicator |
| SG | Service Graph |
| NF-FG | Network Function -Forwarding Graph |
| TG | Troubleshooting Graph |
| SAP | Service Access Point |
| WP | Work Package |
| OVS | OpenV Switch |
| ODL | OpenDayLight |
| VM | Virtual Machine |
| MF | Monitoring Function |
| MMP | Monitoring Management Plugin |
| UN | Universal Node |
| TCP | Troubleshooting Controller App |
| OP | Observability Point |
| SNMP | Simple Network Management Protocol |
| DB | Data Base |
| ACL | Access Control List |
| DNS | Domain Name System |
| REST | Representational State Transfer |
| API | Application Programming Interface |
| BPMN | Business Process Model and Notation |
| ISP | Internet Service Provider |
| SP-DevOps | Service Provider – DevOps |

| ESCAPE | Extensible Service CHAIn Prototyping Enviroment |
|--------|------------------------------------------------|
| MEASURE | Measurements, States, and Reactions |
| RateMon | Rate Monitoring |
| EPLE | Efficient SDN Loss Monitoring |
| IntPro | Integrate Prototype |
| AutoTPG | Automatic Test Packet Generation |
| Cf-Or | Control Function – Resource Orchestration |
| TWAMP | Two-Way Active Measurement Protocol |
| MoS | Mean Opinion Score |
| GoP | Group of Pictures |

# References

[1]    A. Manzalini, R. Minerva, F. Callegati, W. Cerroni and A. Campi, "Clouds of Virtual Machines at the Edge," *IEEE Com Mag Future Carriers Networks*.

[2]    A. Manzalini, "Future Edge ICT Networks," *IEEE COMSOC MMTC E-Letter,* vol. 8, no. 3, pp. 32 – 35, May 2013.

[3]    A. Manzalini, N. Crespi and e. al, "Manifesto of Edge ICT Fabric," in *ICIN2013,* Venice, October 2013.

[4]    R. Saracco and A. Manzalini, "Software Networks at the Edge: a shift of paradigm," in *IEEE Workshop SDN4FNS,* Trento, Nov.2013.

[5]    [Online]. Available: http://sdn.ieee.org/.

[6]    R. Steinert, W. John, P. Sköldström, B. Pechenot, A. Gulyas, I. Pelle, T. Lévai, F. Nemeth, J. Kim, C. Meirosu, X. Cai, C. Fu, K. Pentikousis, S. Sharma, I. Papafili, G. Marchetto, R. Sisto, F. Risso, P. Kreuger, J. Ekman, S. Liu, A. Manzalini, A. Shukla and S. Schmid, "Service Provider DevOps network capabilities and tools," *CoRR,* vol. abs/1510.02818, 2015, [Online]. Available: http://arxiv.org/abs/1510.02818.

[7]    FP7 UNIFY project, "D3.3: Revised framework with functions and semantics," 2015.

[8]    W. John, K. Pentikousis, G. Agapiou, E. Jacob, M. Kind, A. Manzalini, F. Risso, D. Staessens, R. Steinert and C. Meirosu, "Research Directions in Network Service Chaining," in *Future Networks and Services (SDN4FNS), 2013 IEEE SDN for,* 2013.

[9]    5.-P. Association, December 2013. [Online]. Available: http://www.5g-ppp.eu/contract.

[10]    FP7 UNIFY project, "D2.1: Use Cases and Initial Architecture," 2014.

[11]    FP7 UNIFY project, "D2.2: Final Architecture," 2014.

[12]    FP7 UNIFY project, "D2.3: Economic viability of technical solutions," 2016.

[13]    A. Csoma, B. Sonkoly, L. Csikor, F. Nemeth, A. Gulyas, W. Tavernier and S. Sahhaf, "ESCAPE: Extensible Service ChAin Prototyping Environment using Mininet, Click, NETCONF and POX," in *SIGCOMM'14,* Chicago, 2014.

[14]    FP7 UNIFY project, "D2.4: Integration of prototypes," 2016.

[15]    FP7 UNIFY project, "D3.1: Programmability framework," 2014.

[16]    FP7 UNIFY project, "D3.2: Detailed functional specification and algorithm description," 2015.

[17]    FP7 UNIFY project, "D3.4: Prototype deliverable," 2015.

[18]   FP7 UNIFY project, "D3.5: Programmability framework prototype report," 2016.

[19]   FP7 UNIFY project, "D5.1 Universal Node functional specification and use case requirements on data plane," 2014.

[20]   FP7 UNIFY project, "D5.2: Universal Node Interfaces and Software Architecture," 2014.

[21]   FP7 UNIFY project, "D5.3: Prototype description and implementation plan," 2014.

[22]   FP7 UNIFY project, "D5.5: Universal Node Prototype," 2015.

[23]   FP7 UNIFY project, "D5.4: Initial Benchmarking Documentation," 2015.

[24]   FP7 UNIFY project, "D5.6: Final benchmarking documentation," 2016.

[25]   W. John, C. Meirosu, P. Sköldström, F. Nemeth, A. Gulyas, M. Kind, S. Sharma, I. Papafili, G. Agapiou, G. Marchetto, R. Sisto, R. Steinert, P. Kreuger, H. Abrahamsson, A. Manzalini and N. Sarrar, "Initial Service Provider DevOps concept, capabilities and proposed," *CoRR*, vol. abs/1510.02220, 2015, Available: http://arxiv.org/abs/1510.02220.

[26]   FP7 UNIFY project, "M4.1: SP-DevOps concept evolution and initial plans for prototyping," 2014.

[27]   F. Nèmeth, R. Steinert, P. Kreuger and P. Sköldström, "Roles of DevOps tools in an automated, dynamic service creation architecture,," in *Integrated Network Management (IM), 2015 IFIP/IEEE International Symposium on*, Ottawa, 2015.

[28]   S. Inbar, S. Yaniv, P. Gil, S. Eran, S. Ilan, K. Olga and S. Ravi, "DevOps and OpsDev: How Maturity Model Works," April 2013. [Online]. Available: http://h30499.www3.hp.com/t5/Business-Service-Management-BAC/DevOps-and-OpsDev-How-Maturity-Model-Works/ba-p/6042901#.UqXiVMu9KSM.

[29]   PuppetLabs, "2015 State of DevOps Report," 2015.

[30]   DZone, "The guide to continuous delivery," 2015.

[31]   Spirent, "Heavy Reading Report Finds Congestion and Network Failures are Growing Causes of Mobile Network Outages and Service Degradations," Barcelona, 2016.

[32]   W. John, C. Meirosu, B. Pechenot, P. Skoldstrom, P. Kreuger and R. Steinert, "Scalable Software Defined Monitoring for Service Provider DevOps," in *Fourth European Workshop on Software Defined Networks*, Bilbao, 2015.

[33]   Ericsson, "Handling of signaling storms in mobile networks," February 2016. [Online]. Available: http://www.ericsson.com/res/docs/2015/handling-of-signaling-storms-in-mobile-networks-august.pdf.

[34]   Citrix, February 2016. [Online]. Available:

http://74.205.123.199/ctx_landing/xs_whitepapers/XS_LowVirt_TCO_WP_US_attachment.pdf.

[35]  Amazon, 2014. [Online]. Available: https://s3-eu-west-1.amazonaws.com/donovapublic/TCOOutput.pdf.

[36]  Alcatel-Lucent, February 2016. [Online]. Available: http://www.tmcnet.com/tmc/whitepapers/documents/whitepapers/2014/10349-business-case-moving-dns-the-cloud.pdf.

[37]  Enisa, August 2015. [Online]. Available: https://www.enisa.europa.eu/publications/annual-incident-reports-2014.

[38]  C. Fernandez and a. et, "A recursive orchestration and control framework for large-scale, federated SDN experiments: the FELIX architecture and use cases," *International Journal of Parallel, Emergent and Distributed Systems*, vol. 30, no. 6, pp. 428-446.

[39]  E. Al-Shaer and S. Al-Haj, "FlowChecker: configuration analysis and verification of federated openflow infrastructures," in *Proceedings of the 3rd ACM workshop on Assurable and usable security configuration*, New York, 2010.

[40]  M. Canini, D. Venzano, P. Peresini, D. Kostic and J. Rexford, "A NICE Way to Test Openflow Applications," in *9th USENIX Conference on Networked Systems Design and Implementation (NSDI'12)*, San Jose, 2012.

[41]  A. Panda, O. Lahav, K. Argyraki, M. Sagiv and S. Shenker, "Verifying Isolation Properties in the Presence of Middleboxes," http://arxiv.org/abs/1409.7687, 2014.

[42]  L. De Moura and N. Bjørner, "Z3: an efficient SMT solver," in *14th international conference on Tools and algorithms for the construction and analysis of systems*, Berlin, 2008.

[43]  F. Valenza, S. Spinoso, C. Basile, R. Sisto and A. Lioy, "A formal model of network policy analysis," in *Research and Technologies for Society and Industry Leveraging a better tomorrow (RTSI), 2015 IEEE 1st International Forum on*, Turin, 2015, September.

[44]  N. Foster, R. Harrison, M. Freedman, C. Monsanto, J. Rexford, A. Story and D. Walker, "Frenetic: a network programming language," in *Proceedings of the 16th ACM SIGPLAN international conference on Functional programming (ICFP '11)*, New York, 2011.

[45]  A. Ludwig, J. Marcinkowski and S. Schmid , "Scheduling Loop-free Network Updates: It's Good to Relax!," in *Proceedings of the 2015 ACM Symposium on Principles of Distributed Computing*, 2015, July.

[46]  A. Ludwig, M. Rost, D. Foucard and S. Schmid, "Good network updates for bad packets: Waypoint enforcement beyond destination-based routing policies," in *Proceedings of the 13th ACM Workshop on Hot Topics in Networks (p. 15). ACM*, 2014, October.

[47]  S. Dudyez, A. Ludwig and S. Schmid, "Can't Touch This: Consistent Network Updates for Multiple Policies," in

*46th IEEE/IFIP International Conference on Dependable Systems and Networks*, Toulouse, France, 2016.

[48]  A. Ludwig, S. Dudyez, M. Rost and S. Schmid, "Transiently Secure Network Updates," in *ACM SIGMETRICS 2016*, Antibes Juan-les-Pins, France, 2016.

[49]  A. Shukla, L. A. Schuetze, S. Schmid and A. Feldmann, *Consistent Network Updates in Software Defined Networking*, demonstrated in ETSI From Research To Standardization, Sophia Antipolis, France.

[50]  A. Shukla, A. Schuetze, A. Ludwig, S. Dudycz, S. Schmid and A. Feldmann, "Towards Transiently Secure Updates in Asynchronous SDNs," in *ACM SIGCOMM*, Florianopolis, Brazil, 2016.

[51]  [Online]. Available: https://github.com/google/cadvisor.

[52]  R. Civil, A. Morton, L. Zheng, R. Rahman, J. Jethanandani and K. Pentikousis, "Two-Way Active Measurement Protocol (TWAMP) Data Model," in *Internet-Draft draft-ietf-ippm-twamp-yang-00*, IETF IPPM, March 2016..

[53]  2016. [Online]. Available: https://datatracker.ietf.org/doc/draft-ietf-ippm-twamp-yang/.

[54]  P. Kreuger and R. Steinert, "Scalable in-network rate monitoring," in *Proc. of IM 2015*, Ottawa, Canada, May 2015.

[55]  R. Presuhn, "Management information base (MIB) for the simple network management protocol (SNMP)," in *RFC 3418, Internet Engineering Task Force*, 2002.

[56]  P. Phaal, S. Panchen and M. N., "InMon corporation's sFlow: A method for monitoring traffic in switched and routed networks," 2001.

[57]  Cisco IOS, 2008. [Online].

[58]  FP7 UNIFY project, 2016. [Online].

[59]  P. Kreuger, R. Steinert, S. Liu and J. Ekman, "Scalable high precision rate monitoring and congestion detection," in *Submitted to CNSM 2016*.

[60]  FP7 UNIFY project, "M4.2: DevOpsPro status report towards the Integration Plan," 2015.

[61]  A. Hernandez and E. Magana, "One-way Delay Measurement and Characterization," Athens, 2007.

[62]  M. Kalman and B. Girod, "Modeling the delays of successively-transmitted Internet packets," in *Multimedia and Expo, 2004. ICME '04. 2004 IEEE International Conference on*, 2004.

[63]  C. J. Bovj, H. T. Mertodimedjo, G. Hooghiemstra, H. Uijterwaal and P. Van Mieghem, "Analysis of end-to-end delay measurements in internet," *PAM 2002*, 2002.

[64]  A. Hess and R. Steinert, "Observing Software-defined Networks Using a Decentralized Link Monitoring

Approach," in *Proc. IEEE NetSoft*, London, UK, 2015.

[65]  J. Ekman, R. Steinert, S. Liu and P. Kreuger, "Link delay monitoring with probabilistic guarantees on precision," in *Submitted to CONEXT 2016*..

[66]  J. Ekman, R. Steinert, S. Liu and P. Kreuger, "Probabilistic Delay Change Detection for Network Management," *Submitted to Int. J. Network Mgmt*, 2016.

[67]  C. Yu, C. Lumezanu, Y. Zhang, V. Singh, G. Jiang and H. V. Madhyastha, "Flowsense: monitoring network utilisation with zero measurement cost," in *Proceedings of the 14th International Conference on Passive and Active Measurement, PAM'13*, 2013.

[68]  J. C. Mogul, J. Tourrilhes, P. Yalagandula, P. Sharma, A. R. Curtis and S. Banerjee, "DevoFlow: Cost-Effective Flow Management for High Performance Enterprise Networks," in *Ninth ACM Workshop on Hot Topics in Networks (HotNets-IX)*, Monterey, CA, 2010.

[69]  W. John and C. Meirosu, "Low-overhead packet loss and one-way delay measurements in Service Provider SDN," in *ONS*, Santa Clara, 2014.

[70]  C. Fu, W. John and C. Meirosu, "EPLE: an Efficient Passive Lightweight Estimator for SDN Packet Loss Measurement," in *IEEE NFV-SDN*, Palo Alto, CA, USA, 2016.

[71]  WAND Network Research Group, 2016. [Online]. Available: http://wand.net.nz/wits/waikato/8/.

[72]  The Linux Foundation, November 2009. [Online]. Available: http://www.linuxfoundation.org/collaborate/workgroups/networking/netem.

[73]  RFC4656, September 2006. [Online]. Available: https://tools.ietf.org/html/rfc4656.

[74]  Internet2, 2016. [Online]. Available: http://software.internet2.edu/owamp/.

[75]  ietf.org, October 2008. [Online]. Available: https://tools.ietf.org/html/rfc5357.

[76]  B. Hedstrom, A. Watwe and S. Sakthidharan, "Protocol Efficiencies of NETCONF versus SNMP for Configuration Management Functions," 2011.

[77]  Sandvine, September 2015. [Online]. Available: https://www.sandvine.com/downloads/general/global-internet-phenomena/2015/global-internet-phenomena-report-apac-and-europe.pdf.

[78]  Sandvine, May 2015. [Online]. Available: https://www.sandvine.com/downloads/general/global-internet-phenomena/2015/global-internet-phenomena-report-latin-america-and-north-america.pdf.

[79]  M.-N. Garcia, P. List, S. Argyropoulos, D. Lindegren, M. Pettersson, B. Feiten, J. Gustafsson and A. Raake, "Parametric model for audiovisual quality assessment in IPTV: ITU-T Rec. P. 1201.2," in *Multimedia Signal*

*Processing (MMSP), 2013 IEEE 15th International Workshop on*, 2013.

[80]  P. Kazemian, G. Varghese and N. McKeown, " Header Space Analysis: Static Checking For Networks," in *9th USENIX conference on Networked Systems Design and Implementation (NSDI'12)*, 2012.

[81]  H. Mai, A. Khurshid, R. Agarwal, M. Caesar, P. Godfrey and S. T. King, "Debugging the data plane with anteater," in *ACM SIGCOMM Computer Communication Review*, 2011.

[82]  A. Khurshid, W. Zhou, M. Caesar and P. B. Godfrey, "VeriFlow: Verifying Network-Wide Invariants in Real Time," in *First ACM SIGCOMM Workshop on Hot Topics in Software Defined Networking (HotSDN '12)*, 2012.

[83]  H. Zeng, P. Kazemian and G. M. N. Varghese, "Automatic Test Packet Generation," in *8th International Conference on emerging Networking EXperiments and Technologies (CoNEXT)*, Nice, 2012.

[84]  S. Sharma, W. Tavernier, S. Sahhaf, D. Colle, M. Pickavet and P. Demeester, "Verification of Flow Matching Functionality in the Forwarding Plane of OpenFlow Networks," *IEICE Transactions on Communications,* vol. E98.B, pp. 2190-2201, November 2015.

[85]  S. Sharma, D. Staessens, D. Colle, M. Pickavet and P. Demeester, "In-band control, queuing, and failure recovery functionalities for OpenFlow," *IEEE Network*, vol. 30, no. 1, pp. 106-112, January 2016.

[86]  X. Cai, C. Meirosu and G. Mirsky, "Recursive Monitoring Language in Network Function Virtualization (NFV) Infrastructures," in *Internet-Draft draft-cai-nfvrg-recursive-monitor-01*, IRTF NFVRG, March 2016.

[87]  X. Cai, W. John and C. Meirosu, "Recursively Querying Monitoring Data in NFV Environments," in *IEEE Conference on Network Softwarization (NetSoft'16)*, 2016.

[88]  T. J. Green, S. S. Huang, B. T. Loo and W. Zhou, "Datalog and recursive query processing," *Found. Trends databases,*, vol. 5, no. 2, 2013.

[89]  FP7 UNIFY project, "D4.4: DevOpsPro code base," 2015.

[90]  C. Meirosu, F. Nemeth, R. Steinert, S. Sharma, P. Sköldström and W. John, "A DevOps Toolkit for Networks," in *IETF 94, NFVRG*, November 2015.

[91]  C. Meirosu, June 2016. [Online]. Available: http://www.slideshare.net/OPNFV/summit-16-software-defined-operations-the-unify-spdevops-toolkit.

[92]  OMG, 2011. [Online]. Available: http://www.omg.org/spec/BPMN/2.0/.

[93]  A. Khurshid, W. Zhou, M. Caesar and P. B. Godfrey, "VeriFlow Source Code Release Form," [Online]. Available: http://web.engr.illinois.edu/~khurshi1/projects/veriflow/veriflow-code-release.php. [Accessed September 2014].

[94] P. Kazemian, M. Chang, H. Zeng, G. Varghese, N. McKeown and S. Whyte, "Real Time Network Policy Checking using Header Space Analysis," in *10th USENIX Symposium on Networked Systems Design and Implementation (NSDI '13)*, 2013.

[95] M. Kühlewind and B. Trammell, "mPlane Protocol Specification," in *ETF Network working group, Information draft*, 2016 March.

[96] D. Nelson, J. McSweeney, J. Ferguson, U. Masood and J. Berkus, 2016. [Online]. Available: https://www.pipelinedb.com/.

[97] [Online]. Available: http://opentsdb.net/.

[98] I. Pelle, T. Lévai, F. Németh and A. Gulyás, "One tool to rule them all: a modular troubleshooting framework for SDN (and other) networks," in *Proceedings of the 1st ACM SIGCOMM Symposium on Software Defined Networking Research (SOSR2015)*, Santa Clara (CA).

[99] C. Meirosu, A. Manzalini, R. Steinert, G. Marchetto, K. Pentikousis, S. Wright, P. Lynch and W. John, "DevOps for Software-Defined Telecom Infrastructures," in *Internet-Draft draft-unify-nfvrg-devops-06*, IRTF NFVRG, July 2016.

[100] I. Society, 2009. [Online]. Available: http://www.internetsociety.org/internet/how-internet-evolving/internet-evolution/data/atlas-internet-observatory-report.

## Annex 1: Mapping final SP-DevOps results on initially defined challenges

*Table 5 Final SP-DevOps results mapped on the Research Challenges described in D4.1 [25]*

| Research Challenges (identified in D4.1) | D4.3 – final SP-DevOps results |
|---|---|
| RC1: Probabilistic in-network monitoring methods | • RateMon (Section 4.5) <br> • Probabilistic delay monitoring (Section 4.6) and E2E delay and loss monitoring (D4.2 [6]) |
| RC2: Scalable observability data transport and processing | • DoubleDecker + MEASURE (Sections 4.10 and 4.3) <br> • Querying aggregated data: RQL + RQE (Section 4.11) |
| RC3: Low-overhead performance monitoring for SDN | • EPLE (Section 4.7) <br> • RateMon (Section 4.5) <br> • Probabilistic delay monitoring (Section 4.6) and E2E delay and loss monitoring (D4.2 [6]) |
| RC4: Novel metrics in counter structures | • MEASURE aggregation point (Section 4.3) <br> • RateMon (Section 4.5) |
| (RC5: Efficient counter retrieval) | This challenge has not been follow-up due to higher engagement in RC2. |
| RC6: Deploy-time functional verification of dynamic Service Graphs | • Integrated verification of NF-FGs (Section 6.1) <br> • Pre-deployment verification tool VeriGraph (Section 4.1) <br> • Network policy watchpointer tool (D4.2 [6]) |
| RC7: Run-time verification of forwarding configurations by enhanced ATPG | • AutoTPG (Section 4.9) <br> • Additional contributions for non-ATPG-based run-time verification: Path Verification (Section 4.2) |
| RC8: Automated troubleshooting workflows | • Troubleshooting framework: EPOXIDE (Section 4.12) <br> • Troubleshooting process descriptions (Section 5.4) <br> • Troubleshooting process for elastic router demo (Section 6.2.3) |
| RC9: In-network troubleshooting | • RateMon (Section 4.5) <br> • Probabilistic delay monitoring (Section 4.6) and E2E delay and loss monitoring (D4.2 [6]) |
| RC10: Troubleshooting with active measurement methods | • Run-time dataplane verification through AutoTPG (Section 4.9) <br> • Programmable TWAMP (Section 4.4) |
| RC11: VNF development support | • VNF developers have access to monitoring, verification troubleshooting processes (Section 5) and tools. Most relevant tools are EPOXIDE (Section 4.12) and the watchpointer tool (D4.2). |
| Additional covered RC (not listed in D4.1): <br> + RC12: Consistent network updates | • Consistent network updates (part of Path Verification, Section 4.2) |
| Additional covered RC (not listed in D4.1): <br> + RC 13: Application-specific network functions | • IPTV Quality Monitor (Section 4.8) |